\documentclass[fleqn,usenatbib]{mnras}

\usepackage{newtxtext,newtxmath}

\usepackage[T1]{fontenc}

\DeclareRobustCommand{\VAN}[3]{#2}
\let\VANthebibliography\thebibliography
\def\thebibliography{\DeclareRobustCommand{\VAN}[3]{##3}\VANthebibliography}

\usepackage{graphicx}	
\usepackage{amsmath}	
\usepackage{orcidlink}
\usepackage[colorinlistoftodos]{todonotes}

\title[Optical and X-ray luminosities of spiders]{A \textit{Gaia} view of the optical and X-ray luminosities of compact binary millisecond pulsars}

\author[K. I. I. Koljonen \& M. Linares]{
Karri I. I. Koljonen$^{\orcidlink{0000-0002-9677-1533}}$,$^{1}$\thanks{E-mail: karri.koljonen@ntnu.no}
Manuel Linares$^{\orcidlink{0000-0002-0237-1636}}$$^{1,2}$\thanks{E-mail: manuel.linares@ntnu.no}
\\
$^{1}$Institutt for Fysikk, Norwegian University of Science and Technology, H{\o}gskloreringen 5, Trondheim, 7491, Norway\\
$^2$ Departament de F{\'i}sica, EEBE, Universitat Polit{\`e}cnica de Catalunya, Av. Eduard Maristany 16, E-08019 Barcelona, Spain.\\
}

\date{Accepted 2023 August 11. Received 2023 August 11; in original form 2023 June 16}

\pubyear{2015}

\begin{document}
\label{firstpage}
\pagerange{\pageref{firstpage}--\pageref{lastpage}}
\maketitle

\begin{abstract}
In this paper, we study compact binary millisecond pulsars with low- and very low-mass companion stars (spiders) in the Galactic field, using data from the latest \textit{Gaia} data release (DR3). We infer the parallax distances of the optical counterparts to spiders, which we use to estimate optical and X-ray luminosities. We compare the parallax distances to those derived from radio pulse dispersion measures and find that they have systematically larger values, by 40\% on average. We also test the correlation between X-ray and spin-down luminosities, finding that most redbacks have a spin-down to X-ray luminosity conversion efficiency of $\sim$0.1\%, indicating a contribution from the intrabinary shock. On the other hand, most black widows have an efficiency of $\sim$0.01\%, similar to the majority of the pulsar population. Finally, we find that the bolometric optical luminosity significantly correlates with the orbital period, with a large scatter due to different irradiated stellar temperatures and binary properties. We interpret this correlation as the effect of the increasing size of the Roche Lobe radius with the orbital period. With this newly found correlation, an estimate of the optical magnitude can be obtained from the orbital period and a distance estimate. 
\end{abstract}

\begin{keywords}
pulsars: general -- stars: distances -- stars: neutron -- X-rays: binaries
\end{keywords}



\section{Introduction}

Compact binary millisecond pulsars are rapidly spinning neutron stars in close orbits with lighter companion stars, with orbital periods $P \lesssim 1$ d. The pulsar’s relativistic wind can strongly irradiate and ablate the companion star. Such a destructive effect of the pulsar on its companion has inspired cannibalistic spider nicknames: black widows (with companion masses $M_{c} \sim 0.01 M_{\odot}$) and redbacks (with ten times larger $M_{c} \sim 0.1 M_{\odot}$). We collectively refer to both types of pulsars as \textit{spiders}. 

While the pulsar wind is invisible, it manifests indirectly by ramming into the ablated companion star wind or interstellar gas producing intrabinary and termination shocks \citep{phinney88,wadiasingh17,wadiasingh18}. Particles at the shocks produce synchrotron emission and inverse-Compton scattering that can be observed in X-rays and potentially in gamma-rays \citep{harding90}. Other sources of X-ray emission in rotation-powered millisecond pulsars can arise from thermal emission from a polar cap region heated by returning particles accelerated in the outer gap region of the neutron star magnetosphere \citep{harding01}. Additionally, non-thermal X-ray emission can arise from synchrotron radiation of the pairs created in the magnetosphere near the pulsar’s light cylinder \citep{cheng98}.

Distances to compact binary millisecond pulsars are challenging to measure. Typically, a distance estimate can be obtained from the radio pulsar's dispersion measure (DM), which quantifies the amount of pulse dispersion in the interstellar medium. This method relies on modeling the electron density distribution in the line-of-sight (the most widely used models being \textsc{NE2001}; \citealt{cordes02} and \textsc{YMW2016}; \citealt{yao17}). It disregards small-scale structures in the interstellar medium, which can affect the distance estimates. Another method is to model the (irradiated) multi-band optical lightcurves to obtain, among other parameters, the source distance, the Roche-lobe filling factor of the companion star, and the strength of the pulsar wind heating at the inner (day-)side of the companion star (some widely used codes being \textsc{ICARUS}; \citealt{breton12}, and \textsc{ELC}; \citealt{orosz00}). A much more robust (model-independent) distance estimate can be obtained via parallax measurements; either using radio timing parallax \citep{smits11} or geometric parallax (essentially through radio interferometry or high-precision astrometric optical measurements). Also, an association with a population with an accurate distance measurement, such as a globular cluster with a typical distance accuracy of 2\% \citep{baumgardt21}, can yield an accurate distance estimate for the pulsar. 

Many pulsar parameters are not directly tied to the distance, such as the ones obtained through pulsar timing (e.g., the spin period, its derivative, and the spin-down power) or photometric and spectroscopic observations (e.g., orbital periods and component masses). However, the observed luminosities depend strongly on the distance, affecting estimates of emission region sizes and other system parameters. Subsequently, these affect, e.g., the physics of the companion star irradiation and the relativistic intrabinary shocks. 

In this article, we compiled the \textit{Gaia} counterparts of Galactic field spiders (Section 3.1) using the latest data release (DR3; Section 2). From the measured parallax, we inferred distance estimates (Section 3.2) and subsequently optical and X-ray luminosities (Sections 3.3 and 3.4) from \textit{Gaia} magnitudes and X-ray fluxes (Section 2.1). We compared these distances to the ones derived using the radio DM and found that they have systematically larger values for many sources (Section 3.2). We also compared the derived X-ray luminosities with spiders found in globular clusters (with accurate distance estimates) and found that their luminosity distributions are similar to the Galactic field sources (Section 3.4). Subsequently, we checked if the derived luminosities correlate with the orbital period and found that the intrinsic optical luminosity significantly correlates with the orbital period (Section 3.3). We discuss the implications of our results in Section 4 and conclude in Section 5. 

\section{Observations}

We searched for the optical counterparts to the currently known spider population in the \textit{Gaia} data release three (DR3; \citealt{gaia16b,gaia22k}). To the list of redbacks presented in \citet{linares21}, we added three confirmed redbacks from recent surveys: PSR J2039$-$5618 from Einstein@Home survey \citep{clark21,corongiu21}, and two from the Trasients and Pulsars with MeerKAT (TRAPUM) survey (PSR J1036$-$4353 and PSR J1803$-$6707; \citealt{clark23}). Also, recently, radio pulsations were found from the redback candidate 3FGL J0212.5$+$5320 \citep{perez23} bringing the number of confirmed redbacks to 19. For the list of redback candidates, we added six redback candidates from the literature \citep{miller20,kennedy20,swihart21,halpern22} including two `Huntsman pulsars' 1FGL J1417.7$-$4407 and 2FGL J0846.0$+$2820 \citep{strader19} bringing the number of redback candidates to 12. From this list, we decided to exclude 4FGL J0744.1$-$2525, since its spin period of 92 msec in gamma-rays\footnote{As noted in the Einstein@Home pulsar discoveries webpage: https://einsteinathome.org/gammaraypulsar/FGRP1\_discoveries.html} indicates that this source would not qualify for most millisecond pulsar definitions (spin periods typically less than 30 msec). The list of confirmed black widows (37) and candidates (4) was taken from \citet{swihart22}. We restricted the search initially to 1\arcsec\/ from the known radio (in case of confirmed pulsars) or optical position (in the case of candidate pulsars). In a few cases, we had to increase the search radius to 2\arcsec\/ in order to find a counterpart (see Section 3.1). Nevertheless, all currently known redbacks, except PSR J1302$-$3258, and all redback candidates have one \textit{Gaia} counterpart, while for most (29 out of 37) of the black widows, we did not find any. Using the \textit{Gaia} archive\footnote{https://gea.esac.esa.int/archive/} we collated the astrometric \citep{lindegren21b} and photometric \citep{riello21} properties of the found counterparts. When available, we also downloaded and analyzed the epoch photometry across the first 34 months of observations. 

\subsection{X-ray observations}

We collected most of the X-ray flux measurements and spectral parameters from the literature. However, for those sources without any reported values, we checked whether there are any available X-ray data in the High Energy Astrophysics Science Archive Research Center (HEASARC) for these sources. In a few available cases, we reduced and analyzed X-ray data from \textit{Chandra} and \textit{XMM-Newton}. In the following, we briefly go through these data reduction methods. 

\subsubsection{Chandra}

We analyzed the available {\it Chandra} observations of PSR~J1628$-$3205 (ID 13725, PI Roberts) and PSR~J1816+4510 (ID 13724, PI Roberts) available at the time of writing, taken on 2012-05-02 and 2012-12-09 with an exposure time of 20~ksec and 34~ksec, respectively. We extracted source and background ACIS-S spectra with {\sc specextract} (v. 14) using circular regions with 5\arcsec\/ and 15\arcsec\/ radii, respectively. After grouping them to a minimum of 15 counts per spectral bin (channel), we fitted the spectra in the 0.5-10~keV band within {\sc Xspec} (v. 12.11.1), with a simple absorbed power law model. In the case of PSR~J1816+4510, because the source is very faint (about 15 net counts), we use the original channels without rebinning in energy and fit the spectrum using {\sc Xspec}'s C-statistics \citep{arnaud96}. In both sources, the absorbing column density is unconstrained, thus we fix it to the value of 1.3$\times$10$^{21}$~cm$^{-2}$ reported by \citet{roberts15}, and 0.3$\times$10$^{21}$~cm$^{-2}$ expected along the line of sight \citep{kalberla05}, respectively, for PSR~J1628$-$3205 and PSR~J1816+4510, and using the abundances from \citet{wilms00}. The best-fit model for PSR~J1628$-$3205 resulted in reduced Chi-squared statistics of 7.1 with nine degrees of freedom, the power law index of $\Gamma=1.8\pm0.2$, and an X-ray flux of $F_{X}=(9.4\pm1.1)\times 10^{-13}$ erg/s/cm$^{2}$. Similarly, for PSR~J1816+4510, we find a power law index of $\Gamma=2.6\pm0.5$ and an X-ray flux of $F_{X}=(3\pm1)\times 10^{-15}$ erg/s/cm$^{2}$ using Cash statistics due to the low number of source counts (C-stat $\simeq$ 7.1 for 15 degrees of freedom).  

\subsubsection{XMM-Newton} \label{sec:xmm}

We reduced the {\it XMM-Newton}/EPIC data of PSR J1431$-$4715 (ID 0860430101, PI de Martino) and PSR J2339$-$0533 (ID 0790800101, PI Tendulkar, and ID 0721130101, PI Kong) with SAS version 20.0.0 using the standard procedures. The observation of PSR J1431$-$4715 started at 2022-05-12 21:03:23, while the two pointings for PSR J2339$-$0533 started on 2013-12-14 at 23:31:51 and on 2016-12-23 at 08:45:11. We filtered the background flares by excluding the times when the 10$-$12 keV count rate surpassed 0.4 cps for EPIC-pn or 0.35 cps for EPIC-MOS. This left a usable exposure time of 83540 seconds for PSR J1431$-$4715, and 44510 and 85710 seconds for the two pointings of PSR J2339$-$0533, respectively. For PSR J1431$-$4715, the EPIC-pn data were taken in the imaging mode with background-subtracted counts of 638$\pm$42 or a count rate of 0.0076$\pm$0.0005 cps. We extracted a light curve with 2.5 ksec bins and found that the count rate remained constant over the exposure (spanning two orbits; see Appendix C, Fig. \ref{fig:J1431_XMM}). We extracted also a spectrum from 0.3 to 10 keV, grouping it so that each bin has a minimum S/N=3 resulting in eight spectral bins. We fitted this spectrum with an absorbed powerlaw model using Cash statistics and subplex optimization. The best-fit model has Cash statistics of 8.5 with six degrees of freedom, with the power law index of $\Gamma=1.2\pm0.2$ and an X-ray flux of $F_{X}=(2\pm0.5)\times10^{-14}$ erg s$^{-1}$ cm$^{-2}$. For PSR J2339$-$0533, both pointings were heavily affected by soft proton radiation from the Earth's magnetosphere, thus rendering the EPIC-pn camera data unusable. Nevertheless, we could partly use the MOS camera data. We analyzed the EPIC-MOS imaging data with background-subtracted counts of 686$\pm$33 (0.0177$\pm$0.0009 cps) and 1667$\pm$50 (0.0176$\pm$0.0005 cps) for the two pointings. We extracted the spectra from both observations similar to those above. Due to being very similar in shape, we combined the spectra using \textsc{epicspeccombine} routine in SAS. We fitted the combined spectrum with an absorbed blackbody and powerlaw components using Chi-squared statistics and Levenburg-Marquardt optimization. The best-fit model has chi-squared statistics of 63.7 with 64 degrees of freedom, the power law index of $\Gamma=1.1\pm0.1$, blackbody temperature of $T_{\mathrm{bb}}=0.14\pm0.02$ keV, and an X-ray flux of $F_{\mathrm{X}}=(1.8\pm0.2)\times10^{-13}$ erg s$^{-1}$ cm$^{-2}$.

\section{Results}

\subsection{\textit{Gaia} counterparts}

\begin{table*} \centering
\footnotesize
  \caption{\textit{Gaia} DR3 location, its distance from the radio position, the probability of chance coincidence, \textit{Gaia} proper motion, \textit{Gaia} parallax and its error, \textit{Gaia} G-band magnitude (square brackets show a range of values for those sources with multiple detections), and orbital period (sources marked with `*' have a period in \textit{Gaia} data) of the Galactic field spiders and candidates.}
  \label{tab:gaiaDR3}
\begin{tabular}{lllllllllll}
\\[-1.8ex]\hline
\hline \\[-1.8ex]
Source name & Gaia source ID & RA & Dec & dR & Pr.$^{a}$ & Pm & Px & Err & G-band & Per \\
& & (deg) & (deg) & (\arcsec) & (\%) & (\arcsec/yr) & (mas) & (mas) & (mag) & (hr) \\
\hline \\[-1.8ex]
\textbf{Redbacks} \\
\hline \\[-1.8ex]
J0212$+$5320   & 455282205716288384  & 33.043636404(5)  & 53.360780822(5)  & --$^{b}$ & -- & 3.3 &  0.86 & 0.02  & [14.2--14.4] & 20.9* \\
J1023$+$0038 & 3831382647922429952 & 155.94870458(2) & 0.64464705(2)    & 0.12 & 2e-3 & 17.9 & 0.69 & 0.07  & [15.8--16.7] & 4.8 \\
J1036$-$4353 & 5367876720979404288 & 159.12589636(6) & -43.88575701(7)  & --$^{c}$   & --   & 12.0 & 0.36 & 0.33  & 19.7 & 6.2 \\
J1048$+$2339 & 3990037124929068032 & 162.18089988(9) & 23.6648310(1)    & 0.04 & 3e-4 & 19.3 & 0.49 & 0.44  & 19.6 & 6.0 \\
J1227$-$4853 & 6128369984328414336 & 186.99465511(3) & -48.89519662(2)  & 0.07 & 6e-3 & 20.1 & 0.46 & 0.13  & 18.1 & 6.9 \\
J1306$-$40   & 6140785016794586752 & 196.73446720(4) & -40.58983303(3)  & 0.33$^{d}$ & 8e-2 & 7.5 & 0.31 & 0.15  & 18.1 & 26.3 \\
J1431$-$4715 & 6098156298150016768 & 217.93588661(2) & -47.25767551(3)  & 0.08 & 0.01 & 18.7 & 0.53 & 0.13  & [17.7--17.8] & 10.8* \\
J1622$-$0315 & 4358428942492430336 & 245.74844491(7) & -3.26035180(5)   & 0.07 & 1e-3 & 13.4 & 0.62 & 0.30  & 19.2 & 3.9 \\
J1628$-$3205 & 6025344817107454464 & 247.02917619(9) & -32.09695036(7)  & 0.12 & 0.05 & 22.3 & 0.68 & 0.41  & [19.3--19.7] & 5.0* \\
J1723$-$2837 & 4059795674516044800 & 260.84658121(1) & -28.632665334(7) & 0.43 & 3.0 & 26.8 & 1.07 & 0.04  & [15.4--15.7] & 14.8* \\
J1803$-$6707 & 6436867623955512064 & 270.76764725(5) & -67.12671046(5)  & --$^{c}$   & --   & 10.6 & 0.18 & 0.27  & 19.4 & 9.1 \\
J1816$+$4510 & 2115337192179377792 & 274.14972586(3) & 45.17606865(3)   & 0.01 & 4e-5 & 4.4 & 0.22 & 0.10  & [18.1--18.3] & 8.7* \\
J1908$+$2105 & 4519819661567533696 & 287.2387173(2)  & 21.0839198(5)    & 0.62 & 2.0 & 8.4 & -2.17 & 1.05 & 20.8 & 3.5 \\
J1910$-$5320 & 6644467032871428992 & 287.7046688(5)  & -53.34920015(5)  & --$^{c}$   & --   & 7.0 & -0.42 & 0.26 & 19.1 & 8.4 \\
J1957$+$2516 & 1834595731470345472 & 299.3942099(1)  & 25.2672362(2)    & 0.03 & 6e-3 & 13.1 & 2.15 & 0.85  & 20.3 & 5.7 \\
J2039$-$5618 & 6469722508861870080 & 309.89570154(3) & -56.28590999(3)  & 0.01 & 1e-4 & 15.7 & 0.49 & 0.17  & 18.5 & 5.4 \\
J2129$-$0429 & 2672030065446134656 & 322.43774948(2) & -4.48522514(2)   & 1.26 & 0.4 & 15.8 & 0.48 & 0.07  & [16.6--17.0] & 15.2* \\
J2215$+$5135 & 2001168543319218048 & 333.88619558(5) & 51.59345464(6)   & 0.10 & 0.03 & 2.2 & 0.30 & 0.23  & [18.7--20.0] & 4.1* \\
J2339$-$0533 & 2440660623886405504 & 354.91143951(4) & -5.55146474(4)   & 0.09 & 7e-4 & 11.0 & 0.53 & 0.18  & [17.8--20.6] & 4.6* \\
\hline \\[-1.8ex]
\textbf{RB candidates} \\
\hline \\[-1.8ex]
J0407.7$-$5702 & 4682464743003293312 & 61.8821756(1)    & -57.0070199(1)   & -- & -- & 1.3 &  -0.04 & 0.42 & 20.1 & -- \\
J0427.9$-$6704 & 4656677385699742208 & 66.95685947(2)   & -67.07640585(2)  & -- & -- & 12.5 &  0.37 & 0.07  & [17.1--18.9] & 8.8 \\
J0523$-$2529   & 2957031626919939456 & 80.820547134(8)  & -25.460313037(9) & -- & -- & 5.4 &  0.45 & 0.04  & 16.5 & 16.5 \\
J0838.8$-$2829 & 5645504747023158400 & 129.71007559(6)  & -28.46582581(8)  & -- & -- & 12.4 &  0.43 & 0.40  & [19.4--20.5] & 5.1* \\
J0846.0$+$2820 & 705098703608575744  & 131.591146456(9) & 28.144662921(5)  & -- & -- & 2.9 &  0.22 & 0.04  & [15.6--15.7] & 195 \\
J0935.3$+$0901 & 588191888537402112  & 143.8363297(4)   & 9.0099717(3)     & -- & -- & 8.2 &  0.71 & 1.12  & 20.6 & 2.5 \\
J0940.3$-$7610 & 5203822684102798592 & 145.09911146(6)  & -76.16670037(6)  & -- & -- & 14.4 &  0.63 & 0.23  & 19.3 & 6.5 \\
J0954.8$-$3948 & 5419965878188457984 & 148.86586954(2)  & -39.79785919(3)  & -- & -- & 10.9 &  0.27 & 0.13  & 18.5 & 9.3 \\
J1417.5$-$4402 & 6096705840454620800 & 214.37735925(1)  & -44.049327679(9) & -- & -- & 7.0 &  0.20 & 0.05  & [15.6--15.9] & 129* \\
J1544$-$1128   & 6268529198286308224 & 236.16412032(5)  & -11.46802219(3)  & -- & -- & 23.4 &  0.39 & 0.18  & 18.6 & 5.8 \\
J2333.1$-$5527 & 6496325574947304448 & 353.31653167(9)  & -55.4391960(1)   & -- & -- & 3.5 &  -0.34 & 0.52 & 20.2 & 6.9 \\
\hline \\[-1.8ex]
\textbf{Black widows} \\
\hline \\[-1.8ex]
J1311$-$3430 & 6179115508262195200 & 197.9405055(3)  & -34.5084379(2) & 0.04 & 4e-4 & 8.0 & 1.93  & 0.97 & 20.4 & 1.56 \\
J1555$-$2908 & 6041127310076589056 & 238.9194106(3) & -29.1412286(2) & 0.003 & 1e-5 & -- & -- & -- & 20.4 & 5.6 \\
J1653$-$0158 & 4379227476242700928 & 253.4085552(2)  & -1.97691527(1) & 0.01 & 6e-5 & 17.6 & 1.77  & 0.78 & 20.4 & 1.25 \\
J1731$-$1847$^{e}$ & 4121864828231575168 & 262.8229115(2)  & -18.7925468(1) & 1.66 & 30.6 & 8.0 & 0.71  & 0.72 & 19.5 & 7.5 \\
J1810$+$1744 & 4526229058440076288 & 272.6553648(1)  & 17.7437108(1)  & 0.11 & 0.01 & 8.6 & 0.65  & 0.54 & 20.0 & 3.6 \\
J1928$+$1245 & 4316237348443952128 & 292.18909203(3) & 12.76481021(4) & 0.18 & 0.4 & 4.6 & 0.15  & 0.17 & 18.2 & 3.3 \\
J1959$+$2048 & 1823773960079216896 & 299.9030930(2)  & 20.8040198(2)  & 0.76 & 3.9 & 32.8 & 1.19  & 1.36 & 20.2 & 9.2 \\
J2055$+$3829 & 1872588462410154240 & 313.7931442(5)  & 38.491495(1)   & 1.62 & 9.8 & -- & -- & -- & 21.0 & 3.1 \\
\hline \\[-1.8ex]
\textbf{BW candidates} \\
\hline \\[-1.8ex]
J0336.0$+$7505 & 544927450310303104 & 54.0424214(5) & 75.0547967(3) & -- & -- & 9.9 & -0.85 & 1.64 & 20.6 & 3.7 \\
J0935.3$+$0901 & 588191888537402112 & 143.8363297(4) & 9.0099717(3) & -- & -- & 8.2 & 0.71 & 1.12 & 20.6 & 2.4 \\
J1406$+$1222$^{e}$ & 1226507282368609152 & 211.7342086(2) & 12.3786882(3) & -- & -- & 74.5 & 2.11 & 1.59 & 20.1 & 1.0 \\
\hline \\[-1.8ex]
\multicolumn{11}{|p{0.95\linewidth}|}{\textit{$^{a}$} We estimated the chance coincidence by taking the number of \textit{Gaia} sources inside an arcminute radius from the source in question and calculating the probability that they would randomly coincide with the radio position.} \\
\multicolumn{11}{|p{0.95\linewidth}|}{\textit{$^{b}$} The radio position was fixed to the location of the \textit{Gaia} source in \citet{perez23}.} \\
\multicolumn{11}{|p{0.95\linewidth}|}{\textit{$^{c}$} For these sources found in the TRAPUM survey, the radio positional accuracy is a few arcseconds containing the \textit{Gaia} source.} \\
\multicolumn{11}{|p{0.95\linewidth}|}{\textit{$^{d}$} PSR J1306$-$40 has a very uncertain radio position, and the value displayed corresponds to the  distance from the optical counterpart found by \citet{linares18a}.} \\
\multicolumn{11}{|p{0.95\linewidth}|}{\textit{$^{e}$} These sources exhibit significant astrometric noise and therefore the \textit{Gaia} parameters are not reliable.}   
\end{tabular}
\end{table*}

All currently known redbacks and redback candidates except PSR J1302$-$3258 have a \textit{Gaia} counterpart. Most of the known black widows do not have a \textit{Gaia} counterpart. This is not surprising as the optical magnitudes of black widow counterparts are known to be much fainter due to cooler companion stars. The \textit{Gaia} DR3 counterparts for the Galactic field spiders are listed in Table~\ref{tab:gaiaDR3}. Black widows PSR J1555$-$2908 and PSR J2055$+$3829 have a \textit{Gaia} counterpart but no reported parallax. All counterparts exhibit astrometrically good fits\footnote{Defined as the \textit{Gaia} archive goodness-of-fit parameter, \textsc{astrometric\_gof\_al}, less than three.}, except for the black widow PSR J1731$-$1847 and the black widow candidate ZTF J1406$+$1222. Consequently, in these two cases, the \textit{Gaia} parameters are considered unreliable and we exclude these sources from subsequent analyses. Furthermore, apart from the aforementioned two sources, only PSR J1816$+$4510 exhibit significant astrometric excess noise\footnote{Defined as the \textit{Gaia} archive parameter, \textsc{astrometric\_excess\_noise\_sig}, greater than two.}. Therefore, the orbital wobble of the companion stars does not significantly affect the astrometric solutions for most systems studied here, possibly due to the distances being approximately 2 kpc or greater \citep{gandhi22}. 


For known spiders, the median distance from the radio counterpart is 0.1\arcsec (radio timing locations are from ATNF \textsc{psrcat}\footnote{https://www.atnf.csiro.au/research/pulsar/psrcat/}). Three redbacks that were identified in the TRAPUM survey\footnote{http://www.trapum.org/discoveries} by \citet{clark23} and \citet{au23} have a radio position accuracy of a few arcseconds with one \textit{Gaia} source in the beam. Six \textit{Gaia} candidates have larger than 0.2\arcsec\/ separation from the radio position: PSR J1306$-$40, PSR J1723$-$2837, PSR J1908$+$2105, PSR J2129$-$0429, PSR J1959$+$2048, and PSR J2055$+$3829. PSR J1306$-$40 has a very uncertain radio position \citep[7\arcsec;][]{keane18}, but \citet{linares18a} found an optical counterpart candidate with the orbital period found from radio timing (in fact, the ATNF location corresponds to the optical coordinates). The separation of the \textit{Gaia} source to this optical candidate is 0.33\arcsec, but the location of the optical candidate has an uncertainty of 0.2\arcsec. Redbacks PSR J1723$-$2837 and PSR J2129$-$0429 have \textit{Gaia} epoch photometry with detected periods corresponding to the one given by the radio timing observations (see Appendix D) confirming the association (although we note that the period peak for PSR J1723$-$2837 is significant at $\sim 2\sigma$ level). Therefore, we deem these three \textit{Gaia} counterparts as the likely optical counterparts of these pulsars. This leaves the redback PSR J1908$+$2105 and the two black widows only as tentative counterparts (PSR J1959$+$2048 has a high proper motion). PSR J1908$+$2105 is a peculiar source with a very low mass companion for a redback \citep{cromartie16,deneva21}. In addition, the black widow PSR J1928$+$1245 has a counterpart that is two magnitudes brighter than the others. This source lies in a crowded field, so there is a chance that another star is in the line of sight, although we estimate this to be relatively small (0.4\%). We estimated this chance coincidence by taking the number of \textit{Gaia} sources inside an arcminute radius from the source and calculated the probability that they would randomly coincide with the radio position (see Table \ref{tab:gaiaDR3}).

\subsubsection{Phase-folded light curve of PSR J1816$+$4510}

\begin{figure}
    \includegraphics[width=1.0\columnwidth]{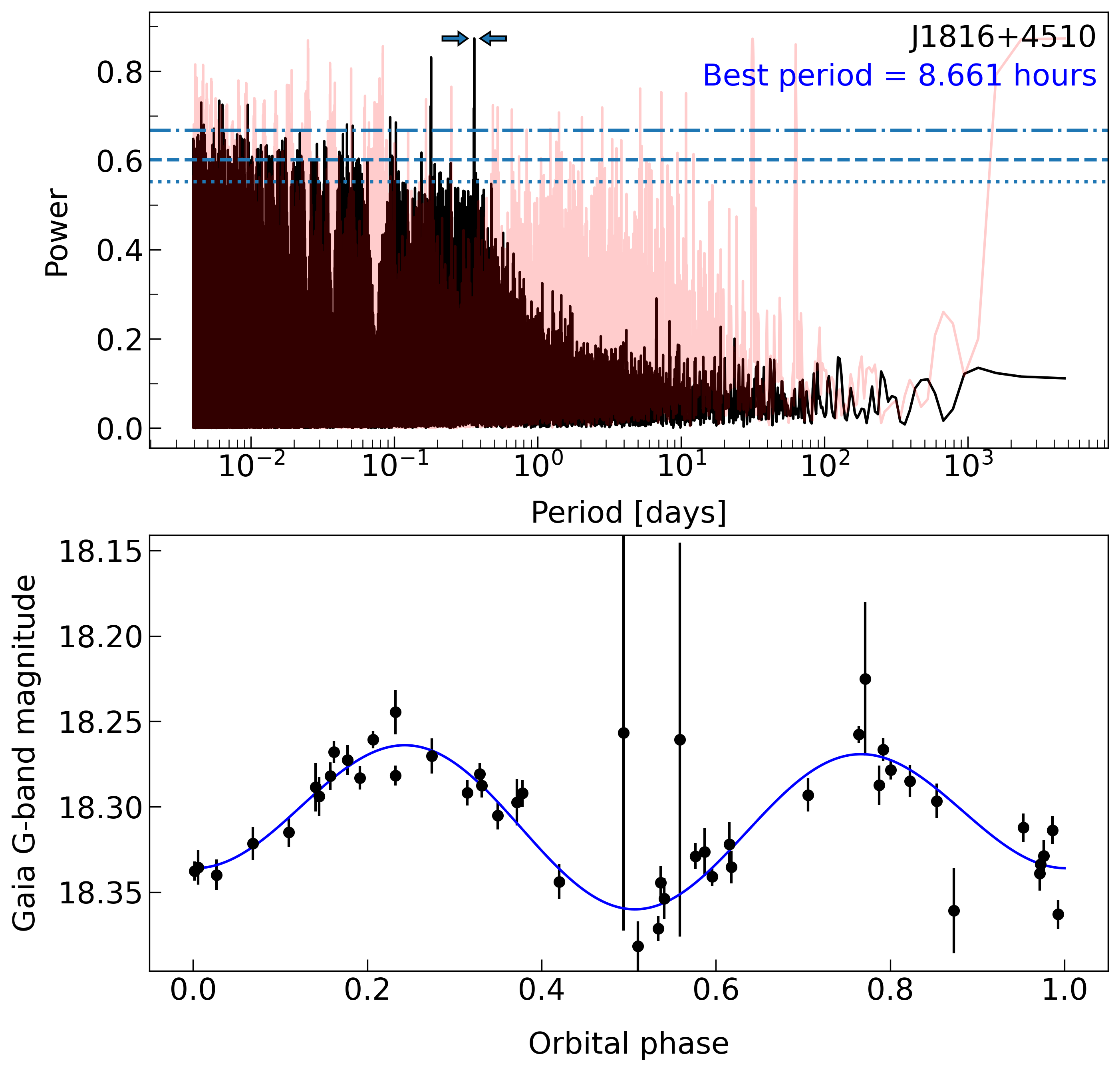}
    \caption{\textit{Upper panel:} The Lomb-Scargle periodogram of PSR J1816$+$4510 (black line) with the periodogram of the window function (red transparent line) and the best period indicated with arrows. The horizontal dotted, dashed, and dot-dashed lines mark the local false alarm probability levels of 5\%, 1\%, and 0.1\%, respectively. \textit{Lower panel:} \textit{Gaia} G-band light curve folded at the best period. The solid line is the least squares fit of the two-term Fourier model to the data.}
    \label{fig:gaia_epoch_1816}
\end{figure}

For all the sources with \textit{Gaia} epoch photometry, we performed a period search (see Appendix D). Most sources already have similar or better quality optical light curves presented in the literature apart from PSR J1816$+$4510. In this case, the light curve presented by \citet{kaplan12} barely shows any variability over the orbital phase. In Fig. \ref{fig:gaia_epoch_1816} (upper panel), we show the Lomb-Scargle periodogram of the \textit{Gaia} epoch photometry of PSR J1816$+$4510 (black line) together with the periodogram of the window function (red, transparent line) and the best period indicated with arrows. In addition, we plot the local false alarm probabilities at the period peak (5\%, 1\%, and 0.1\% probability levels marked as dotted, dashed, and dot-dashed horizontal lines) assuming a null hypothesis of non-varying data with Gaussian noise and the probability based on bootstrap resamplings of the input data. The period that we find (8.661 hours) matches well with the one obtained from radio timing analysis \citep{stovall14}. The bottom panel of Fig. \ref{fig:gaia_epoch_1816} shows the light curve folded at this period. The phase-folded light curve shows two maxima at phases 0.25 and 0.75 indicating ellipsoidal modulation of the companion star radiation with an amplitude of $\sim$0.1 mag.

\subsection{Distance estimates of the Galactic field spiders}

\begin{figure*}
    \includegraphics[width=2.0\columnwidth]{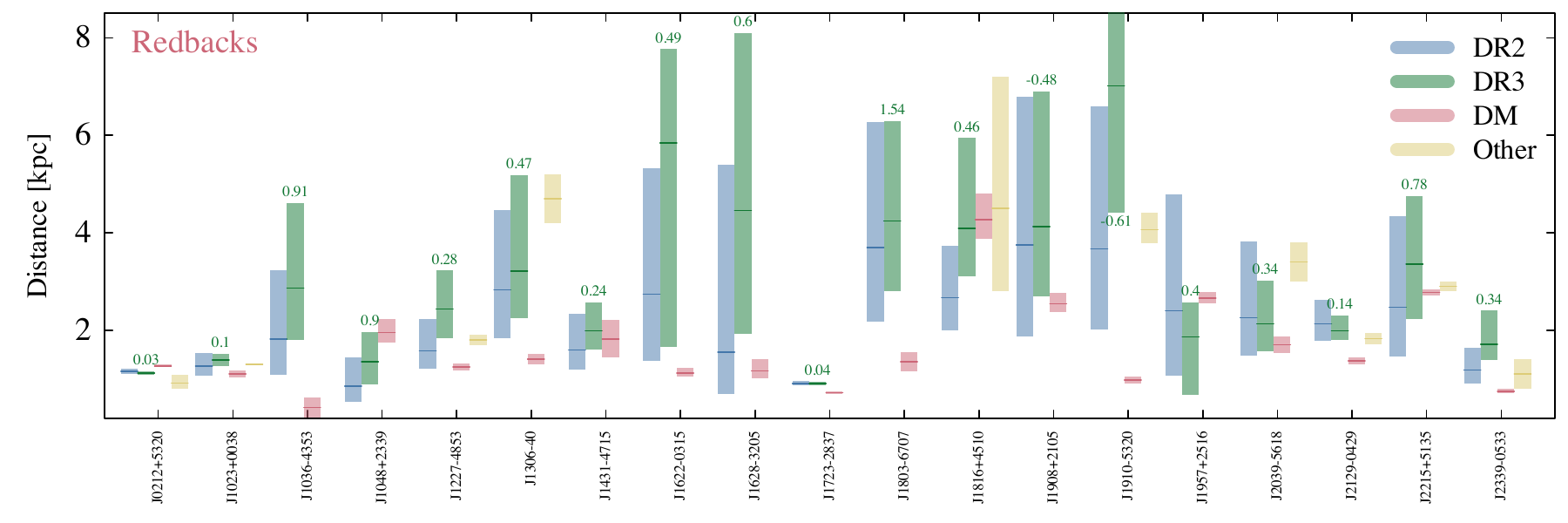}
    \caption{The DM, \textit{Gaia} parallax, and other distance estimates (radio parallax, optical studies) for Galactic field redbacks are shown side-by-side. An error-to-parallax-ratio is marked for each \textit{Gaia} DR3 distance estimate. All values are tabulated in Table~\ref{tab:distance}.}
    \label{fig:distance_rb}
\end{figure*}

\begin{figure*}
    \includegraphics[width=2.0\columnwidth]{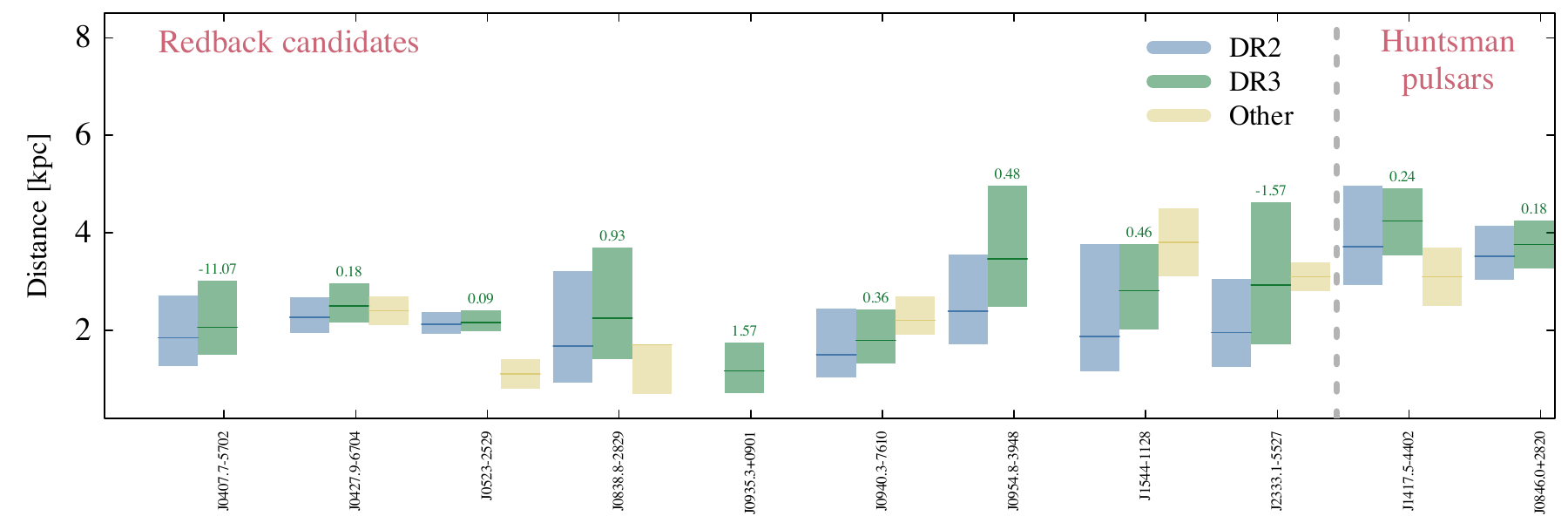}
    \caption{The parallax and other distance estimates (optical studies) for Galactic field  redback candidates shown side-by-side. So-called \textit{Huntsman pulsars} with massive companion stars are delineated. An error-to-parallax-ratio is marked for each \textit{Gaia} DR3 distance estimate. All values are tabulated in Table~\ref{tab:distance_cand}.}
    \label{fig:distance_rbc}
\end{figure*}

\begin{figure*}
    \includegraphics[width=1.3\columnwidth]{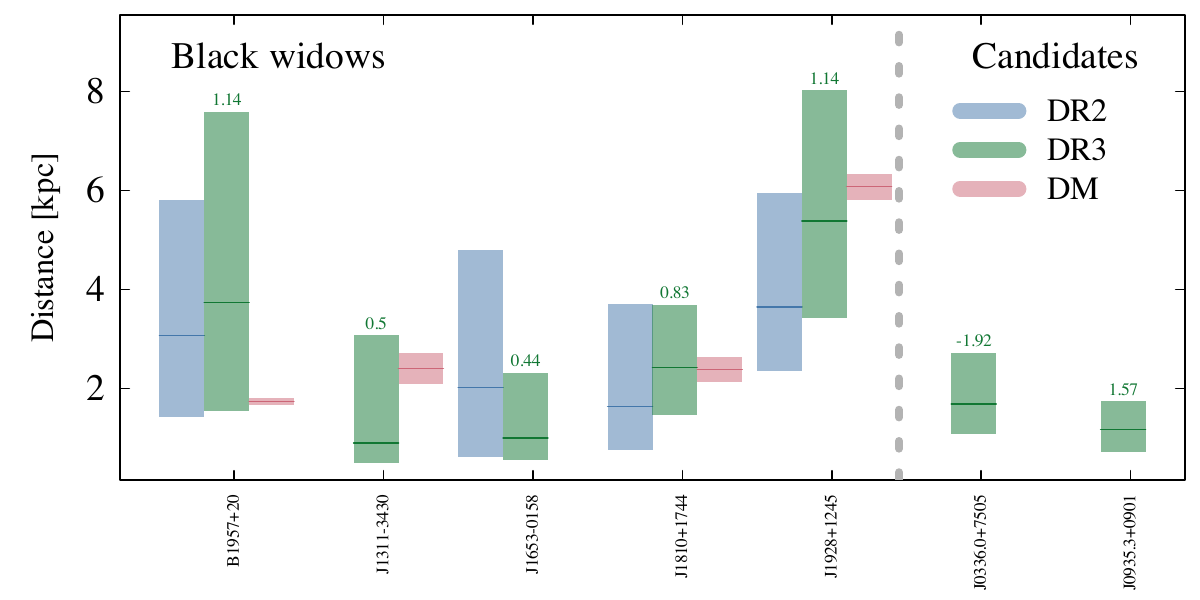}
    \caption{DM and \textit{Gaia} parallax distance estimates for Galactic field black widows and candidates (last two) shown side-by-side. Error-to-parallax-ratio is marked for the DR3 distance. Values are tabulated in Tables~\ref{tab:distance} and \ref{tab:distance_cand}.}
    \label{fig:distance_bw}
\end{figure*}

We calculate the geometric distance estimates using \textit{Gaia} DR3 parallaxes and the distance prior from \citet{bailerjones21} (see Appendix A for details and a comparison with an alternative prior, which employs the model distribution of Galactic pulsars as proposed by \citealt{lorimer06}). We also compare the current values to those obtained in DR2 (and using the distance prior from \citealt{bailerjones18}) and other methods such as radio parallax and optical studies. Figs.~\ref{fig:distance_rb}--\ref{fig:distance_bw} show the distance estimates for spiders and spider candidates. All the distance values with 68\% confidence intervals (quantiles 0.159 and 0.841) are tabulated in Appendix B (Table \ref{tab:distance} for the Galactic field spiders and in Table \ref{tab:distance_cand} for spider candidates).

Seven sources have precise \textit{Gaia} parallaxes with error-to-parallax-ratios lower than 0.2 (see Figs. \ref{fig:distance_rb} and \ref{fig:distance_rbc}), thanks to a combination of small distances and bright companion stars. In these seven cases, the distances obtained using the \textit{Gaia} data releases DR2 and DR3 agree within 10\%. In other cases, the \textit{Gaia} parallax has become smaller when going from DR2 to DR3. However, the error on the parallax has not decreased in proportion, making the DR3 posterior distribution wider in many cases. The prior has also changed slightly between \citet{bailerjones18} and \citet{bailerjones21}. In the former, they used a one-parameter exponential decreasing space density distance prior. In the latter, this is expanded to a more flexible three-parameter generalized gamma distribution where the previous prior is a special case (see Appendix A). 

We collected the up-to-date DMs of spiders from the Australia Telescope National Facility (ATNF) and TRAPUM web pages. We calculated the distance estimates from the DM using \textsc{PSRdist}\footnote{https://github.com/tedwards2412/PSRdist} \citep{bartels18} and the Galactic electron density model of \citet{yao17}. \textsc{PSRdist} calculates the errors on the distances to the pulsars assuming that the Galactic electron density model parameters of \citet{yao17} are uncertain and performs a grid scan to produce an array of distances to each source. Out of this distance distribution, we take the highest peak and 68\% confidence interval around the peak to mark the 1$\sigma$ error on the most probable distance.

\begin{figure}
    \includegraphics[width=\columnwidth]{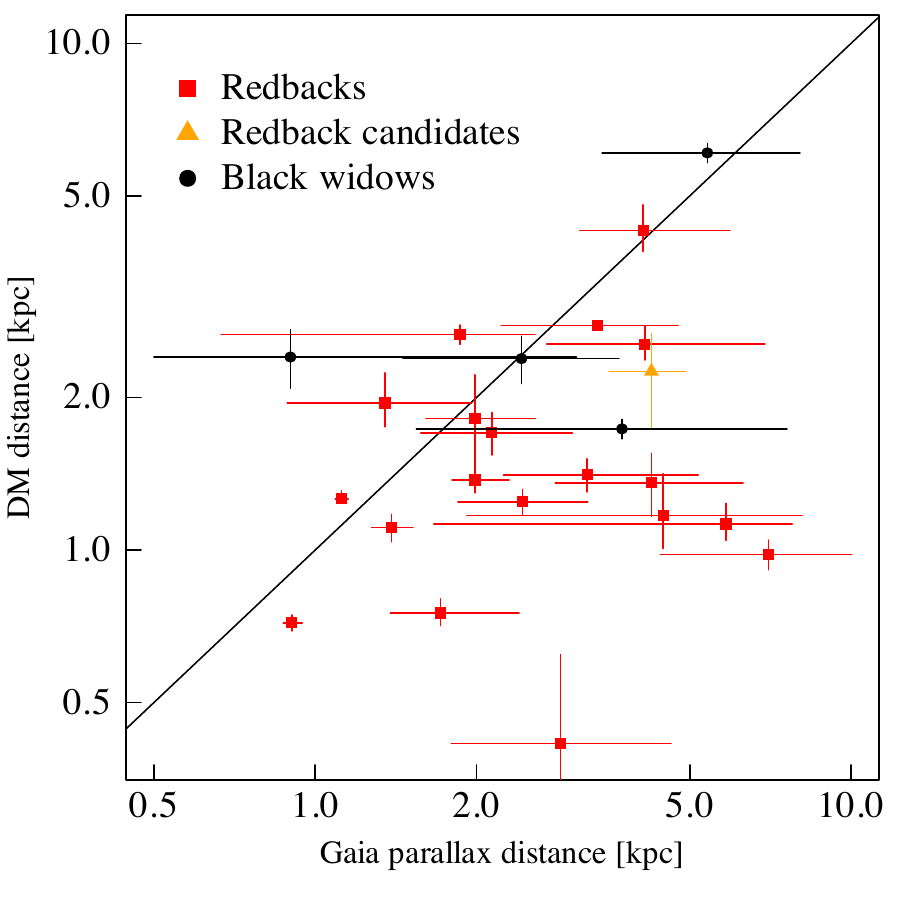}
    \caption{Comparison of the DM distance and \textit{Gaia} DR3 parallax distance estimates for the Galactic field spiders. DM distances tend to have systematically lower values.}
    \label{fig:DM_vs_DR3}
\end{figure}

\begin{figure}
    \includegraphics[width=\columnwidth]{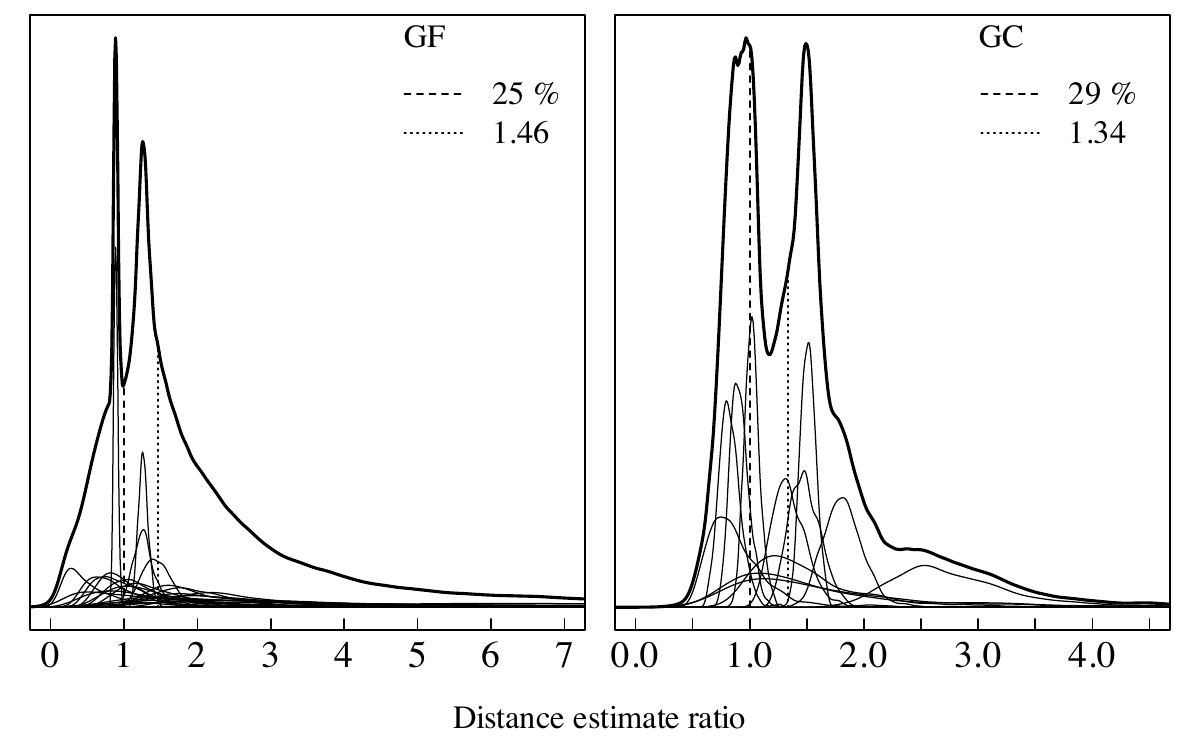}
    \caption{KDEs for the ratios of the derived \textit{Gaia} DR3 and DM distances over the Galactic field (\textit{left}) and globular cluster spiders (\textit{right}). The solid lines show the KDEs for single sources; the thick solid line is their sum. Vertical dashed lines mark the ratio of unity with the legend showing the percentage of the area of the summed KDE below it (for a ratio randomly distributed around unity, this would be $\sim$50\%). The dotted vertical lines show the mean of the summed KDE with the corresponding ratio in the legend.}    
    \label{fig:dist_ratio}
\end{figure}    

Comparing the DM distances to the DR3 parallax distances shows that DM distances have systematically lower values (Table~\ref{tab:distance}; Fig.~\ref{fig:DM_vs_DR3}). This was already noticed by \citet{jennings18} for Galactic binary pulsars, although the effect was somewhat reduced when using the \citet{yao17} electron density model as opposed to using an older model. In particular, for PSR J1023$+$0038, the Gaia DR3 parallax distance agrees well with the radio parallax distance from VLBA \citep{deller12}, while the DM distance estimate is lower. Similarly, for PSR J1227$-$4853, PSR J1306$-$40, and PSR J2129$-$0429, the DR3 parallax distance agrees well with the distance estimates from optical studies \citep{demartino14,swihart19,bellm16}. These are all bright sources with optical magnitudes of $\sim$16--18 and error-to-parallax-ratios below 0.5.   

To gauge the magnitude of the systematic error on the DM-derived distances, we formed kernel density estimates (KDEs) of the derived DR3 and DM distance ratios. First, we sample the DR3 geometric distance and DM distributions to form distance ratio distributions (DR3/DM) for each Galactic field spider. Out of these, we form the KDEs, including a combined one that is plotted in Fig.~\ref{fig:dist_ratio} (left panel). The combined KDE has a majority (75\%) of factors above unity with a 50\% quantile at 1.46. We also calculated the DM distances of globular cluster spiders and compared them to their literature distance from the compilation of \citet{baumgardt21}. They estimated the distances to globular clusters using various methods based on \textit{Gaia} EDR3 data, but essentially all methods are within 2\% of each other. Here, we use their mean distance values. The results are similar to the Galactic field sources, with the combined KDE having 71\% of factors above unity and the 50\% quantile at 1.34. This means that, on average, the DM distance estimate is a factor of $\sim$1.4 lower than the parallax distance. If this is an effect of the Galactic electron density model, it would imply low-density voids in the line of sight. We did not find any correlation between the distance ratio and the Galactic coordinates. 

\subsection{Orbital period -- optical luminosity correlation in spiders}

\begin{table*} \centering
  \caption{Orbital period -- optical luminosity correlations for different distance measurements used (\textit{Gaia} DR3 or DM). The columns display the median values and their 1$\sigma$ errors for Pearson correlation coefficients, the probability of no correlation, and the parameters for the best-fit least squares regression lines for each sample.}
  \label{tab:spider_orb_lum}
\begin{tabular}{@{\extracolsep{5pt}} llccl}
\\[-1.8ex]\hline
\hline \\[-1.8ex]
Distance method & Sample & Pearson corr. & Null & Least squares regression \\
\hline \\[-1.8ex]
(DR3) & Full sample & 0.71 & 3$\times 10^{-5}$ & log(L/(erg s$^{-1}$)) = 1.34($\pm$0.22) $\times$ log(P/hr) + 31.80($\pm$0.28) \\
             & & (0.57, 0.78) \\
             & Parallax-selected sample$^{a}$ & 0.835 & 7$\times 10^{-6}$ & log(L/(erg s$^{-1}$)) = 1.34($\pm$0.21) $\times$ log(P/hr) + 31.78($\pm$0.28) \\
             & & (0.72, 0.87) \\
(DM) & Full sample & 0.50 & 0.035 & log(L/(erg s$^{-1}$)) = 0.94($\pm$0.32) $\times$ log(P/hr) + 31.82($\pm$0.41) \\
             & & (0.20, 0.61) \\
             & Outliers removed$^{b}$ & 0.675 & 0.0011 & log(L/(erg s$^{-1}$)) = 1.12($\pm$0.19) $\times$ log(P/hr) + 31.64($\pm$0.24) \\
             & & (0.48, 0.73) \\   
\hline \\[-1.8ex]
\multicolumn{5}{|p{0.6\linewidth}|}{\textbf{Notes:} $^{a}$Sources with error-to-parallax-ratio between $0<\Delta\omega/\omega<1$.} \\ 
\multicolumn{5}{|p{0.6\linewidth}|}{\hspace{0.825cm} $^{b}$PSR J1036$-$4353 and PSR J1928$+$1245 excluded.} \\
\end{tabular}
\end{table*}

\begin{figure}
    \includegraphics[width=\columnwidth]{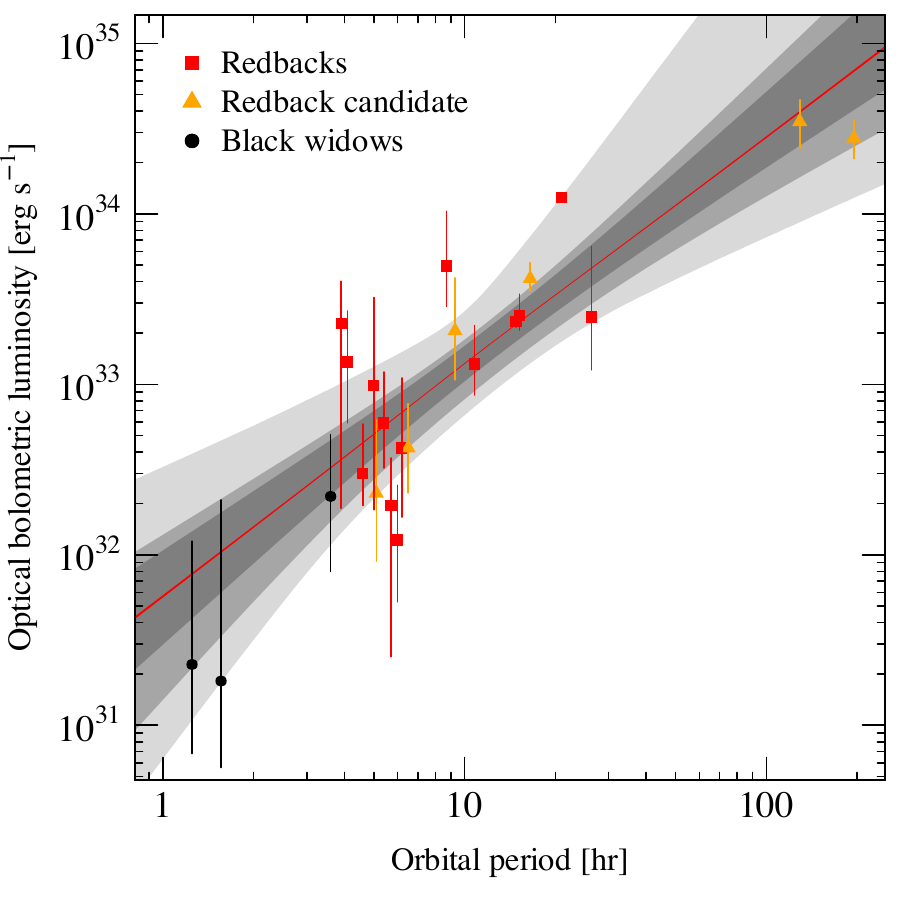}
    \caption{Optical bolometric luminosity derived from \textit{Gaia} G-band magnitude as a function of the orbital period of the Galactic field spiders with the error-to-parallax-ratio between 0 and 1. The red line shows the best-fit linear regression of the data and the shaded grey regions show its 95\% (darkest), 99\%, and 99.9\% (lightest) confidence intervals.}
    \label{fig:spider_orb_opt}
\end{figure}

To derive the intrinsic optical magnitudes of Galactic field spiders, we used the average \textit{Gaia} magnitudes dereddened using estimates for V-band extinction (A$_V$) from \citet{amores05,amores21} and transferred to \textit{Gaia} passbands using the description presented in \citet{riello21}. Then we converted the dereddened magnitudes to absolute magnitude distributions and subsequently to luminosity distributions by sampling either the \textit{Gaia} DR3 parallax or DM posterior distances, using temperature-dependent bolometric corrections from \citet{creevey22} which are derived using the MARCS synthetic stellar spectra \citep{gustafsson08}. The temperatures were either taken from the literature or the \textit{Gaia} archive or if these were not available we used a solar temperature (see Section \ref{sec:model} and Table \ref{tab:spider_temp}). In any case, the bolometric correction is small ($\pm$0.1 mag) for the temperature range 4700-9000 K.       

Fig. \ref{fig:spider_orb_opt} shows the bolometric luminosity derived using the \textit{Gaia} G-band magnitudes and the DR3 distances of those Galactic field spiders that have error-to-parallax-ratio between zero and one plotted as a function of the orbital period. We implemented this range based on the recommendation provided by \citet{bailerjones21}. They point out that sources with a negative error-to-parallax ratio or a ratio greater than one are typically dominated by priors. In such cases, geometric distances might not provide particularly meaningful results. Here, we have also excluded the sources suggested to be transitional millisecond pulsars (namely PSR J1023$+$0038, PSR J1227$-$4853, 3FGL J0427.9$-$6704, 3FGL J0407.7$-$5702, and 3FGL J1544.6$-$1125). We discover a clear correlation between the optical luminosity and orbital period, with a Pearson correlation coefficient of 0.835 and a $7 \times 10^{-6}$ probability of no correlation. The best-fit linear least squares regression of the sample is log(L/erg s$^{-1}$) = 1.34($\pm$0.21) $\times$ log(P/hr) + 31.78($\pm$0.28). We find also statistically significant correlations between the G-band \textit{Gaia} luminosity and the orbital period using the full sample or using the DM-derived distances (see Table \ref{tab:spider_orb_lum}). In the case of using the DM-derived distances, we excluded the black widow PSR J1928$+$1245 and the redback PSR J1036$-$4353 as the former has an anomalously luminous counterpart and the latter has very low DM distance improving the overall significance (see Table \ref{tab:spider_orb_lum}). We also tested the correlations using different \textit{Gaia} filters with similar results. Given the above correlations, an estimate of the dereddened magnitude can be obtained knowing the distance and the orbital period (neglecting bolometric correction) via 

\begin{equation}
m_{\mathrm{G}} \approx M_{\odot, \mathrm{bol}} - 2.5 \, \mathrm{log} \frac{L/L_{\odot}}{(d/10)^2} \, ,
\end{equation}

\noindent where $L$ is given in erg/s using the least squares regression results from Table \ref{tab:spider_orb_lum} and $d$ in parsecs. We derive using $M_{\odot, \mathrm{bol}}$=4.74 mag and $L_{\odot}$=3.85$\times10^{33}$ erg/s a slightly simplified version:

\begin{equation}
m_{\mathrm{G}} \approx 16.26 + 2.5 \, \mathrm{log} \frac{d_{\mathrm{kpc}}^2}{L_{33}} \, ,
\end{equation}

\noindent where $d_{\mathrm{kpc}}$ is given in kiloparsecs and $L_{33}$ in $10^{33}$ erg/s, respectively. Inserting to the above equation, e.g., the luminosity estimate from the least squares regression using the \textit{Gaia} DR3 distances of the parallax-selected sample (Table \ref{tab:spider_orb_lum}), gives an estimate for the expected spider G-band magnitude given the distance and orbital period:

\begin{equation}
m_{\mathrm{G}} \approx 19.3(\pm 0.7) + 2.5 \, \mathrm{log} \frac{d_{\mathrm{kpc}}^2}{P_{\mathrm{hr}}^{1.34(\pm0.21)}} \, .
\end{equation}

In summary, we discover a correlation between the optical luminosity and orbital period of spiders, which we further discuss and interpret in Section \ref{sec:model}.   

\subsection{X-ray properties of the Galactic field and globular cluster spiders}

\begin{table*} \centering
  \caption{Properties of spiders in globular clusters.}
  \label{tab:gc_spider_x}
\begin{tabular}{@{\extracolsep{5pt}} llccccccccl}
\\[-1.8ex]\hline
\hline \\[-1.8ex]
Cluster & Source name & Type & Cl. dist.$^{a}$ & Per. & DM & DM dist. & $\dot{E}$ & $L_{X}^b$ & $\Gamma$ & Ref$^{d}$ \\
& & & (kpc) & (hr) & (cm$^{-3}$ pc) & (kpc) & (10$^{34}$ erg/s) & (10$^{31}$ erg/s) \\
\hline \\[-1.8ex]
47Tuc    & J0024$-$7204I  & BW  & 4.52(3) & 5.5 & 24.43 & 2.5(2) & -4.3 & 0.5$\pm$0.1 & -- & 1 \\
47Tuc    & J0024$-$7204J  & BW  & 4.52(3) & 2.9 & 24.5932 & 2.6(3) & -4.2 & 1.3$\pm$0.4 & 1.0$\pm$0.6 & 1 \\
47Tuc    & J0024$-$7204O  & BW  & 4.52(3) & 3.3 & 24.356 & 2.5(2) & 6.5 & 1.0$\pm$0.3 & 1.3$\pm$0.8 & 1 \\
47Tuc    & J0024$-$7204R  & BW  & 4.52(3) & 1.6 & 24.361 & 2.5(2) & 13.9 & 0.6$\pm$0.1 & -- & 1 \\
47Tuc    & J0024$-$7204W  & RB  & 4.52(3) & 3.2 & 24.367 & 2.5(2) & -26.3 & 3.3$\pm$0.5 & 1.2$\pm$0.2 & 2 \\
OCen     & J1326$-$4728B  & BW  & 5.43(5) & 2.2 & 100.273 & 6.6(1.7) & -2.0 & 0.9$\pm$0.3 & 2.6$\pm$0.5 & 3 \\
M5       & J1518$+$0204C  & BW  & 7.48(6) & 2.1 & 29.3146 & 4.7(2.8) & 6.7 & 0.8$\pm$0.3 & 4.3$\pm$0.8 & 3 \\
M13      & J1641$+$3627E  & BW  & 7.42(8) & 2.7 & 30.54 & 5.3(2.7) & 4.5 & 1.1$\pm$0.4 & 2.2$\pm$0.6 & 4 \\
M22      & J1836$-$2354A  & BW  & 3.30(4) & 4.9 & 89.107 & 3.2(2) & 0.2 & 0.4$\pm$0.1 & 1.5$\pm$0.7 & 5 \\
M28      & J1824$-$2452G  & BW  & 5.37(10) & 2.5 & 119.4 & 3.6(3) & 3.4 & 0.17$\pm$0.06 & 3.5$\pm$0.7 & 6 \\
M28      & J1824$-$2452H  & RB  & 5.37(10) & 10.4 & 121.5 & 3.8(3) & 3.3 & 2.3$\pm$0.4 & 1.0$\pm$0.2 & 6 \\
M28      & J1824$-$2452I  & RB  & 5.37(10) & 11.0 & 119 & 3.6(3) & -- & 14$\pm$3 & 1.1$\pm$0.2 & 6 \\
M28      & J1824$-$2452J  & BW  & 5.37(10) & 2.3 & 119.2 & 3.6(3) & -4.5 & 0.5$\pm$0.1 & 1.0$\pm$0.8 & 6 \\
M28      & J1824$-$2452M  & BW  & 5.37(10) & 5.8 & 119.35 & 3.6(3) & 4.4 & 0.3$\pm$0.07 & 3.6$\pm$1.3 & 6 \\
M30      & J2140$-$2310A  & RB  & 8.46(9) & 4.2 & 25.0630 & 3.1(6) & -0.2 & 0.7$\pm$0.3 & 2.9$\pm$0.9 & 7 \\
M62      & J1701$-$3006B  & RB  & 6.41(10) & 3.5 & 115.21 & 4.8(5) & -29.8 & 10$\pm$2 & 1.9$\pm$0.5$^{c}$ & 8 \\
M71      & J1953$+$1846A  & BW  & 4.00(5) & 4.2 & 117 & 4.4(4) & 1.6 & 1.1$\pm$0.2 & 1.9$\pm$0.3 & 9 \\
M92      & J1717$+$4308A  & RB  & 8.50(7) & 4.8 & 35.45 & 6.1(1.8) & 7.7 & 7.5$\pm$1.4 & 1.8$\pm$0.3 & 3 \\
NGC 6397 & J1740$-$5340A  & RB  & 2.48(2) & 32.5 & 71.8 & 3.0(3) & 13.6 & 2.4$\pm$0.3 & 1.7$\pm$0.1 & 10 \\
NGC 6397 & J1740$-$5340B  & RB  & 2.48(2) & 47.5 & 72.2 & 3.1(3) & -0.1 & 8.08$\pm$0.07 & 1.38$\pm$0.05 & 10 \\
Ter5     & J1748$-$2446A  & RB  & 6.62(15) & 1.8 & 242.15 & 4.4(2) & -0.07 & 13$\pm$4  & 1.2$\pm$0.7 & 11 \\
Ter5     & J1748$-$2446O  & BW  & 6.62(15) & 6.2 & 236.38 & 4.4(2) & -57.8 & 2.8$\pm$0.5 & 1.5$\pm$0.5 & 11 \\
Ter5     & J1748$-$2446P  & RB  & 6.62(15) & 10.6 & 238.79 & 4.4(2) & 198.7 & 53$\pm$2 & 0.72$\pm$0.08 & 11 \\
Ter5     & J1748$-$2446ad & RB  & 6.62(15) & 26.3 & 235.6 & 4.4(2) & -49.3 & 20.3$\pm$1.4 & 1.11$\pm$0.12 & 11 \\
\hline \\[-1.8ex]
\multicolumn{10}{|p{0.9\linewidth}|}{\textit{$^{a}$} Cluster distance estimates from \citet{baumgardt21}.} \\
\multicolumn{10}{|p{0.9\linewidth}|}{\textit{$^{b}$} Unabsorbed 0.5--10 keV luminosity at the cluster distance.} \\
\multicolumn{10}{|p{0.9\linewidth}|}{\textit{$^{c}$} Error not given in the reference.} \\
\multicolumn{10}{|p{0.9\linewidth}|}{\textit{$^{d}$} \textbf{References:} 1) \citet{bogdanov06}, 2) \citet{hebbar21} 3) \citet{zhao22}, 4) \citet{zhao21} 5) \citet{amato19}, 6) \citet{vurgun22}, 7) \citet{zhao20}, 8) \citet{oh20}, 9) \citet{elsner08}, 10) \citet{bogdanov10}, 11) \citet{bogdanov21}.}
\end{tabular}
\end{table*}

\begin{figure}
    \includegraphics[width=\columnwidth]{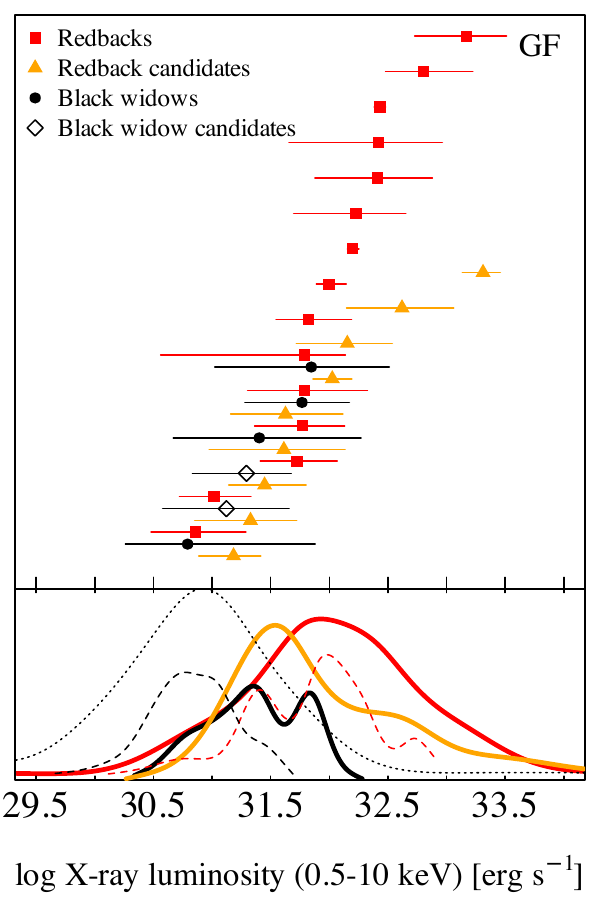}
    \caption{\textit{Top:} Unabsorbed X-ray luminosities (0.5--10 keV) of the Galactic field spiders (red squares: redbacks, orange triangles: redback candidates, black circles: black widows, black diamond: black widow candidates). \textit{Bottom:} Combined KDEs for the Galactic spider population X-ray luminosities (solid lines). Dashed lines show the X-ray luminosity KDEs for globular cluster populations shown in Fig. \ref{fig:spider_lum_gc}). Dotted black lines mark the black widow X-ray luminosity KDE calculated from luminosities given in \citet{swihart22}.}
    \label{fig:spider_lum_gal}
\end{figure}

\begin{figure}
    \includegraphics[width=\columnwidth]{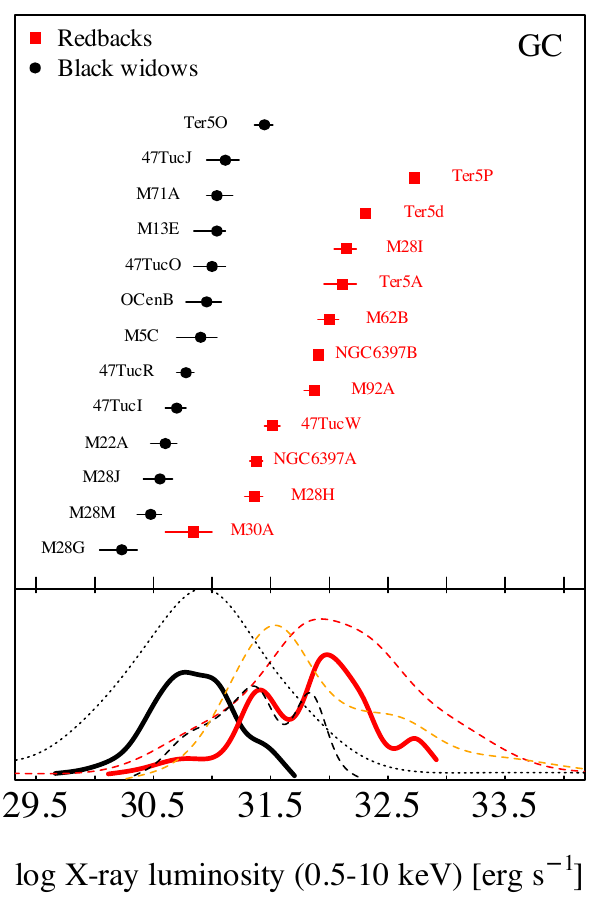}
    \caption{\textit{Top:} Unabsorbed X-ray luminosities (0.5--10 keV) of globular cluster spiders (red squares: redbacks, black circles: black widows). \textit{Bottom:} Combined KDEs for globular cluster population (solid lines). Dashed lines show the KDEs for the Galactic field spider populations shown in Fig. \ref{fig:spider_lum_gal}). Dotted black lines mark the black widow distribution from \citet{swihart22}.}
    \label{fig:spider_lum_gc}
\end{figure}

\begin{figure}
    \includegraphics[width=\columnwidth]{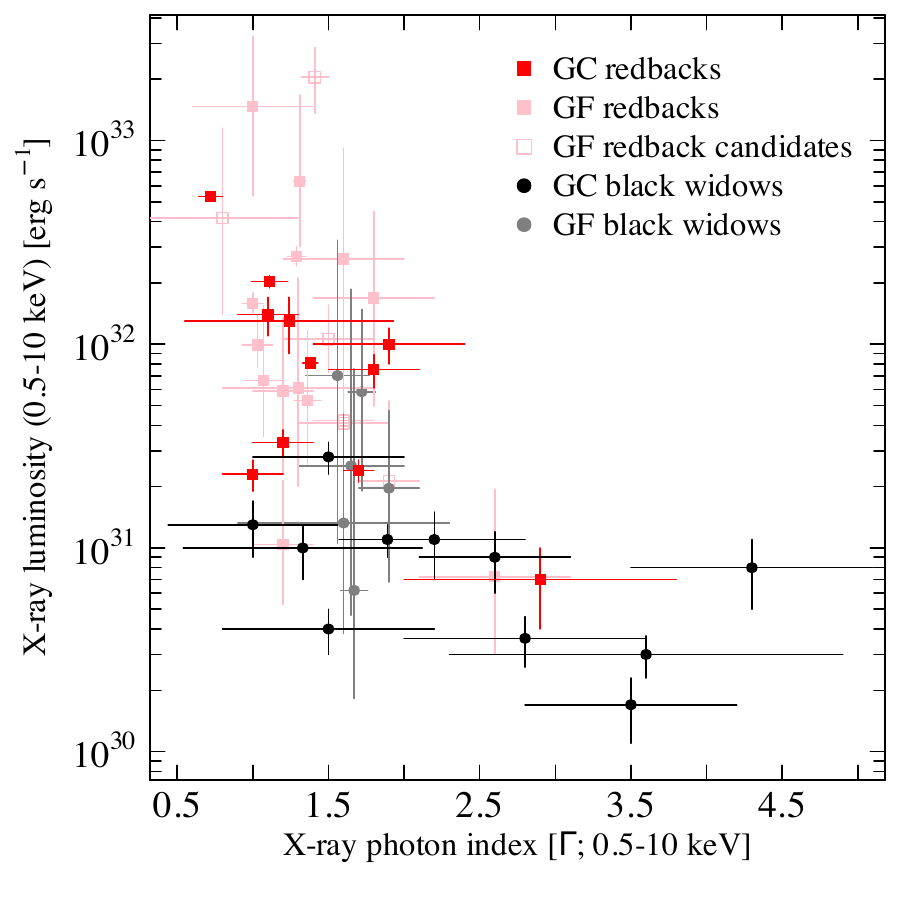}
    \caption{X-ray luminosity as a function of the photon power law index for the globular cluster (GC; red and black symbols) and the Galactic field (GF; pink and grey symbols) spiders (squares for redbacks and circles for black widows).}
    \label{fig:spider_pho_ind}
\end{figure}

\begin{figure}
    \includegraphics[width=\columnwidth]{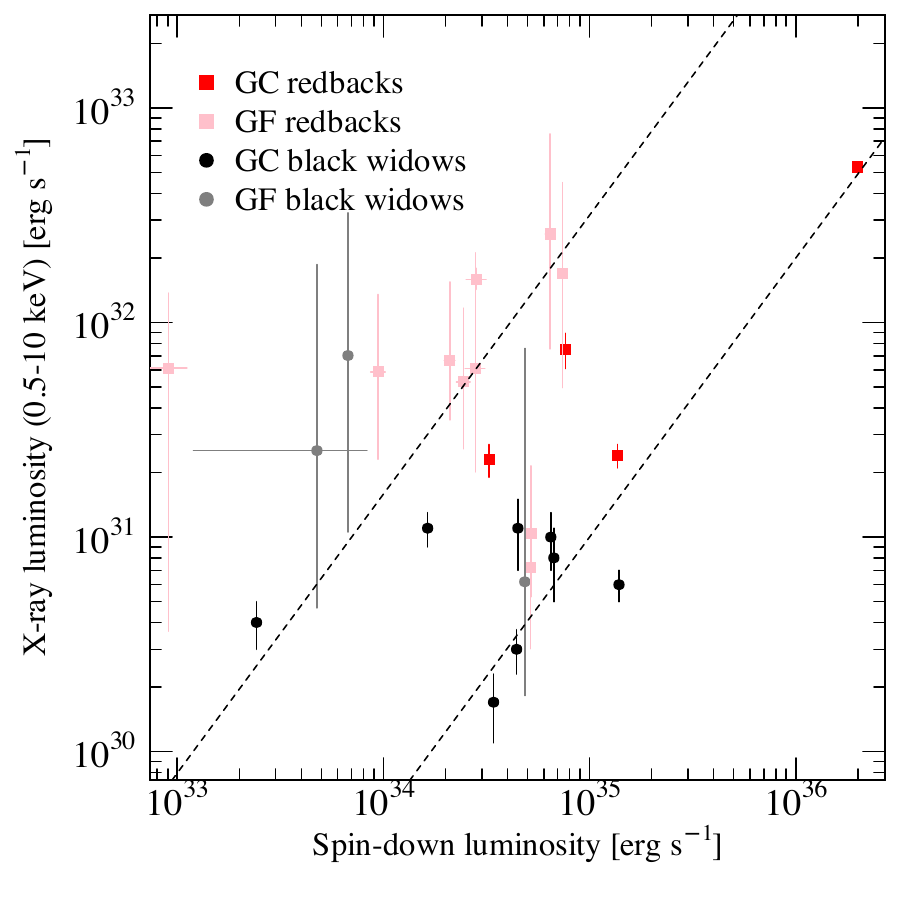}
    \caption{X-ray luminosity as a function of the spin-down luminosity for the globular cluster (GC; red and black symbols) and Galactic field (GF; pink and grey symbols) spiders (squares for redbacks and circles for black widows). The dashed lines show a relation of $\mathrm{log}_{10} L_{X} = 1.3\mathrm{ log}_{10} \dot{E} - 13$ (upper, `efficient' track) and $\mathrm{log}_{10} L_{X} = 1.3\mathrm{ log}_{10} \dot{E} - 14.5$ (lower `inefficient' track).}
    \label{fig:spider_edot}
\end{figure}

We calculated the X-ray luminosities of the Galactic field spiders using the \textit{Gaia} DR3 parallax-derived distance estimates. Fig.~\ref{fig:spider_lum_gal} shows the resulting X-ray luminosities and combined KDEs for the Galactic field spiders. These values are tabulated in Table \ref{tab:gal_spider_x} for confirmed spiders and Table \ref{tab:gal_spider_cand_x} for spider candidates. Similar to the above, we have excluded the transitional sources from the analysis.

The X-ray luminosity distribution peaks of the Galactic field redbacks, redback candidates, and black widows are $L_{\mathrm{X,RB}} = 8\times10^{31}$ erg/s, $L_{\mathrm{X,RBC}} = 4\times10^{31}$ erg/s, and $L_{\mathrm{X,BW}} = 6\times10^{31}$ erg/s, respectively. While these values are similar, the true peak for the black widow distribution likely lies much lower. Due to the low optical luminosities, the \textit{Gaia} parallaxes are not particularly precise for the optical counterparts of black widows. The distance posteriors reflect more the prior distribution and, therefore, can be overestimated. In addition, only six draws from a parent population can result easily in skewed distributions. Indeed, \citet{swihart22} found that the average X-ray luminosity of 18 black widows is $L_{X}=1.4\times10^{31}$ erg/s. As found by \citet{lee18} and \citet{swihart22}, the redback distribution reaches brighter X-ray luminosities by an order of magnitude. These authors argue that in redbacks the companion winds produce a larger surface for the pulsar wind to shock against and that the companion star winds or magnetospheres are stronger than for black widows, based on the modeling work done by \citet{romani16,wadiasingh17,wadiasingh18,vandermerwe20}, resulting in higher X-ray luminosities of the intrabinary shock.   

The more accurately measured globular cluster spider luminosities (due to better distance estimates towards the clusters; \citealt{baumgardt21}) show a clear difference between the black widow and redback luminosities (Fig. ~\ref{fig:spider_lum_gc}, Table \ref{tab:gc_spider_x}). The mean X-ray luminosity for our sample of 13 globular cluster black widows is $L_{X} = 7\times10^{30}$ erg/s, while for our sample of 11 redbacks, it is $L_{X} = 10^{32}$ erg/s. The bimodal distribution of the X-ray luminosity of globular cluster spiders confirms that seen in Galactic field spiders (using the black widow distribution from \citealt{swihart22}), with independent and more accurate distance measurements. This agrees with the result reached by \citet{lee23}, who found that the cumulative distribution functions of the X-ray luminosities of the globular cluster and Galactic field millisecond pulsars are similar.    

We also studied whether the X-ray luminosity correlates with the orbital period; however, there seems to be no correlation for the whole samples of the Galactic field or globular cluster spiders (see Appendix C). As discussed above, the X-ray luminosity distribution for spiders is bimodal. However, the black widow and redback populations have sources with similar orbital periods, producing large scatter on similar orbital periods. 

Fig.~\ref{fig:spider_pho_ind} shows the X-ray luminosity as a function of the photon power law index. Most spiders lie below $\Gamma<2$ but below X-ray luminosities of about $10^{31}$ erg/s spiders have $\Gamma>2$. The X-ray softening is likely due to a larger contribution from the heated polar cap compared to the intrabinary shock in the soft X-rays \citep{harding02,bogdanov09,lee18,vurgun22}. In our joint analysis of the Galactic field and globular cluster spiders, we find that their locations in the plane of X-ray luminosity and photon power law index are consistent. 

Two notable sources are PSR J2140$-$2310A in M30 and PSR J1816$+$4510. Both sources are redbacks with black widow-like X-ray properties, with luminosities of the order of $10^{31}$ erg/s and power law indices above two. PSR J1816$+$4510 hosts a peculiar companion; a hot proto-white dwarf \citep{kaplan12} which may have a weaker wind leading to a fainter intrabinary shock. On the other hand, the companion of PSR J2140$-$2310A seems to be an ordinary main-sequence star \citep{zhao20}, thus similar reasoning as for PSR J1816$+$4510 cannot be applied in this case. \citep{zhao20} modeled the \textit{Chandra} X-ray spectrum of PSR J2140$-$2310A with both a power law and a black body model with equally good fit statistics. The X-ray observations did not show any variability, thus pointing towards emission from the neutron star surface. However, it remains unclear why the intrabinary shock would be faint in this redback. A possible reason could be an atypically slow spin period of the neutron star \citep[$P_s$ = 11 ms;][]{ransom04}.   
 
Finally, we compare the X-ray luminosity with the spin-down luminosity for spiders (Fig. \ref{fig:spider_edot}). We estimate the spin-down luminosity using the standard value for the moment of inertia; 10$^{45}$ g cm$^{2}$, and correct for the Shklovskii effect for sources where the proper motion is available (see Table \ref{tab:gaiaDR3}) and using the \textit{Gaia} DR3 distances. Previous studies have found a relation of $L_{X} \propto \dot{E}^{1.3}$ for millisecond pulsars \citep{possenti02,lee18}. Fig. \ref{fig:spider_edot} shows that the spiders can be arranged along two $L_{X} \propto \dot{E}^{1.3}$ tracks but with a seemingly different normalization. The upper one is populated by redbacks (but not exclusively) with roughly 0.1\% efficiency in converting the spin-down power to X-ray luminosity. In contrast, the lower one is populated by black widows (but not exclusively) with roughly 0.01\% efficiency in converting the spin-down power to X-ray luminosity. The difference in efficiency is generally attributed to the intrabinary shocks dominating the X-ray emission in redbacks, while in black widows the X-ray emission is likely dominated by the thermal polar cap emission as we discuss in Section \ref{sec:efficiency}. We can state that, on average, redbacks are more efficient in turning spin-down power to X-ray luminosity. 

The `inefficient' redbacks are PSR J1816$+$4510 and PSR J1431$-$4715 in the Galactic field. The peculiar nature of PSR J1816$+$4510 was already discussed above, and for PSR J1431$-$4715 our {\it XMM-Newton} analysis did not show any X-ray variability over the orbit indicating faint intrabinary shock emission (see Appendix C). However, the spectral index is rather hard with $\Gamma=1.2\pm0.2$, which is more in line with non-thermal emission. In addition, two globular cluster redbacks PSR J1748$-$2446P and PSR J1740$-$5340A are found on the `inefficient' track. PSR J1748$-$2446P is an X-ray bright redback with a very high spin-down luminosity showing X-ray orbital modulation and spectral properties consistent with intrabinary shock emission \citep{bogdanov21}. Similarly, the X-ray properties of PSR J1740$-$5340A are indicative of intrabinary shock emission \citep{bogdanov10}. In these cases, the lower efficiency could arise from orbital inclination or beamed/less effective pulsar wind.     

On the other hand, the `efficient' black widows are PSR J1653$-$0158 and PSR J1959$+$2048 in the Galactic field and globular cluster sources PSR J1836$-$2354A and PSR J1953$+$1846A. PSR J1653$-$0158 is the most compact black widow system known so far with a 75-minute orbit, and its broadband X-ray spectrum can be fitted with a power law model indicating emission solely from the intrabinary shock \citep{long22}. Also, the orbitally-modulated X-ray emission from the original black widow system PSR J1959$+$2048 indicates a strong contribution from the intrabinary shock \citep{huang07,kandel21}. The X-ray observations of PSR J1953$+$1846A indicate also emission from the intrabinary shock with variable X-ray emission and a best-fit power law spectrum \citep{elsner08}. While the X-ray observations are not conclusive of the origin of the X-ray emission for PSR J1836$-$2354A \citep{amato19}, a power law model is preferred.     

To summarize, we find that the X-ray properties of spiders in the Galactic field and globular clusters are similar to each other and that the X-ray efficiency is on average an order of magnitude higher for redbacks than for black widows. However, there are some exceptions indicating that the difference in the X-ray efficiency does not fully arise from the different sizes of the companion stars. We will further discuss the X-ray efficiency in Section \ref{sec:efficiency}.     

\section{Discussion}

\subsection{What drives the orbital period -- optical luminosity correlation?} \label{sec:model}

\begin{figure}
    \includegraphics[width=\columnwidth]{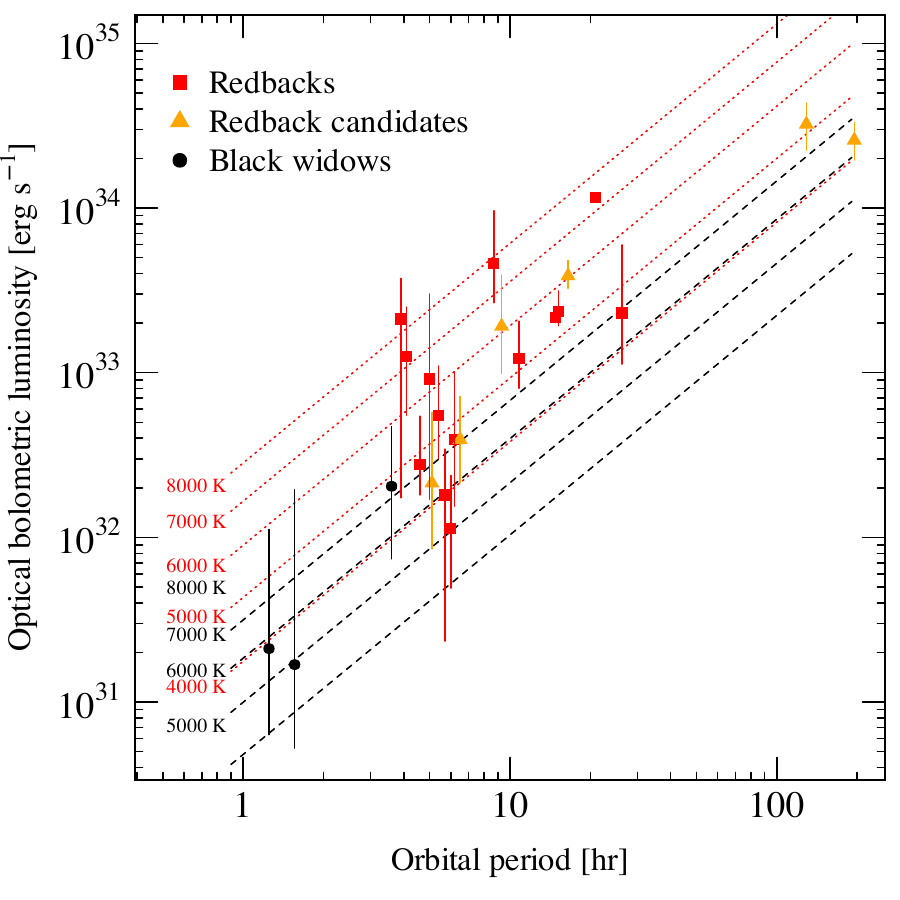}
    \caption{Optical bolometric luminosities derived from the \textit{Gaia} G-band magnitude as a function of the orbital period of Galactic field spiders (parallax-selected sample). Dotted and dashed lines show theoretical relations assuming Roche lobe filling companion stars, a neutron star mass of 1.8 M$_{\odot}$, different mass ratios, and stellar temperatures for redbacks (red dotted lines and a mass ratio of 0.1) and black widows (black dashed lines and a mass ratio of 0.01).}
    \label{fig:spider_orb_opt_mod}
\end{figure}

\begin{figure}
    \includegraphics[width=\columnwidth]{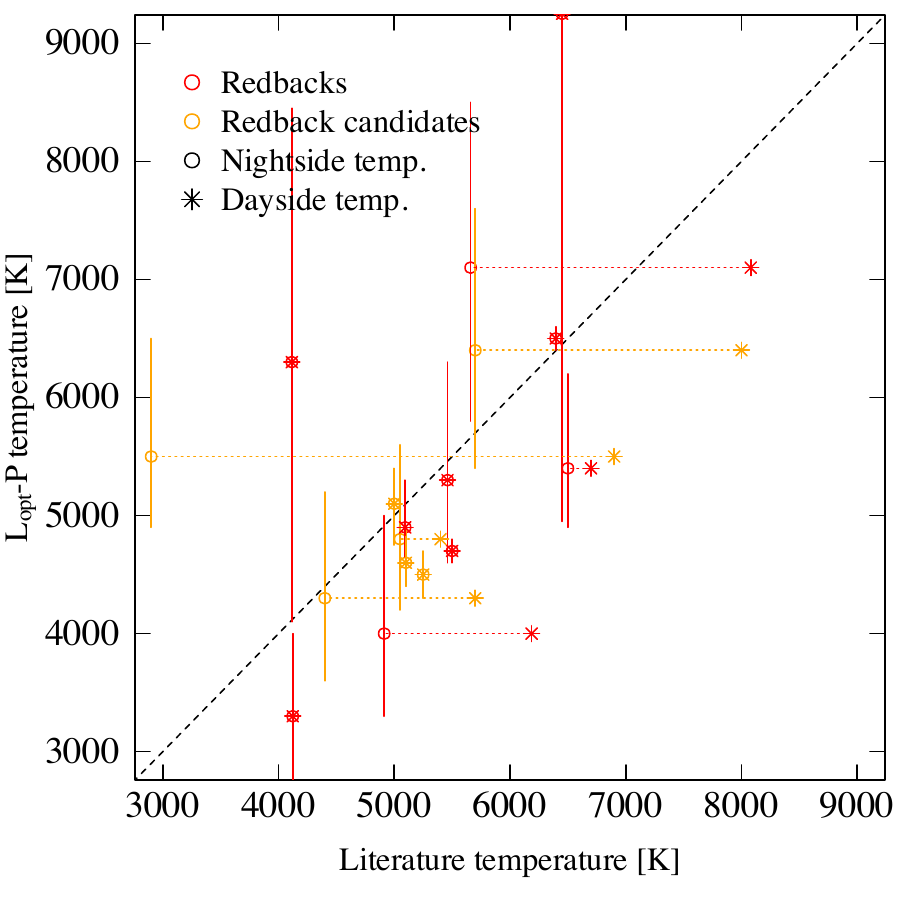}
    \caption{Temperature estimates of the companion star in Galactic field redbacks (red data points) and redback candidates (orange data points) from the literature plotted as a function of the model temperatures described in the text. The night-side/non-irradiated temperatures are marked as empty circles and the day-side/irradiated temperatures as stars that are united with dotted lines. One-to-one correspondence is drawn as a dashed line.}
    \label{fig:spider_mod_temp}
\end{figure}

\begin{table*} \centering
  \caption{Temperature estimates of the companion star in Galactic field redbacks and redback candidates from the literature (both night- and day-side temperatures if available), using \textit{Gaia} BP/RP spectra, and those derived from our empirical L$_{\mathrm{opt}}$--P relation assuming a Roche lobe filling companion and a mass ratio, $q$, from the literature (or $q=0.1$, in case no mass ratio has been reported).}
  \label{tab:spider_temp}
\begin{tabular}{lcccclcl}
\\[-1.8ex]\hline
\hline \\[-1.8ex]
Source name & Night & Day & Ref. & Gaia & L$_{\mathrm{opt}}$--P & q & Ref. \\
& (K) & (K) & & (K) & (K) \\
\hline \\[-1.8ex]
J0212$+$5320 & 6395 & 6395 & 17 & 6240--6304 & 6500$^{+100}_{-100}$ & 0.26 & 1 \\
J1036$-$4353 & --   & --   & -- & --        & 5000$^{+1250}_{-1100}$ & 0.1 & -- \\
J1048$+$2339 & 4123 & 4123 & 2 & --         & 3300$^{+700}_{-700}$ & 0.178 & 2 \\ 
J1306$-$40   & 4913 & 6187 & 3 & 4476--4763 & 4000$^{+1000}_{-700}$ & 0.290 & 3 \\
J1431$-$4715 & 6500 & 6700 & 4 & 6252--6289 & 5400$^{+800}_{-500}$ & 0.1   & 4 \\
J1622$-$0315 & 6450 & 6450 & 5 & --         & 9250$^{+1500}_{-4300}$ & 0.07 & 4 \\
J1628$-$3205 & 4115 & 4115 & 6 & --         & 6300$^{+2150}_{-2200}$ & 0.12 & 6 \\
J1723$-$2837 & 5500 & 5500 & 7 & 5787--5875 & 4700$^{+100}_{-100}$ & 0.3 & 8 \\
J1803$-$6707 & --   & --   & -- & --         & 5400$^{+1100}_{-1000}$ & 0.1 & -- \\
J1816$+$4510 & 16000 & 16000 & 9 & 10594--10641 & 8100$^{+1900}_{-1000}$ & 0.1 & 9 \\
J1908$+$2105 & --   & --   & --  & --         & 6100$^{+1800}_{-1100}$ & 0.1 & -- \\
J1910$-$5320 & --   & --   & --  & --         & 7500$^{+1500}_{-1500}$ & 0.1 & -- \\
J1957$+$2516 & --   & --   & --  & 5352--5518 & 4200$^{+700}_{-1800}$ & 0.1 & -- \\
J2039$-$5618 & 5436 & 5700 & 10 & 5062--5092 & 5300$^{+1000}_{-700}$ & 0.137 & 10 \\
J2129$-$0429 & 5094 & 5094 & 11 & 4862--4928 & 4900$^{+400}_{-250}$ & 0.254 & 11 \\
J2215$+$5135 & 5660 & 8080 & 12 & --         & 7100$^{+1400}_{-1300}$ & 0.144 & 12 \\
J2339$-$0533 & 2900 & 6900 & 13 & --         & 5500$^{+1000}_{-600}$ & 0.054 & 13 \\
\hline \\[-1.8ex]
J0523$-$2529 & 5100 & 5100 & 14 & 4690--4722 & 4600$^{+300}_{-200}$ & 0.61 & 15 \\
J0838.8$-$2829 & --   & --   & --  & --         & 4500$^{+1300}_{-900}$ & 0.1 & -- \\
J0846.0$+$2820 & 5250 & 5250 & 16 & 5390--5502 & 3500$^{+200}_{-200}$ & 0.40 & 16 \\
J0935.3$+$0901 & --   & --   & --  & --         & 3400$^{+600}_{-900}$ & 0.1 & -- \\
J0940.3$-$7610 & 5050 & 5400 & 17 & 4177--4363 & 4800$^{+800}_{-600}$ & 0.1 & -- \\
J0954.8$-$3948 & 5700 & 8000 & 18 & 5508--5557 & 6400$^{+1200}_{-1000}$ & 0.1 & -- \\
J1417.5$-$4402 & 5000 & 5000 & 19 & 5004--5047 & 5100$^{+300}_{-350}$ & 0.171 & 20 \\
J2333.1$-$5527 & 4400 & 5700 & 21 & --         & 4300$^{+900}_{-900}$ & 0.1 & -- \\
\hline \\[-1.8ex]
\multicolumn{8}{|p{0.55\linewidth}|}{\textbf{References:} 1) \citet{linares17}, 2) \citet{yap19}, 3) \citet{swihart19}, 4) \citet{strader19}, 5) Turchetta et al. 2023 (in prep.), 6) \citet{li14}, 7) \citet{vanstaden16}, 8) \citet{crawford13}, 9) \citet{kaplan13}, 10) \citet{clark21}, 11) \citet{bellm16}, 12) \citet{linares18b}, 13) \citet{romani11}, 14) \citet{halpern22}, 15) \citet{strader14}, 16) \citet{swihart17}, 17) \citet{swihart21}, 18) \citet{li18}, 19) \citet{strader15}, 20) \citet{camilo16}, 21) \citet{swihart20}} \\
\end{tabular}
\end{table*}

To estimate the bolometric luminosity of the companion, we use Planck's and Stefan Boltzmann's laws with different surface temperatures and radii. To estimate the stellar radii, we use the formula of \citet{eggleton83} for the volume-equivalent Roche lobe radius:  

\begin{equation}
    R = af \frac{q^{2/3}}{0.6 q^{2/3} + \mathrm{log}(1+q^{1/3})} \, ,
\end{equation}

\noindent where the binary separation $a$ is estimated using Kepler's third law with a neutron star mass of 1.8 M$_{\odot}$ and a mass ratio from the literature (Table \ref{tab:spider_temp}, or $q=0.1$ for a redback and $q=0.01$ for a black widow in case no mass ratio is reported). We assume that the companion fills its Roche lobe, i.e., the filling factor is $f=1$. Figure \ref{fig:spider_orb_opt_mod} shows theoretical relations between the optical bolometric luminosity and the orbital period (`L$_{\mathrm{opt}}$--P relations') using the above values for the radius and stellar temperatures varying from 4000 K to 8000 K. The individual relations have a slope of 1.33, as the luminosity is proportional to the orbital period in the above formalism; $L \propto P^{4/3}$. Interestingly, this matches well with the best-fitting value of 1.38$\pm$0.20 for the optical luminosity and orbital period relation of our parallax-selected distance sample. 

In addition to the distance, we attribute the observed scatter in optical luminosity to a combination of the exact Roche Lobe filling factor, average stellar temperature (taking into account the effect of irradiation which can be substantial), and the mass ratio of the system. This can be an order of magnitude difference for a given orbital period. However, this scatter is small enough to retain the overall relation. Therefore, a rough estimate of the bolometric luminosity of the companion star can be obtained if the orbital period is known, e.g., from radio timing observations, according to the relations shown in Figures \ref{fig:spider_orb_opt} and \ref{fig:spider_orb_opt_mod}, and tabulated in Table \ref{tab:spider_orb_lum}. Similarly, if the measured optical luminosity lies above or below the relation could mean that the temperature of the companion is either higher or lower, respectively, than the population mean for a given orbital period. Obviously, for irradiated systems, the temperature strongly depends on the orbital phase of the observation. However, the number of \textit{Gaia} transits for a given source varies between 17 and 90 with a median of 44, which likely averages out the obtained magnitudes and therefore the optical bolometric luminosities over the orbit.   

Using the above formalism, we estimate the average temperatures of the companion stars in redbacks, assuming the DR3 parallax distance, a filling factor of unity, and a mass ratio taken from the literature or, if not available, using the canonical value of 0.1 solar mass. We compared these model temperatures to the literature values gained from optical spectroscopy or light curve modeling (for both day- and night-side temperatures, when available), and to effective temperatures given in the \textit{Gaia} archive, when available, and that are produced by the General Stellar Parametrizer from Photometry (GSP-Phot) Aeneas algorithm using the low-resolution \textit{Gaia} BP/RP spectra. All the different temperature estimates are tabulated in Table \ref{tab:spider_temp} for each source. Since the \textit{Gaia} magnitudes correspond to average values through the orbit, we can expect these temperatures to lie between day- and night-side temperatures. Fig. \ref{fig:spider_mod_temp} shows our estimated temperatures plotted as a function of the day- and night-side temperatures. The error in the estimated temperature reflects the uncertainty of the intrinsic optical luminosity, i.e., it is calculated by translating the optical luminosity error shown in Figure \ref{fig:spider_orb_opt_mod} to the corresponding temperature error using the formalism described above with the corresponding orbital periods and mass ratios for each source. On average, there is a one-to-one correspondence between our estimated and literature temperatures assuming that they can be anything between the day- and night-side temperatures. Sources producing lower temperatures compared to the lightcurve analyses presented in the literature are one of the Huntsman pulsars; J0846.0$+$2820, and redbacks PSR J1816$+$4510, PSR J1431$-$2837, and PSR J1723$-$2837. In these cases, it might be that the temperature is underestimated due to assuming a Roche lobe filling companion. The needed filling factor to reach the literature temperatures for these sources is $f=0.45$, $f=0.25$, $f=0.7$, and $f=0.75$, respectively. The light curve analyses support this assumption with estimates for the filling factor of $f=0.86$ for J0846.0$+$2820 \citep{swihart17}, $f=0.35$ for J1816$+$4510 \citep{kaplan13}, and $f=0.65$ for J1431$-$2837 \citep{strader19}. In the case of J1723$-$2837, the literature temperature is not very precise with a standard deviation of 600 K \citep{vanstaden16}.

On the other hand, we estimate a higher temperature for PSR J2339$-$0533 (also for PSR J1622$-$0315 and PSR J1628$-$3205, but these are highly uncertain values). J2339$-$0533 has an irradiated companion with a much hotter day-side temperature. Thus, in these cases, the average \textit{Gaia} magnitude, and therefore the implied model temperature, likely correspond to the day-side temperature. Finally, we place estimates for the temperatures of the companion stars in five redbacks (J1036$-$4353, J1803$-$6707, J1908$+$2105, J1910$-$5320, J1957$+$2516) and two redback candidates (J0838.8$-$2829, J0935.3$+$0901); see values in Table \ref{tab:spider_temp}. 

\begin{table} \centering
  \caption{Black widow systems with radio timing parallax-derived distances and confirmed optical counterparts with optical periods.}
  \label{tab:low_lum_bws}
\begin{tabular}{llclclc}
\\[-1.8ex]\hline
\hline \\[-1.8ex]
Source & $d_{R}$ & Ref. & R-band & Per. & Night & Ref. \\
& (kpc) & & (mag) & (hr) & (K) & \\
\hline \\[-1.8ex]
J0023$+$0923 & 1.1$\pm$0.2 & 1 & 22.6--23.9 & 3.3 & 3340 & 2 \\
J0610$-$2100 & 1.5$^{+0.3}_{-0.2}$ & 3 & 25.5--27.0 & 6.9 & 1600 & 4 \\ 
J2256$-$1024 & 2.0$\pm$0.6 & 5 & 21.4--24.1$^{a}$ & 5.1 & 2450 & 6 \\ 
\hline \\[-1.8ex]
\multicolumn{7}{|p{0.9\linewidth}|}{\textbf{Notes:} $^{a}$$i$-band magnitude.} \\
\multicolumn{7}{|p{0.9\linewidth}|}{\textbf{References:} 1) \citet{matthews16}, 2) \citet{draghis19}, 3) \citet{ding23}, 4) \citet{vanderwateren22}, 5) \citet{crowter20}, 6) \citet{breton13}. } \\
\end{tabular}
\end{table} 

Since the limiting magnitude of \textit{Gaia} is quite high ($\sim$21), it is not possible to probe the majority of the black widow population with it since their optical counterparts are typically much dimmer. However, for three systems, there are radio timing parallax distances available together with the optical counterparts detected using large telescopes (Table \ref{tab:low_lum_bws}). The tabulated magnitudes correspond roughly to luminosities $10^{29} - 10^{30}$ erg/s, $5 \times 10^{27} - 10^{28}$ erg/s, and $5 \times 10^{28} - 10^{29}$ erg/s, for PSR J0023$+$0923, PSR J0610$-$2100, and PSR J2256$-$1024, respectively. These values are orders of magnitude lower than for the spiders plotted in Fig. \ref{fig:spider_orb_opt} for similar orbital periods. However, based on the optical lightcurve modeling, the companion stars present very low temperatures (Table \ref{tab:low_lum_bws}) pushing them out of the correlation (see Fig. \ref{fig:spider_orb_opt_mod}). Unless the effect of the low stellar temperature on the optical luminosity can be compensated somehow, the correlations presented above most likely do not give correct luminosity estimates for dim black widows at a given orbital period.    

\subsection{Efficiency of X-ray emission from spiders} \label{sec:efficiency}

The relation between the spin-down power ($\dot{E}$) and the X-ray luminosity in pulsars has been the focus of several studies \citep[e.g.][]{seward88,becker97,possenti02,cheng04,li08,kargaltsev08,marelli11,kargaltsev12,posselt12,spiewak16,lee18,vahdat22,lee23}. A correlation is expected assuming that the pulsar spin down is powering all the emission processes that can be connected to it (also gamma-rays are correlated to $\dot{E}$; e.g., \citealt{abdo10,sazparkinson10,pletsch12,kargaltsev12}). However, the X-ray luminosity correlation to $\dot{E}$ is relatively weak due to a large, four orders of magnitude scatter in the X-ray efficiencies ($\eta = L_{X}/\dot{E}$). The scatter likely arises from a combination of uncertainties in the distances and spin-down powers, different beaming factors, and differences in the underlying emission process and spectrum. 

Observing a similar trend in spiders (see Fig. \ref{fig:spider_edot} and Section 3.4) is not surprising, assuming that the spin-down power is the primary driver in producing the observed emission. However, the sources at the efficient track ($\eta \sim 10^{-3}$; mainly redbacks) present higher X-ray efficiencies/luminosities than other ordinary or millisecond pulsars at the corresponding spin-down power ($\eta \sim 10^{-4}$; e.g., \citealt{vahdat22}, see also Fig. 2 in \citealt{kargaltsev12}). This likely arises from the more efficient conversion of the spin-down power to radiation in the intrabinary shock. In rare cases, some redbacks have even larger X-ray efficiencies: PSR J1023+0038 has an X-ray efficiency of $\eta \sim 0.02$ with a non-thermal spectrum \citep[$\Gamma=1.17$;][]{tendulkar14} that occurred before the transition to an accretion disk state. In this case, the enhanced efficiency could arise from shocks forming closer to the neutron star interacting with the infalling matter building up the accretion disk \citep{stappers14}. On the other hand, the X-ray efficiencies in the spiders located in the inefficient track ($\eta \sim 10^{-4}$) are compatible with the other pulsar populations. This can be understood if the contribution from the intrabinary shock in these systems is small, and the emission arises close to the neutron star.  

\section{Conclusions}

This paper presents the optical and X-ray luminosities of the Galactic field and globular cluster spiders using the {\it Gaia} parallax and dispersion measure distances. Similar to earlier studies, we found that the dispersion measure distances derived using the electron density model of \cite{yao17} underestimate the spider distances compared to parallax-derived values. On average, this is a factor of 1.4 difference corresponding to a factor of two difference in luminosity. 

We compared the X-ray luminosities using the parallax distances of the Galactic field spiders to spiders found in globular clusters (using accurate cluster distance estimates based on {\it Gaia} and {\it HST} data given in \citealt{baumgardt21}) and found that the luminosity distributions of redbacks agree in both samples. Due to the dim optical counterparts of black widows, {\it Gaia} can only observe the brightest ones, producing a flux-limited and thus biased sample. This is evident when comparing to the full sample of black widow X-ray luminosities \citep{swihart22} that agree well with the sample from globular clusters.   

We also studied the trend between the X-ray and spin-down luminosities for millisecond pulsars. Spiders seem to follow this trend, with the majority of redbacks located at an `efficient' track with spin-down to X-ray luminosity conversion rate of $\sim$0.1\%, and the majority of black widows located at an `inefficient' track with a spin-down to X-ray luminosity conversion rate of $\sim$0.01\%. This is likely due to the larger angular sizes of the redback companion winds that the pulsar wind can shock against. However, some redbacks are found at the `inefficient' track while some black widows are found at the `efficient' track indicating that other parameters than the companion star size affect the efficiency.   

Finally, we studied the relation of the derived luminosities with the orbital periods of the Galactic field spiders. We found that the intrinsic optical luminosity significantly correlates with the period (independent of the distance method used). We interpret this correlation as the effect of the increasing size of the Roche Lobe radius with the orbital period. Therefore, using the orbital period, an estimate of the optical luminosity (and magnitude together with a distance estimate) can be obtained. The correlation between the optical luminosity and orbital period that we discovered has a large scatter likely due to different (irradiated) stellar temperatures, binary mass ratios, and Roche lobe filling factors. Assuming that the latter two are known, the source displacement from the relation can indicate lower or higher (irradiated) stellar temperatures to the population mean for a given orbital period.        

\section*{Acknowledgements}

This project has received funding from the European Research Council (ERC) under the European Union’s Horizon 2020 research and innovation programme (grant agreement No. 101002352). This work has made use of data from the European Space Agency (ESA) mission {\it Gaia} (\url{https://www.cosmos.esa.int/gaia}), processed by the {\it Gaia} Data Processing and Analysis Consortium (DPAC, \url{https://www.cosmos.esa.int/web/gaia/dpac/consortium}). Funding for the DPAC has been provided by national institutions, in particular the institutions participating in the {\it Gaia} Multilateral Agreement. This product makes use of public auxiliary models provided by ESA/Gaia/DPAC/CU8 and prepared by Christophe Ordenovic, Orlagh Creevey, Andreas Korn, Bengt Edvardsson, Oleg Kochukhov, and Fr\'ed\'eric Th\'evenin. This research has made use of data and/or software provided by the High Energy Astrophysics Science Archive Research Center (HEASARC), which is a service of the Astrophysics Science Division at NASA/GSFC. We gratefully acknowledge the use of \textit{astropy} \citep{astropy2013,astropy2018} and \textit{gatspy} \citep{vanderplas15,gatspy}.

\section*{Data Availability}

The \textit{Gaia} data analyzed here are available at the European Space Agency's \textit{Gaia} archive (https://gea.esac.esa.int/archive). X-ray data are available at HEASARC (https://heasarc.gsfc.nasa.gov).




\bibliographystyle{mnras}
\bibliography{bibliography} 

\begin{thebibliography}{}
\makeatletter
\relax
\def\mn@urlcharsother{\let\do\@makeother \do\$\do\&\do\#\do\^\do\_\do\%\do\~}
\def\mn@doi{\begingroup\mn@urlcharsother \@ifnextchar [ {\mn@doi@}
  {\mn@doi@[]}}
\def\mn@doi@[#1]#2{\def\@tempa{#1}\ifx\@tempa\@empty \href
  {http://dx.doi.org/#2} {doi:#2}\else \href {http://dx.doi.org/#2} {#1}\fi
  \endgroup}
\def\mn@eprint#1#2{\mn@eprint@#1:#2::\@nil}
\def\mn@eprint@arXiv#1{\href {http://arxiv.org/abs/#1} {{\tt arXiv:#1}}}
\def\mn@eprint@dblp#1{\href {http://dblp.uni-trier.de/rec/bibtex/#1.xml}
  {dblp:#1}}
\def\mn@eprint@#1:#2:#3:#4\@nil{\def\@tempa {#1}\def\@tempb {#2}\def\@tempc
  {#3}\ifx \@tempc \@empty \let \@tempc \@tempb \let \@tempb \@tempa \fi \ifx
  \@tempb \@empty \def\@tempb {arXiv}\fi \@ifundefined
  {mn@eprint@\@tempb}{\@tempb:\@tempc}{\expandafter \expandafter \csname
  mn@eprint@\@tempb\endcsname \expandafter{\@tempc}}}

\bibitem[\protect\citeauthoryear{{Abdo} et~al.,}{{Abdo} et~al.}{2010}]{abdo10}
{Abdo} A.~A.,  et~al., 2010, \mn@doi [\apjs] {10.1088/0067-0049/187/2/460},
  \href {https://ui.adsabs.harvard.edu/abs/2010ApJS..187..460A} {187, 460}

\bibitem[\protect\citeauthoryear{{Al Noori} et~al.,}{{Al Noori}
  et~al.}{2018}]{alnoori18}
{Al Noori} H.,  et~al., 2018, \mn@doi [\apj] {10.3847/1538-4357/aac828}, \href
  {https://ui.adsabs.harvard.edu/abs/2018ApJ...861...89A} {861, 89}

\bibitem[\protect\citeauthoryear{{Amato}, {D'A{\i}}, {Del Santo}, {de Martino},
  {Marino}, {Di Salvo}, {Iaria}  \& {Mineo}}{{Amato} et~al.}{2019}]{amato19}
{Amato} R.,  {D'A{\i}} A.,  {Del Santo} M.,  {de Martino} D.,  {Marino} A.,
  {Di Salvo} T.,  {Iaria} R.,   {Mineo} T.,  2019, \mn@doi [\mnras]
  {10.1093/mnras/stz1100}, \href
  {https://ui.adsabs.harvard.edu/abs/2019MNRAS.486.3992A} {486, 3992}

\bibitem[\protect\citeauthoryear{{Am{\^o}res} \& {L{\'e}pine}}{{Am{\^o}res} \&
  {L{\'e}pine}}{2005}]{amores05}
{Am{\^o}res} E.~B.,  {L{\'e}pine} J.~R.~D.,  2005, \mn@doi [\aj]
  {10.1086/430957}, \href
  {https://ui.adsabs.harvard.edu/abs/2005AJ....130..659A} {130, 659}

\bibitem[\protect\citeauthoryear{{Am{\^o}res} et~al.,}{{Am{\^o}res}
  et~al.}{2021}]{amores21}
{Am{\^o}res} E.~B.,  et~al., 2021, \mn@doi [\mnras] {10.1093/mnras/stab2248},
  \href {https://ui.adsabs.harvard.edu/abs/2021MNRAS.508.1788A} {508, 1788}

\bibitem[\protect\citeauthoryear{{An}, {Romani}, {Johnson}, {Kerr}  \&
  {Clark}}{{An} et~al.}{2017}]{an17}
{An} H.,  {Romani} R.~W.,  {Johnson} T.,  {Kerr} M.,   {Clark} C.~J.,  2017,
  \mn@doi [\apj] {10.3847/1538-4357/aa947f}, \href
  {https://ui.adsabs.harvard.edu/abs/2017ApJ...850..100A} {850, 100}

\bibitem[\protect\citeauthoryear{{Arnaud}}{{Arnaud}}{1996}]{arnaud96}
{Arnaud} K.~A.,  1996, in {Jacoby} G.~H.,  {Barnes} J.,  eds,  Astronomical
  Society of the Pacific Conference Series Vol. 101, Astronomical Data Analysis
  Software and Systems V. pp 17--+

\bibitem[\protect\citeauthoryear{{Astropy Collaboration} et~al.,}{{Astropy
  Collaboration} et~al.}{2013}]{astropy2013}
{Astropy Collaboration} et~al., 2013, \mn@doi [\aap]
  {10.1051/0004-6361/201322068}, \href
  {https://ui.adsabs.harvard.edu/abs/2013A&A...558A..33A} {558, A33}

\bibitem[\protect\citeauthoryear{{Astropy Collaboration} et~al.,}{{Astropy
  Collaboration} et~al.}{2018}]{astropy2018}
{Astropy Collaboration} et~al., 2018, \mn@doi [\aj] {10.3847/1538-3881/aabc4f},
  \href {https://ui.adsabs.harvard.edu/abs/2018AJ....156..123A} {156, 123}

\bibitem[\protect\citeauthoryear{{Au} et~al.,}{{Au} et~al.}{2023}]{au23}
{Au} K.-Y.,  et~al., 2023, \mn@doi [\apj] {10.3847/1538-4357/acae8a}, \href
  {https://ui.adsabs.harvard.edu/abs/2023ApJ...943..103A} {943, 103}

\bibitem[\protect\citeauthoryear{{Bailer-Jones}, {Rybizki}, {Fouesneau},
  {Mantelet}  \& {Andrae}}{{Bailer-Jones} et~al.}{2018}]{bailerjones18}
{Bailer-Jones} C.~A.~L.,  {Rybizki} J.,  {Fouesneau} M.,  {Mantelet} G.,
  {Andrae} R.,  2018, \mn@doi [\aj] {10.3847/1538-3881/aacb21}, \href
  {https://ui.adsabs.harvard.edu/abs/2018AJ....156...58B} {156, 58}

\bibitem[\protect\citeauthoryear{{Bailer-Jones}, {Rybizki}, {Fouesneau},
  {Demleitner}  \& {Andrae}}{{Bailer-Jones} et~al.}{2021}]{bailerjones21}
{Bailer-Jones} C.~A.~L.,  {Rybizki} J.,  {Fouesneau} M.,  {Demleitner} M.,
  {Andrae} R.,  2021, \mn@doi [\aj] {10.3847/1538-3881/abd806}, \href
  {https://ui.adsabs.harvard.edu/abs/2021AJ....161..147B} {161, 147}

\bibitem[\protect\citeauthoryear{{Bartels}, {Edwards}  \& {Weniger}}{{Bartels}
  et~al.}{2018}]{bartels18}
{Bartels} R.~T.,  {Edwards} T.~D.~P.,   {Weniger} C.,  2018, \mn@doi [\mnras]
  {10.1093/mnras/sty2529}, \href
  {https://ui.adsabs.harvard.edu/abs/2018MNRAS.481.3966B} {481, 3966}

\bibitem[\protect\citeauthoryear{{Baumgardt} \& {Vasiliev}}{{Baumgardt} \&
  {Vasiliev}}{2021}]{baumgardt21}
{Baumgardt} H.,  {Vasiliev} E.,  2021, \mn@doi [\mnras]
  {10.1093/mnras/stab1474}, \href
  {https://ui.adsabs.harvard.edu/abs/2021MNRAS.505.5957B} {505, 5957}

\bibitem[\protect\citeauthoryear{{Becker} \& {Truemper}}{{Becker} \&
  {Truemper}}{1997}]{becker97}
{Becker} W.,  {Truemper} J.,  1997, \mn@doi [\aap]
  {10.48550/arXiv.astro-ph/9708169}, \href
  {https://ui.adsabs.harvard.edu/abs/1997A&A...326..682B} {326, 682}

\bibitem[\protect\citeauthoryear{{Bellm} et~al.,}{{Bellm}
  et~al.}{2016}]{bellm16}
{Bellm} E.~C.,  et~al., 2016, \mn@doi [\apj] {10.3847/0004-637X/816/2/74},
  \href {https://ui.adsabs.harvard.edu/abs/2016ApJ...816...74B} {816, 74}

\bibitem[\protect\citeauthoryear{{Bogdanov} \& {Grindlay}}{{Bogdanov} \&
  {Grindlay}}{2009}]{bogdanov09}
{Bogdanov} S.,  {Grindlay} J.~E.,  2009, \mn@doi [\apj]
  {10.1088/0004-637X/703/2/1557}, \href
  {https://ui.adsabs.harvard.edu/abs/2009ApJ...703.1557B} {703, 1557}

\bibitem[\protect\citeauthoryear{{Bogdanov}, {Grindlay}, {Heinke}, {Camilo},
  {Freire}  \& {Becker}}{{Bogdanov} et~al.}{2006}]{bogdanov06}
{Bogdanov} S.,  {Grindlay} J.~E.,  {Heinke} C.~O.,  {Camilo} F.,  {Freire} P.
  C.~C.,   {Becker} W.,  2006, \mn@doi [\apj] {10.1086/505133}, \href
  {https://ui.adsabs.harvard.edu/abs/2006ApJ...646.1104B} {646, 1104}

\bibitem[\protect\citeauthoryear{{Bogdanov}, {van den Berg}, {Heinke}, {Cohn},
  {Lugger}  \& {Grindlay}}{{Bogdanov} et~al.}{2010}]{bogdanov10}
{Bogdanov} S.,  {van den Berg} M.,  {Heinke} C.~O.,  {Cohn} H.~N.,  {Lugger}
  P.~M.,   {Grindlay} J.~E.,  2010, \mn@doi [\apj]
  {10.1088/0004-637X/709/1/241}, \href
  {https://ui.adsabs.harvard.edu/abs/2010ApJ...709..241B} {709, 241}

\bibitem[\protect\citeauthoryear{{Bogdanov}, {Bahramian}, {Heinke}, {Freire},
  {Hessels}, {Ransom}  \& {Stairs}}{{Bogdanov} et~al.}{2021}]{bogdanov21}
{Bogdanov} S.,  {Bahramian} A.,  {Heinke} C.~O.,  {Freire} P. C.~C.,  {Hessels}
  J. W.~T.,  {Ransom} S.~M.,   {Stairs} I.~H.,  2021, \mn@doi [\apj]
  {10.3847/1538-4357/abee78}, \href
  {https://ui.adsabs.harvard.edu/abs/2021ApJ...912..124B} {912, 124}

\bibitem[\protect\citeauthoryear{{Boztepe}, {G{\"o}{\u{g}}{\"u}{\c{s}}},
  {G{\"u}ver}  \& {Schwenzer}}{{Boztepe} et~al.}{2020}]{boztepe20}
{Boztepe} T.,  {G{\"o}{\u{g}}{\"u}{\c{s}}} E.,  {G{\"u}ver} T.,   {Schwenzer}
  K.,  2020, \mn@doi [\mnras] {10.1093/mnras/staa2503}, \href
  {https://ui.adsabs.harvard.edu/abs/2020MNRAS.498.2734B} {498, 2734}

\bibitem[\protect\citeauthoryear{{Breton}, {Rappaport}, {van Kerkwijk}  \&
  {Carter}}{{Breton} et~al.}{2012}]{breton12}
{Breton} R.~P.,  {Rappaport} S.~A.,  {van Kerkwijk} M.~H.,   {Carter} J.~A.,
  2012, \mn@doi [\apj] {10.1088/0004-637X/748/2/115}, \href
  {https://ui.adsabs.harvard.edu/abs/2012ApJ...748..115B} {748, 115}

\bibitem[\protect\citeauthoryear{{Breton} et~al.,}{{Breton}
  et~al.}{2013}]{breton13}
{Breton} R.~P.,  et~al., 2013, \mn@doi [\apj] {10.1088/0004-637X/769/2/108},
  \href {https://ui.adsabs.harvard.edu/abs/2013ApJ...769..108B} {769, 108}

\bibitem[\protect\citeauthoryear{{Britt}, {Strader}, {Chomiuk}, {Tremou},
  {Peacock}, {Halpern}  \& {Salinas}}{{Britt} et~al.}{2017}]{britt17}
{Britt} C.~T.,  {Strader} J.,  {Chomiuk} L.,  {Tremou} E.,  {Peacock} M.,
  {Halpern} J.,   {Salinas} R.,  2017, \mn@doi [\apj]
  {10.3847/1538-4357/aa8e41}, \href
  {https://ui.adsabs.harvard.edu/abs/2017ApJ...849...21B} {849, 21}

\bibitem[\protect\citeauthoryear{{Camilo} et~al.,}{{Camilo}
  et~al.}{2016}]{camilo16}
{Camilo} F.,  et~al., 2016, \mn@doi [\apj] {10.3847/0004-637X/820/1/6}, \href
  {https://ui.adsabs.harvard.edu/abs/2016ApJ...820....6C} {820, 6}

\bibitem[\protect\citeauthoryear{{Cheng}, {Gil}  \& {Zhang}}{{Cheng}
  et~al.}{1998}]{cheng98}
{Cheng} K.~S.,  {Gil} J.,   {Zhang} L.,  1998, \mn@doi [\apjl]
  {10.1086/311117}, \href
  {https://ui.adsabs.harvard.edu/abs/1998ApJ...493L..35C} {493, L35}

\bibitem[\protect\citeauthoryear{{Cheng}, {Taam}  \& {Wang}}{{Cheng}
  et~al.}{2004}]{cheng04}
{Cheng} K.~S.,  {Taam} R.~E.,   {Wang} W.,  2004, \mn@doi [\apj]
  {10.1086/425295}, \href
  {https://ui.adsabs.harvard.edu/abs/2004ApJ...617..480C} {617, 480}

\bibitem[\protect\citeauthoryear{{Cho}, {Halpern}  \& {Bogdanov}}{{Cho}
  et~al.}{2018}]{cho18}
{Cho} P.~B.,  {Halpern} J.~P.,   {Bogdanov} S.,  2018, \mn@doi [\apj]
  {10.3847/1538-4357/aade92}, \href
  {https://ui.adsabs.harvard.edu/abs/2018ApJ...866...71C} {866, 71}

\bibitem[\protect\citeauthoryear{{Clark} et~al.,}{{Clark}
  et~al.}{2021}]{clark21}
{Clark} C.~J.,  et~al., 2021, \mn@doi [\mnras] {10.1093/mnras/staa3484}, \href
  {https://ui.adsabs.harvard.edu/abs/2021MNRAS.502..915C} {502, 915}

\bibitem[\protect\citeauthoryear{{Clark} et~al.,}{{Clark}
  et~al.}{2023}]{clark23}
{Clark} C.~J.,  et~al., 2023, \mn@doi [Nature Astronomy]
  {10.1038/s41550-022-01874-x}, \href
  {https://ui.adsabs.harvard.edu/abs/2023NatAs.tmp...31C} {}

\bibitem[\protect\citeauthoryear{{Cordes} \& {Lazio}}{{Cordes} \&
  {Lazio}}{2002}]{cordes02}
{Cordes} J.~M.,  {Lazio} T.~J.~W.,  2002, arXiv e-prints, \href
  {https://ui.adsabs.harvard.edu/abs/2002astro.ph..7156C} {pp
  astro--ph/0207156}

\bibitem[\protect\citeauthoryear{{Corongiu} et~al.,}{{Corongiu}
  et~al.}{2021}]{corongiu21}
{Corongiu} A.,  et~al., 2021, \mn@doi [\mnras] {10.1093/mnras/staa3463}, \href
  {https://ui.adsabs.harvard.edu/abs/2021MNRAS.502..935C} {502, 935}

\bibitem[\protect\citeauthoryear{{Crawford} et~al.,}{{Crawford}
  et~al.}{2013}]{crawford13}
{Crawford} F.,  et~al., 2013, \mn@doi [\apj] {10.1088/0004-637X/776/1/20},
  \href {https://ui.adsabs.harvard.edu/abs/2013ApJ...776...20C} {776, 20}

\bibitem[\protect\citeauthoryear{{Creevey} et~al.,}{{Creevey}
  et~al.}{2022}]{creevey22}
{Creevey} O.~L.,  et~al., 2022, \mn@doi [arXiv e-prints]
  {10.48550/arXiv.2206.05864}, \href
  {https://ui.adsabs.harvard.edu/abs/2022arXiv220605864C} {p. arXiv:2206.05864}

\bibitem[\protect\citeauthoryear{{Cromartie} et~al.,}{{Cromartie}
  et~al.}{2016}]{cromartie16}
{Cromartie} H.~T.,  et~al., 2016, \mn@doi [\apj] {10.3847/0004-637X/819/1/34},
  \href {https://ui.adsabs.harvard.edu/abs/2016ApJ...819...34C} {819, 34}

\bibitem[\protect\citeauthoryear{{Crowter} et~al.,}{{Crowter}
  et~al.}{2020}]{crowter20}
{Crowter} K.,  et~al., 2020, \mn@doi [\mnras] {10.1093/mnras/staa933}, \href
  {https://ui.adsabs.harvard.edu/abs/2020MNRAS.495.3052C} {495, 3052}

\bibitem[\protect\citeauthoryear{{Deller} et~al.,}{{Deller}
  et~al.}{2012}]{deller12}
{Deller} A.~T.,  et~al., 2012, \mn@doi [\apjl] {10.1088/2041-8205/756/2/L25},
  \href {https://ui.adsabs.harvard.edu/abs/2012ApJ...756L..25D} {756, L25}

\bibitem[\protect\citeauthoryear{{Deneva} et~al.,}{{Deneva}
  et~al.}{2021}]{deneva21}
{Deneva} J.~S.,  et~al., 2021, \mn@doi [\apj] {10.3847/1538-4357/abd7a1}, \href
  {https://ui.adsabs.harvard.edu/abs/2021ApJ...909....6D} {909, 6}

\bibitem[\protect\citeauthoryear{{Ding} et~al.,}{{Ding} et~al.}{2023}]{ding23}
{Ding} H.,  et~al., 2023, \mn@doi [\mnras] {10.1093/mnras/stac3725}, \href
  {https://ui.adsabs.harvard.edu/abs/2023MNRAS.519.4982D} {519, 4982}

\bibitem[\protect\citeauthoryear{{Draghis}, {Romani}, {Filippenko}, {Brink},
  {Zheng}, {Halpern}  \& {Camilo}}{{Draghis} et~al.}{2019}]{draghis19}
{Draghis} P.,  {Romani} R.~W.,  {Filippenko} A.~V.,  {Brink} T.~G.,  {Zheng}
  W.,  {Halpern} J.~P.,   {Camilo} F.,  2019, \mn@doi [\apj]
  {10.3847/1538-4357/ab378b}, \href
  {https://ui.adsabs.harvard.edu/abs/2019ApJ...883..108D} {883, 108}

\bibitem[\protect\citeauthoryear{{Eggleton}}{{Eggleton}}{1983}]{eggleton83}
{Eggleton} P.~P.,  1983, \mn@doi [\apj] {10.1086/160960}, \href
  {https://ui.adsabs.harvard.edu/abs/1983ApJ...268..368E} {268, 368}

\bibitem[\protect\citeauthoryear{{Elsner} et~al.,}{{Elsner}
  et~al.}{2008}]{elsner08}
{Elsner} R.~F.,  et~al., 2008, \mn@doi [\apj] {10.1086/591899}, \href
  {https://ui.adsabs.harvard.edu/abs/2008ApJ...687.1019E} {687, 1019}

\bibitem[\protect\citeauthoryear{{GRAVITY Collaboration} et~al.,}{{GRAVITY
  Collaboration} et~al.}{2018}]{gravity18}
{GRAVITY Collaboration} et~al., 2018, \mn@doi [\aap]
  {10.1051/0004-6361/201833718}, \href
  {https://ui.adsabs.harvard.edu/abs/2018A&A...615L..15G} {615, L15}

\bibitem[\protect\citeauthoryear{{Gaia Collaboration} et~al.,}{{Gaia
  Collaboration} et~al.}{2016}]{gaia16b}
{Gaia Collaboration} et~al., 2016, \mn@doi [\aap]
  {10.1051/0004-6361/201629272}, \href
  {https://ui.adsabs.harvard.edu/abs/2016A&A...595A...1G} {595, A1}

\bibitem[\protect\citeauthoryear{{Gaia Collaboration} et~al.,}{{Gaia
  Collaboration} et~al.}{2022}]{gaia22k}
{Gaia Collaboration} et~al., 2022, arXiv e-prints, \href
  {https://ui.adsabs.harvard.edu/abs/2022arXiv220800211G} {p. arXiv:2208.00211}

\bibitem[\protect\citeauthoryear{{Gandhi} et~al.,}{{Gandhi}
  et~al.}{2022}]{gandhi22}
{Gandhi} P.,  et~al., 2022, \mn@doi [\mnras] {10.1093/mnras/stab3771}, \href
  {https://ui.adsabs.harvard.edu/abs/2022MNRAS.510.3885G} {510, 3885}

\bibitem[\protect\citeauthoryear{{Gentile}}{{Gentile}}{2018}]{gentile18}
{Gentile} P.~A.,  2018, PhD thesis, West Virginia University

\bibitem[\protect\citeauthoryear{{Gustafsson}, {Edvardsson}, {Eriksson},
  {J{\o}rgensen}, {Nordlund}  \& {Plez}}{{Gustafsson}
  et~al.}{2008}]{gustafsson08}
{Gustafsson} B.,  {Edvardsson} B.,  {Eriksson} K.,  {J{\o}rgensen} U.~G.,
  {Nordlund} {\r{A}}.,   {Plez} B.,  2008, \mn@doi [\aap]
  {10.1051/0004-6361:200809724}, \href
  {https://ui.adsabs.harvard.edu/abs/2008A&A...486..951G} {486, 951}

\bibitem[\protect\citeauthoryear{{Halpern}, {Strader}  \& {Li}}{{Halpern}
  et~al.}{2017}]{halpern17}
{Halpern} J.~P.,  {Strader} J.,   {Li} M.,  2017, \mn@doi [\apj]
  {10.3847/1538-4357/aa7cff}, \href
  {https://ui.adsabs.harvard.edu/abs/2017ApJ...844..150H} {844, 150}

\bibitem[\protect\citeauthoryear{{Halpern}, {Perez}  \& {Bogdanov}}{{Halpern}
  et~al.}{2022}]{halpern22}
{Halpern} J.~P.,  {Perez} K.~I.,   {Bogdanov} S.,  2022, \mn@doi [\apj]
  {10.3847/1538-4357/ac8161}, \href
  {https://ui.adsabs.harvard.edu/abs/2022ApJ...935..151H} {935, 151}

\bibitem[\protect\citeauthoryear{{Harding} \& {Gaisser}}{{Harding} \&
  {Gaisser}}{1990}]{harding90}
{Harding} A.~K.,  {Gaisser} T.~K.,  1990, \mn@doi [\apj] {10.1086/169009},
  \href {https://ui.adsabs.harvard.edu/abs/1990ApJ...358..561H} {358, 561}

\bibitem[\protect\citeauthoryear{{Harding} \& {Muslimov}}{{Harding} \&
  {Muslimov}}{2001}]{harding01}
{Harding} A.~K.,  {Muslimov} A.~G.,  2001, \mn@doi [\apj] {10.1086/321589},
  \href {https://ui.adsabs.harvard.edu/abs/2001ApJ...556..987H} {556, 987}

\bibitem[\protect\citeauthoryear{{Harding} \& {Muslimov}}{{Harding} \&
  {Muslimov}}{2002}]{harding02}
{Harding} A.~K.,  {Muslimov} A.~G.,  2002, \mn@doi [\apj] {10.1086/338985},
  \href {https://ui.adsabs.harvard.edu/abs/2002ApJ...568..862H} {568, 862}

\bibitem[\protect\citeauthoryear{{Hebbar}, {Heinke}, {Kandel}, {Romani}  \&
  {Freire}}{{Hebbar} et~al.}{2021}]{hebbar21}
{Hebbar} P.~R.,  {Heinke} C.~O.,  {Kandel} D.,  {Romani} R.~W.,   {Freire}
  P.~C.~C.,  2021, \mn@doi [\mnras] {10.1093/mnras/staa3072}, \href
  {https://ui.adsabs.harvard.edu/abs/2021MNRAS.500.1139H} {500, 1139}

\bibitem[\protect\citeauthoryear{{Huang} \& {Becker}}{{Huang} \&
  {Becker}}{2007}]{huang07}
{Huang} H.~H.,  {Becker} W.,  2007, \mn@doi [\aap]
  {10.1051/0004-6361:20066568}, \href
  {https://ui.adsabs.harvard.edu/abs/2007A&A...463L...5H} {463, L5}

\bibitem[\protect\citeauthoryear{{Hui}, {Tam}, {Takata}, {Kong}, {Cheng}, {Wu},
  {Lin}  \& {Wu}}{{Hui} et~al.}{2014}]{hui14}
{Hui} C.~Y.,  {Tam} P.~H.~T.,  {Takata} J.,  {Kong} A.~K.~H.,  {Cheng} K.~S.,
  {Wu} J.~H.~K.,  {Lin} L.~C.~C.,   {Wu} E.~M.~H.,  2014, \mn@doi [\apjl]
  {10.1088/2041-8205/781/1/L21}, \href
  {https://ui.adsabs.harvard.edu/abs/2014ApJ...781L..21H} {781, L21}

\bibitem[\protect\citeauthoryear{{Jennings}, {Kaplan}, {Chatterjee}, {Cordes}
  \& {Deller}}{{Jennings} et~al.}{2018}]{jennings18}
{Jennings} R.~J.,  {Kaplan} D.~L.,  {Chatterjee} S.,  {Cordes} J.~M.,
  {Deller} A.~T.,  2018, \mn@doi [\apj] {10.3847/1538-4357/aad084}, \href
  {https://ui.adsabs.harvard.edu/abs/2018ApJ...864...26J} {864, 26}

\bibitem[\protect\citeauthoryear{{Kalberla}, {Burton}, {Hartmann}, {Arnal},
  {Bajaja}, {Morras}  \& {P{\"o}ppel}}{{Kalberla} et~al.}{2005}]{kalberla05}
{Kalberla} P.~M.~W.,  {Burton} W.~B.,  {Hartmann} D.,  {Arnal} E.~M.,  {Bajaja}
  E.,  {Morras} R.,   {P{\"o}ppel} W.~G.~L.,  2005, \mn@doi [\aap]
  {10.1051/0004-6361:20041864}, \href
  {http://adsabs.harvard.edu/abs/2005A\%26A...440..775K} {440, 775}

\bibitem[\protect\citeauthoryear{{Kandel}, {Romani}  \& {An}}{{Kandel}
  et~al.}{2021}]{kandel21}
{Kandel} D.,  {Romani} R.~W.,   {An} H.,  2021, \mn@doi [\apjl]
  {10.3847/2041-8213/ac15f7}, \href
  {https://ui.adsabs.harvard.edu/abs/2021ApJ...917L..13K} {917, L13}

\bibitem[\protect\citeauthoryear{{Kaplan} et~al.,}{{Kaplan}
  et~al.}{2012}]{kaplan12}
{Kaplan} D.~L.,  et~al., 2012, \mn@doi [\apj] {10.1088/0004-637X/753/2/174},
  \href {https://ui.adsabs.harvard.edu/abs/2012ApJ...753..174K} {753, 174}

\bibitem[\protect\citeauthoryear{{Kaplan}, {Bhalerao}, {van Kerkwijk},
  {Koester}, {Kulkarni}  \& {Stovall}}{{Kaplan} et~al.}{2013}]{kaplan13}
{Kaplan} D.~L.,  {Bhalerao} V.~B.,  {van Kerkwijk} M.~H.,  {Koester} D.,
  {Kulkarni} S.~R.,   {Stovall} K.,  2013, \mn@doi [\apj]
  {10.1088/0004-637X/765/2/158}, \href
  {https://ui.adsabs.harvard.edu/abs/2013ApJ...765..158K} {765, 158}

\bibitem[\protect\citeauthoryear{{Kargaltsev} \& {Pavlov}}{{Kargaltsev} \&
  {Pavlov}}{2008}]{kargaltsev08}
{Kargaltsev} O.,  {Pavlov} G.~G.,  2008, in {Bassa} C.,  {Wang} Z.,  {Cumming}
  A.,   {Kaspi} V.~M.,  eds,  American Institute of Physics Conference Series
  Vol. 983, 40 Years of Pulsars: Millisecond Pulsars, Magnetars and More. pp
  171--185 (\mn@eprint {arXiv} {0801.2602}), \mn@doi{10.1063/1.2900138}

\bibitem[\protect\citeauthoryear{{Kargaltsev}, {Durant}, {Pavlov}  \&
  {Garmire}}{{Kargaltsev} et~al.}{2012}]{kargaltsev12}
{Kargaltsev} O.,  {Durant} M.,  {Pavlov} G.~G.,   {Garmire} G.,  2012, \mn@doi
  [\apjs] {10.1088/0067-0049/201/2/37}, \href
  {https://ui.adsabs.harvard.edu/abs/2012ApJS..201...37K} {201, 37}

\bibitem[\protect\citeauthoryear{{Keane} et~al.,}{{Keane}
  et~al.}{2018}]{keane18}
{Keane} E.~F.,  et~al., 2018, \mn@doi [\mnras] {10.1093/mnras/stx2126}, \href
  {https://ui.adsabs.harvard.edu/abs/2018MNRAS.473..116K} {473, 116}

\bibitem[\protect\citeauthoryear{{Kennedy} et~al.,}{{Kennedy}
  et~al.}{2020}]{kennedy20}
{Kennedy} M.~R.,  et~al., 2020, \mn@doi [\mnras] {10.1093/mnras/staa912}, \href
  {https://ui.adsabs.harvard.edu/abs/2020MNRAS.494.3912K} {494, 3912}

\bibitem[\protect\citeauthoryear{{Lee}, {Hui}, {Takata}, {Kong}, {Tam}  \&
  {Cheng}}{{Lee} et~al.}{2018}]{lee18}
{Lee} J.,  {Hui} C.~Y.,  {Takata} J.,  {Kong} A.~K.~H.,  {Tam} P.~H.~T.,
  {Cheng} K.~S.,  2018, \mn@doi [\apj] {10.3847/1538-4357/aad284}, \href
  {https://ui.adsabs.harvard.edu/abs/2018ApJ...864...23L} {864, 23}

\bibitem[\protect\citeauthoryear{{Lee}, {Hui}, {Takata}, {Kong}, {Tam}, {Li}
  \& {Cheng}}{{Lee} et~al.}{2023}]{lee23}
{Lee} J.,  {Hui} C.~Y.,  {Takata} J.,  {Kong} A.~K.~H.,  {Tam} P.-H.~T.,  {Li}
  K.-L.,   {Cheng} K.~S.,  2023, \mn@doi [arXiv e-prints]
  {10.48550/arXiv.2302.08776}, \href
  {https://ui.adsabs.harvard.edu/abs/2023arXiv230208776L} {p. arXiv:2302.08776}

\bibitem[\protect\citeauthoryear{{Li}, {Lu}  \& {Li}}{{Li} et~al.}{2008}]{li08}
{Li} X.-H.,  {Lu} F.-J.,   {Li} Z.,  2008, \mn@doi [\apj] {10.1086/589495},
  \href {https://ui.adsabs.harvard.edu/abs/2008ApJ...682.1166L} {682, 1166}

\bibitem[\protect\citeauthoryear{{Li}, {Halpern}  \& {Thorstensen}}{{Li}
  et~al.}{2014}]{li14}
{Li} M.,  {Halpern} J.~P.,   {Thorstensen} J.~R.,  2014, \mn@doi [\apj]
  {10.1088/0004-637X/795/2/115}, \href
  {https://ui.adsabs.harvard.edu/abs/2014ApJ...795..115L} {795, 115}

\bibitem[\protect\citeauthoryear{{Li} et~al.,}{{Li} et~al.}{2018}]{li18}
{Li} K.-L.,  et~al., 2018, \mn@doi [\apj] {10.3847/1538-4357/aad243}, \href
  {https://ui.adsabs.harvard.edu/abs/2018ApJ...863..194L} {863, 194}

\bibitem[\protect\citeauthoryear{{Li}, {Jane Yap}, {Hui}  \& {Kong}}{{Li}
  et~al.}{2021}]{li21}
{Li} K.-L.,  {Jane Yap} Y.~X.,  {Hui} C.~Y.,   {Kong} A. K.~H.,  2021, \mn@doi
  [\apj] {10.3847/1538-4357/abeb76}, \href
  {https://ui.adsabs.harvard.edu/abs/2021ApJ...911...92L} {911, 92}

\bibitem[\protect\citeauthoryear{{Linares}}{{Linares}}{2014}]{linares14}
{Linares} M.,  2014, \mn@doi [\apj] {10.1088/0004-637X/795/1/72}, \href
  {https://ui.adsabs.harvard.edu/abs/2014ApJ...795...72L} {795, 72}

\bibitem[\protect\citeauthoryear{{Linares}}{{Linares}}{2018}]{linares18a}
{Linares} M.,  2018, \mn@doi [\mnras] {10.1093/mnrasl/slx153}, \href
  {https://ui.adsabs.harvard.edu/abs/2018MNRAS.473L..50L} {473, L50}

\bibitem[\protect\citeauthoryear{{Linares} \& {Kachelrie{\ss}}}{{Linares} \&
  {Kachelrie{\ss}}}{2021}]{linares21}
{Linares} M.,  {Kachelrie{\ss}} M.,  2021, \mn@doi [\jcap]
  {10.1088/1475-7516/2021/02/030}, \href
  {https://ui.adsabs.harvard.edu/abs/2021JCAP...02..030L} {2021, 030}

\bibitem[\protect\citeauthoryear{{Linares}, {Miles-P{\'a}ez},
  {Rodr{\'\i}guez-Gil}, {Shahbaz}, {Casares}, {Fari{\~n}a}  \&
  {Karjalainen}}{{Linares} et~al.}{2017}]{linares17}
{Linares} M.,  {Miles-P{\'a}ez} P.,  {Rodr{\'\i}guez-Gil} P.,  {Shahbaz} T.,
  {Casares} J.,  {Fari{\~n}a} C.,   {Karjalainen} R.,  2017, \mn@doi [\mnras]
  {10.1093/mnras/stw3057}, \href
  {https://ui.adsabs.harvard.edu/abs/2017MNRAS.465.4602L} {465, 4602}

\bibitem[\protect\citeauthoryear{{Linares}, {Shahbaz}  \& {Casares}}{{Linares}
  et~al.}{2018}]{linares18b}
{Linares} M.,  {Shahbaz} T.,   {Casares} J.,  2018, \mn@doi [\apj]
  {10.3847/1538-4357/aabde6}, \href
  {https://ui.adsabs.harvard.edu/abs/2018ApJ...859...54L} {859, 54}

\bibitem[\protect\citeauthoryear{{Lindegren} et~al.,}{{Lindegren}
  et~al.}{2021a}]{lindegren21b}
{Lindegren} L.,  et~al., 2021a, \mn@doi [\aap] {10.1051/0004-6361/202039653},
  \href {https://ui.adsabs.harvard.edu/abs/2021A&A...649A...4L} {649, A4}

\bibitem[\protect\citeauthoryear{{Lindegren} et~al.,}{{Lindegren}
  et~al.}{2021b}]{lindegren21a}
{Lindegren} L.,  et~al., 2021b, \mn@doi [\aap] {10.1051/0004-6361/202039653},
  \href {https://ui.adsabs.harvard.edu/abs/2021A&A...649A...4L} {649, A4}

\bibitem[\protect\citeauthoryear{{Long}, {Kong}, {Wu}, {Takata}, {Han}, {Hui}
  \& {Li}}{{Long} et~al.}{2022}]{long22}
{Long} J.~S.,  {Kong} A. K.~H.,  {Wu} K.,  {Takata} J.,  {Han} Q.,  {Hui} D.
  C.~Y.,   {Li} K.~L.,  2022, \mn@doi [\apj] {10.3847/1538-4357/ac7720}, \href
  {https://ui.adsabs.harvard.edu/abs/2022ApJ...934...17L} {934, 17}

\bibitem[\protect\citeauthoryear{{Lorimer} et~al.,}{{Lorimer}
  et~al.}{2006}]{lorimer06}
{Lorimer} D.~R.,  et~al., 2006, \mn@doi [\mnras]
  {10.1111/j.1365-2966.2006.10887.x}, \href
  {https://ui.adsabs.harvard.edu/abs/2006MNRAS.372..777L} {372, 777}

\bibitem[\protect\citeauthoryear{{Marelli}, {De Luca}  \& {Caraveo}}{{Marelli}
  et~al.}{2011}]{marelli11}
{Marelli} M.,  {De Luca} A.,   {Caraveo} P.~A.,  2011, \mn@doi [\apj]
  {10.1088/0004-637X/733/2/82}, \href
  {https://ui.adsabs.harvard.edu/abs/2011ApJ...733...82M} {733, 82}

\bibitem[\protect\citeauthoryear{{Matthews} et~al.,}{{Matthews}
  et~al.}{2016}]{matthews16}
{Matthews} A.~M.,  et~al., 2016, \mn@doi [\apj] {10.3847/0004-637X/818/1/92},
  \href {https://ui.adsabs.harvard.edu/abs/2016ApJ...818...92M} {818, 92}

\bibitem[\protect\citeauthoryear{{Miller} et~al.,}{{Miller}
  et~al.}{2020}]{miller20}
{Miller} J.~M.,  et~al., 2020, \mn@doi [\apj] {10.3847/1538-4357/abbb2e}, \href
  {https://ui.adsabs.harvard.edu/abs/2020ApJ...904...49M} {904, 49}

\bibitem[\protect\citeauthoryear{{Nieder} et~al.,}{{Nieder}
  et~al.}{2020}]{nieder20}
{Nieder} L.,  et~al., 2020, \mn@doi [\apjl] {10.3847/2041-8213/abbc02}, \href
  {https://ui.adsabs.harvard.edu/abs/2020ApJ...902L..46N} {902, L46}

\bibitem[\protect\citeauthoryear{{Oh}, {Hui}, {Li}  \& {Kong}}{{Oh}
  et~al.}{2020}]{oh20}
{Oh} K.,  {Hui} C.~Y.,  {Li} K.~L.,   {Kong} A.~K.~H.,  2020, \mn@doi [\mnras]
  {10.1093/mnras/staa2462}, \href
  {https://ui.adsabs.harvard.edu/abs/2020MNRAS.498..292O} {498, 292}

\bibitem[\protect\citeauthoryear{{Orosz} \& {Hauschildt}}{{Orosz} \&
  {Hauschildt}}{2000}]{orosz00}
{Orosz} J.~A.,  {Hauschildt} P.~H.,  2000, \aap, \href
  {https://ui.adsabs.harvard.edu/abs/2000A&A...364..265O} {364, 265}

\bibitem[\protect\citeauthoryear{{Perez}, {Bogdanov}, {Halpern}  \&
  {Gajjar}}{{Perez} et~al.}{2023}]{perez23}
{Perez} K.~I.,  {Bogdanov} S.,  {Halpern} J.~P.,   {Gajjar} V.,  2023, \mn@doi
  [arXiv e-prints] {10.48550/arXiv.2306.04951}, \href
  {https://ui.adsabs.harvard.edu/abs/2023arXiv230604951P} {p. arXiv:2306.04951}

\bibitem[\protect\citeauthoryear{{Phinney}, {Evans}, {Blandford}  \&
  {Kulkarni}}{{Phinney} et~al.}{1988}]{phinney88}
{Phinney} E.~S.,  {Evans} C.~R.,  {Blandford} R.~D.,   {Kulkarni} S.~R.,  1988,
  \mn@doi [\nat] {10.1038/333832a0}, \href
  {https://ui.adsabs.harvard.edu/abs/1988Natur.333..832P} {333, 832}

\bibitem[\protect\citeauthoryear{{Pletsch} et~al.,}{{Pletsch}
  et~al.}{2012}]{pletsch12}
{Pletsch} H.~J.,  et~al., 2012, \mn@doi [\apj] {10.1088/0004-637X/744/2/105},
  \href {https://ui.adsabs.harvard.edu/abs/2012ApJ...744..105P} {744, 105}

\bibitem[\protect\citeauthoryear{{Posselt}, {Pavlov}, {Manchester},
  {Kargaltsev}  \& {Garmire}}{{Posselt} et~al.}{2012}]{posselt12}
{Posselt} B.,  {Pavlov} G.~G.,  {Manchester} R.~N.,  {Kargaltsev} O.,
  {Garmire} G.~P.,  2012, \mn@doi [\apj] {10.1088/0004-637X/749/2/146}, \href
  {https://ui.adsabs.harvard.edu/abs/2012ApJ...749..146P} {749, 146}

\bibitem[\protect\citeauthoryear{{Possenti}, {Cerutti}, {Colpi}  \&
  {Mereghetti}}{{Possenti} et~al.}{2002}]{possenti02}
{Possenti} A.,  {Cerutti} R.,  {Colpi} M.,   {Mereghetti} S.,  2002, \mn@doi
  [\aap] {10.1051/0004-6361:20020472}, \href
  {https://ui.adsabs.harvard.edu/abs/2002A&A...387..993P} {387, 993}

\bibitem[\protect\citeauthoryear{{Ransom}, {Stairs}, {Backer}, {Greenhill},
  {Bassa}, {Hessels}  \& {Kaspi}}{{Ransom} et~al.}{2004}]{ransom04}
{Ransom} S.~M.,  {Stairs} I.~H.,  {Backer} D.~C.,  {Greenhill} L.~J.,  {Bassa}
  C.~G.,  {Hessels} J. W.~T.,   {Kaspi} V.~M.,  2004, \mn@doi [\apj]
  {10.1086/381730}, \href
  {https://ui.adsabs.harvard.edu/abs/2004ApJ...604..328R} {604, 328}

\bibitem[\protect\citeauthoryear{{Rea} et~al.,}{{Rea} et~al.}{2017}]{rea17}
{Rea} N.,  et~al., 2017, \mn@doi [\mnras] {10.1093/mnras/stx1560}, \href
  {https://ui.adsabs.harvard.edu/abs/2017MNRAS.471.2902R} {471, 2902}

\bibitem[\protect\citeauthoryear{{Riello} et~al.,}{{Riello}
  et~al.}{2021}]{riello21}
{Riello} M.,  et~al., 2021, \mn@doi [\aap] {10.1051/0004-6361/202039587}, \href
  {https://ui.adsabs.harvard.edu/abs/2021A&A...649A...3R} {649, A3}

\bibitem[\protect\citeauthoryear{{Roberts}, {McLaughlin}, {Gentile}, {Ray},
  {Ransom}  \& {Hessels}}{{Roberts} et~al.}{2015}]{roberts15}
{Roberts} M.~S.~E.,  {McLaughlin} M.~A.,  {Gentile} P.~A.,  {Ray} P.~S.,
  {Ransom} S.~M.,   {Hessels} J.~W.~T.,  2015, arXiv:1502.07208; 2014 Fermi
  Symposium proceedings - eConf C14102.1, \href
  {http://adsabs.harvard.edu/abs/2015arXiv150207208R} {}

\bibitem[\protect\citeauthoryear{{Romani} \& {Sanchez}}{{Romani} \&
  {Sanchez}}{2016}]{romani16}
{Romani} R.~W.,  {Sanchez} N.,  2016, \mn@doi [\apj]
  {10.3847/0004-637X/828/1/7}, \href
  {https://ui.adsabs.harvard.edu/abs/2016ApJ...828....7R} {828, 7}

\bibitem[\protect\citeauthoryear{{Romani} \& {Shaw}}{{Romani} \&
  {Shaw}}{2011}]{romani11}
{Romani} R.~W.,  {Shaw} M.~S.,  2011, \mn@doi [\apjl]
  {10.1088/2041-8205/743/2/L26}, \href
  {https://ui.adsabs.harvard.edu/abs/2011ApJ...743L..26R} {743, L26}

\bibitem[\protect\citeauthoryear{{Romani}, {Filippenko}  \& {Cenko}}{{Romani}
  et~al.}{2014}]{romani14}
{Romani} R.~W.,  {Filippenko} A.~V.,   {Cenko} S.~B.,  2014, \mn@doi [\apjl]
  {10.1088/2041-8205/793/1/L20}, \href
  {https://ui.adsabs.harvard.edu/abs/2014ApJ...793L..20R} {793, L20}

\bibitem[\protect\citeauthoryear{{Salvetti} et~al.,}{{Salvetti}
  et~al.}{2015}]{salvetti15}
{Salvetti} D.,  et~al., 2015, \mn@doi [\apj] {10.1088/0004-637X/814/2/88},
  \href {https://ui.adsabs.harvard.edu/abs/2015ApJ...814...88S} {814, 88}

\bibitem[\protect\citeauthoryear{{Saz Parkinson} et~al.,}{{Saz Parkinson}
  et~al.}{2010}]{sazparkinson10}
{Saz Parkinson} P.~M.,  et~al., 2010, \mn@doi [\apj]
  {10.1088/0004-637X/725/1/571}, \href
  {https://ui.adsabs.harvard.edu/abs/2010ApJ...725..571S} {725, 571}

\bibitem[\protect\citeauthoryear{{Seward} \& {Wang}}{{Seward} \&
  {Wang}}{1988}]{seward88}
{Seward} F.~D.,  {Wang} Z.-R.,  1988, \mn@doi [\apj] {10.1086/166646}, \href
  {https://ui.adsabs.harvard.edu/abs/1988ApJ...332..199S} {332, 199}

\bibitem[\protect\citeauthoryear{{Smits}, {Tingay}, {Wex}, {Kramer}  \&
  {Stappers}}{{Smits} et~al.}{2011}]{smits11}
{Smits} R.,  {Tingay} S.~J.,  {Wex} N.,  {Kramer} M.,   {Stappers} B.,  2011,
  \mn@doi [\aap] {10.1051/0004-6361/201016141}, \href
  {https://ui.adsabs.harvard.edu/abs/2011A&A...528A.108S} {528, A108}

\bibitem[\protect\citeauthoryear{{Spiewak} et~al.,}{{Spiewak}
  et~al.}{2016}]{spiewak16}
{Spiewak} R.,  et~al., 2016, \mn@doi [\apj] {10.3847/0004-637X/822/1/37}, \href
  {https://ui.adsabs.harvard.edu/abs/2016ApJ...822...37S} {822, 37}

\bibitem[\protect\citeauthoryear{{Stappers} et~al.,}{{Stappers}
  et~al.}{2014}]{stappers14}
{Stappers} B.~W.,  et~al., 2014, \mn@doi [\apj] {10.1088/0004-637X/790/1/39},
  \href {https://ui.adsabs.harvard.edu/abs/2014ApJ...790...39S} {790, 39}

\bibitem[\protect\citeauthoryear{{Stovall} et~al.,}{{Stovall}
  et~al.}{2014}]{stovall14}
{Stovall} K.,  et~al., 2014, \mn@doi [\apj] {10.1088/0004-637X/791/1/67}, \href
  {https://ui.adsabs.harvard.edu/abs/2014ApJ...791...67S} {791, 67}

\bibitem[\protect\citeauthoryear{{Strader}, {Chomiuk}, {Sonbas}, {Sokolovsky},
  {Sand}, {Moskvitin}  \& {Cheung}}{{Strader} et~al.}{2014}]{strader14}
{Strader} J.,  {Chomiuk} L.,  {Sonbas} E.,  {Sokolovsky} K.,  {Sand} D.~J.,
  {Moskvitin} A.~S.,   {Cheung} C.~C.,  2014, \mn@doi [\apjl]
  {10.1088/2041-8205/788/2/L27}, \href
  {https://ui.adsabs.harvard.edu/abs/2014ApJ...788L..27S} {788, L27}

\bibitem[\protect\citeauthoryear{{Strader} et~al.,}{{Strader}
  et~al.}{2015}]{strader15}
{Strader} J.,  et~al., 2015, \mn@doi [\apjl] {10.1088/2041-8205/804/1/L12},
  \href {https://ui.adsabs.harvard.edu/abs/2015ApJ...804L..12S} {804, L12}

\bibitem[\protect\citeauthoryear{{Strader}, {Li}, {Chomiuk}, {Heinke},
  {Udalski}, {Peacock}, {Shishkovsky}  \& {Tremou}}{{Strader}
  et~al.}{2016}]{strader16}
{Strader} J.,  {Li} K.-L.,  {Chomiuk} L.,  {Heinke} C.~O.,  {Udalski} A.,
  {Peacock} M.,  {Shishkovsky} L.,   {Tremou} E.,  2016, \mn@doi [\apj]
  {10.3847/0004-637X/831/1/89}, \href
  {https://ui.adsabs.harvard.edu/abs/2016ApJ...831...89S} {831, 89}

\bibitem[\protect\citeauthoryear{{Strader} et~al.,}{{Strader}
  et~al.}{2019}]{strader19}
{Strader} J.,  et~al., 2019, \mn@doi [\apj] {10.3847/1538-4357/aafbaa}, \href
  {https://ui.adsabs.harvard.edu/abs/2019ApJ...872...42S} {872, 42}

\bibitem[\protect\citeauthoryear{{Stringer} et~al.,}{{Stringer}
  et~al.}{2021}]{stringer21}
{Stringer} J.~G.,  et~al., 2021, \mn@doi [\mnras] {10.1093/mnras/stab2167},
  \href {https://ui.adsabs.harvard.edu/abs/2021MNRAS.507.2174S} {507, 2174}

\bibitem[\protect\citeauthoryear{{Swihart} et~al.,}{{Swihart}
  et~al.}{2017}]{swihart17}
{Swihart} S.~J.,  et~al., 2017, \mn@doi [\apj] {10.3847/1538-4357/aa9937},
  \href {https://ui.adsabs.harvard.edu/abs/2017ApJ...851...31S} {851, 31}

\bibitem[\protect\citeauthoryear{{Swihart} et~al.,}{{Swihart}
  et~al.}{2018}]{swihart18}
{Swihart} S.~J.,  et~al., 2018, \mn@doi [\apj] {10.3847/1538-4357/aadcab},
  \href {https://ui.adsabs.harvard.edu/abs/2018ApJ...866...83S} {866, 83}

\bibitem[\protect\citeauthoryear{{Swihart}, {Strader}, {Chomiuk}  \&
  {Shishkovsky}}{{Swihart} et~al.}{2019}]{swihart19}
{Swihart} S.~J.,  {Strader} J.,  {Chomiuk} L.,   {Shishkovsky} L.,  2019,
  \mn@doi [\apj] {10.3847/1538-4357/ab125e}, \href
  {https://ui.adsabs.harvard.edu/abs/2019ApJ...876....8S} {876, 8}

\bibitem[\protect\citeauthoryear{{Swihart} et~al.,}{{Swihart}
  et~al.}{2020}]{swihart20}
{Swihart} S.~J.,  et~al., 2020, \mn@doi [\apj] {10.3847/1538-4357/ab77ba},
  \href {https://ui.adsabs.harvard.edu/abs/2020ApJ...892...21S} {892, 21}

\bibitem[\protect\citeauthoryear{{Swihart}, {Strader}, {Aydi}, {Chomiuk},
  {Dage}  \& {Shishkovsky}}{{Swihart} et~al.}{2021}]{swihart21}
{Swihart} S.~J.,  {Strader} J.,  {Aydi} E.,  {Chomiuk} L.,  {Dage} K.~C.,
  {Shishkovsky} L.,  2021, \mn@doi [\apj] {10.3847/1538-4357/abe1be}, \href
  {https://ui.adsabs.harvard.edu/abs/2021ApJ...909..185S} {909, 185}

\bibitem[\protect\citeauthoryear{{Swihart}, {Strader}, {Chomiuk}, {Aydi},
  {Sokolovsky}, {Ray}  \& {Kerr}}{{Swihart} et~al.}{2022}]{swihart22}
{Swihart} S.~J.,  {Strader} J.,  {Chomiuk} L.,  {Aydi} E.,  {Sokolovsky} K.~V.,
   {Ray} P.~S.,   {Kerr} M.,  2022, arXiv e-prints, \href
  {https://ui.adsabs.harvard.edu/abs/2022arXiv221016295S} {p. arXiv:2210.16295}

\bibitem[\protect\citeauthoryear{{Tendulkar} et~al.,}{{Tendulkar}
  et~al.}{2014}]{tendulkar14}
{Tendulkar} S.~P.,  et~al., 2014, \mn@doi [\apj] {10.1088/0004-637X/791/2/77},
  \href {https://ui.adsabs.harvard.edu/abs/2014ApJ...791...77T} {791, 77}

\bibitem[\protect\citeauthoryear{{Vahdat}, {Posselt}, {Santangelo}  \&
  {Pavlov}}{{Vahdat} et~al.}{2022}]{vahdat22}
{Vahdat} A.,  {Posselt} B.,  {Santangelo} A.,   {Pavlov} G.~G.,  2022, \mn@doi
  [\aap] {10.1051/0004-6361/202141795}, \href
  {https://ui.adsabs.harvard.edu/abs/2022A&A...658A..95V} {658, A95}

\bibitem[\protect\citeauthoryear{{VanderPlas} \& {Ivezi{\'c}}}{{VanderPlas} \&
  {Ivezi{\'c}}}{2015}]{vanderplas15}
{VanderPlas} J.~T.,  {Ivezi{\'c}} {\v{Z}}.,  2015, \mn@doi [\apj]
  {10.1088/0004-637X/812/1/18}, \href
  {https://ui.adsabs.harvard.edu/abs/2015ApJ...812...18V} {812, 18}

\bibitem[\protect\citeauthoryear{Vanderplas}{Vanderplas}{2015}]{gatspy}
Vanderplas J.,  2015, {gatspy: General tools for Astronomical Time Series in
  Python}, \mn@doi{10.5281/zenodo.14833}, \url
  {https://doi.org/10.5281/zenodo.14833}

\bibitem[\protect\citeauthoryear{{Verbiest}, {Weisberg}, {Chael}, {Lee}  \&
  {Lorimer}}{{Verbiest} et~al.}{2012}]{verbiest12}
{Verbiest} J.~P.~W.,  {Weisberg} J.~M.,  {Chael} A.~A.,  {Lee} K.~J.,
  {Lorimer} D.~R.,  2012, \mn@doi [\apj] {10.1088/0004-637X/755/1/39}, \href
  {https://ui.adsabs.harvard.edu/abs/2012ApJ...755...39V} {755, 39}

\bibitem[\protect\citeauthoryear{{Vurgun} et~al.,}{{Vurgun}
  et~al.}{2022}]{vurgun22}
{Vurgun} E.,  et~al., 2022, \mn@doi [\apj] {10.3847/1538-4357/ac9ea0}, \href
  {https://ui.adsabs.harvard.edu/abs/2022ApJ...941...76V} {941, 76}

\bibitem[\protect\citeauthoryear{{Wadiasingh}, {Harding}, {Venter},
  {B{\"o}ttcher}  \& {Baring}}{{Wadiasingh} et~al.}{2017}]{wadiasingh17}
{Wadiasingh} Z.,  {Harding} A.~K.,  {Venter} C.,  {B{\"o}ttcher} M.,   {Baring}
  M.~G.,  2017, \mn@doi [\apj] {10.3847/1538-4357/aa69bf}, \href
  {https://ui.adsabs.harvard.edu/abs/2017ApJ...839...80W} {839, 80}

\bibitem[\protect\citeauthoryear{{Wadiasingh}, {Venter}, {Harding},
  {B{\"o}ttcher}  \& {Kilian}}{{Wadiasingh} et~al.}{2018}]{wadiasingh18}
{Wadiasingh} Z.,  {Venter} C.,  {Harding} A.~K.,  {B{\"o}ttcher} M.,   {Kilian}
  P.,  2018, \mn@doi [\apj] {10.3847/1538-4357/aaed43}, \href
  {https://ui.adsabs.harvard.edu/abs/2018ApJ...869..120W} {869, 120}

\bibitem[\protect\citeauthoryear{{Wilms}, {Allen}  \& {McCray}}{{Wilms}
  et~al.}{2000}]{wilms00}
{Wilms} J.,  {Allen} A.,   {McCray} R.,  2000, \mn@doi [\apj] {10.1086/317016},
  542, 914

\bibitem[\protect\citeauthoryear{{Yao}, {Manchester}  \& {Wang}}{{Yao}
  et~al.}{2017}]{yao17}
{Yao} J.~M.,  {Manchester} R.~N.,   {Wang} N.,  2017, \mn@doi [\apj]
  {10.3847/1538-4357/835/1/29}, \href
  {https://ui.adsabs.harvard.edu/abs/2017ApJ...835...29Y} {835, 29}

\bibitem[\protect\citeauthoryear{{Yap}, {Li}, {Kong}, {Takata}, {Lee}  \&
  {Hui}}{{Yap} et~al.}{2019}]{yap19}
{Yap} Y.~X.,  {Li} K.~L.,  {Kong} A.~K.~H.,  {Takata} J.,  {Lee} J.,   {Hui}
  C.~Y.,  2019, \mn@doi [\aap] {10.1051/0004-6361/201834545}, \href
  {https://ui.adsabs.harvard.edu/abs/2019A&A...621L...9Y} {621, L9}

\bibitem[\protect\citeauthoryear{{Zhao} \& {Heinke}}{{Zhao} \&
  {Heinke}}{2022}]{zhao22}
{Zhao} J.,  {Heinke} C.~O.,  2022, \mn@doi [\mnras] {10.1093/mnras/stac442},
  \href {https://ui.adsabs.harvard.edu/abs/2022MNRAS.511.5964Z} {511, 5964}

\bibitem[\protect\citeauthoryear{{Zhao} et~al.,}{{Zhao} et~al.}{2020}]{zhao20}
{Zhao} Y.,  et~al., 2020, \mn@doi [\mnras] {10.1093/mnras/staa2927}, \href
  {https://ui.adsabs.harvard.edu/abs/2020MNRAS.499.3338Z} {499, 3338}

\bibitem[\protect\citeauthoryear{{Zhao}, {Zhao}  \& {Heinke}}{{Zhao}
  et~al.}{2021}]{zhao21}
{Zhao} J.,  {Zhao} Y.,   {Heinke} C.~O.,  2021, \mn@doi [\mnras]
  {10.1093/mnras/stab117}, \href
  {https://ui.adsabs.harvard.edu/abs/2021MNRAS.502.1596Z} {502, 1596}

\bibitem[\protect\citeauthoryear{{Zheng}, {Wang}, {Xing}  \&
  {Vadakkumthani}}{{Zheng} et~al.}{2022}]{zheng22}
{Zheng} D.,  {Wang} Z.-X.,  {Xing} Y.,   {Vadakkumthani} J.,  2022, \mn@doi
  [Research in Astronomy and Astrophysics] {10.1088/1674-4527/ac3e5e}, \href
  {https://ui.adsabs.harvard.edu/abs/2022RAA....22b5012Z} {22, 025012}

\bibitem[\protect\citeauthoryear{{de Martino} et~al.,}{{de Martino}
  et~al.}{2014}]{demartino14}
{de Martino} D.,  et~al., 2014, \mn@doi [\mnras] {10.1093/mnras/stu1640}, \href
  {https://ui.adsabs.harvard.edu/abs/2014MNRAS.444.3004D} {444, 3004}

\bibitem[\protect\citeauthoryear{{van Staden} \& {Antoniadis}}{{van Staden} \&
  {Antoniadis}}{2016}]{vanstaden16}
{van Staden} A.~D.,  {Antoniadis} J.,  2016, \mn@doi [\apjl]
  {10.3847/2041-8213/833/1/L12}, \href
  {https://ui.adsabs.harvard.edu/abs/2016ApJ...833L..12V} {833, L12}

\bibitem[\protect\citeauthoryear{{van der Merwe}, {Wadiasingh}, {Venter},
  {Harding}  \& {Baring}}{{van der Merwe} et~al.}{2020}]{vandermerwe20}
{van der Merwe} C.~J.~T.,  {Wadiasingh} Z.,  {Venter} C.,  {Harding} A.~K.,
  {Baring} M.~G.,  2020, \mn@doi [\apj] {10.3847/1538-4357/abbdfb}, \href
  {https://ui.adsabs.harvard.edu/abs/2020ApJ...904...91V} {904, 91}

\bibitem[\protect\citeauthoryear{{van der Wateren} et~al.,}{{van der Wateren}
  et~al.}{2022}]{vanderwateren22}
{van der Wateren} E.,  et~al., 2022, \mn@doi [\aap]
  {10.1051/0004-6361/202142741}, \href
  {https://ui.adsabs.harvard.edu/abs/2022A&A...661A..57V} {661, A57}

\makeatother
\end{thebibliography}




\appendix

\section{Gaia geometric distance}

Following the description given by \citet{bailerjones21}, we calculate the posterior probability density function for the distance to each source, where the posterior is a product of the likelihood and prior: 

\begin{equation}
    P (d | \omega, \sigma_{\omega}, p) = P (\omega |d, \sigma_{\omega}) P (d | p), 
\end{equation}

\noindent where $d$ is the distance, $\omega$ is the parallax, $\sigma_{\omega}$ is the parallax error, and $p$ is a HEALpixel (Hierarchical Equal Area isoLatitude Pixelation) number. We considered two distinct priors for distance estimation: the distance prior employed by \citet{bailerjones21}, based on stellar populations, and the prior derived from the distribution of pulsars modeled in \citet{lorimer06}. Although the pulsar-based prior is more fitting for spider systems, it's important to note that this distribution is highly contingent on the assumed distribution of free electrons within the Galaxy (\citealt{lorimer06} used the one from \citealt{cordes02}). Furthermore, together with the substantially lower statistics (around 1000 pulsars versus 1.47 billion stars), this can significantly impact the accuracy of the distribution estimate. 

The prior used in \citet{bailerjones21} is a three-parameter generalized gamma distribution depending on the HEALpixel. Each sky direction has different prior parameters ($\alpha$, $\beta$, L) that are given in the auxiliary information online of \citet{bailerjones21}. The prior can be written as:

\begin{equation}
     P (d | p) = 
     \begin{cases}
       \frac{1}{\Gamma(\frac{\beta+1}{\alpha})} \frac{\alpha}{L^{\beta+1}} d^{b}  e^{-(d/L)^{\alpha}} & \mathrm{if} \, d \geq 0 \\
       0 & \text{otherwise.}
     \end{cases}
\end{equation}

\noindent The Galactic distribution of pulsars utilized in \citet{lorimer06} is characterized by a double-exponential function encompassing both radial (galactocentric) and distance-above-the-Galactic-plane components. In our approach, we adopted the formulation outlined in \citet{verbiest12} , which transforms the prior into an Earth-based coordinate system, utilizing Galactic longitude ($G_{\mathrm{l}}$) and latitude ($G_{\mathrm{b}}$) as parameters:

\begin{equation}
    P (d | G_{\mathrm{b}}, G_{\mathrm{l}}) \propto R^{1.9} \mathrm{exp} \Bigg[ -\frac{|z|}{E} - 5\frac{R-R_{0}}{R_{0}}\Bigg]d^2 ,
\end{equation}

\noindent where 

\begin{equation}
    z(d, G_{\mathrm{b}}) = d\sin{G_{\mathrm{b}}} \, ,
\end{equation}

\noindent and

\begin{equation}
    R(d, G_{\mathrm{b}}, G_{\mathrm{l}}) = \sqrt{R_{0}^{2} + (d\cos{G_{\mathrm{b}}}) - 2 R_{0} d \cos{G_{\mathrm{b}}} \cos{G_{\mathrm{l}}} } \, .
\end{equation}

\noindent The constants $R_{0}$ and $E$ are the distance to the Galactic center and the scale height, for which we use values of $R_{0} = 8.12$ kpc \citep{gravity18} and $E = 0.4$ kpc (the scale height of spiders found by \citealt{linares21}).

For Gaussian parallax uncertainties, the likelihood is:

\begin{equation}
    P (\omega |d, \sigma_{\omega}) = \frac{1}{\sqrt{2\pi}\sigma_{\omega}} \mathrm{exp} \Bigg[-\frac{1}{2\sigma_{\omega}^{2}} \Bigg(\omega-\omega_{\mathrm{zp}}-\frac{1}{d} \Bigg)^{2} \Bigg],  
\end{equation}

\noindent where $\omega_{\mathrm{zp}}$ is the parallax bias (zero point). We used the method described in \citet{lindegren21a} to estimate the zero point for each source.

\begin{figure}
    \includegraphics[width=\columnwidth]{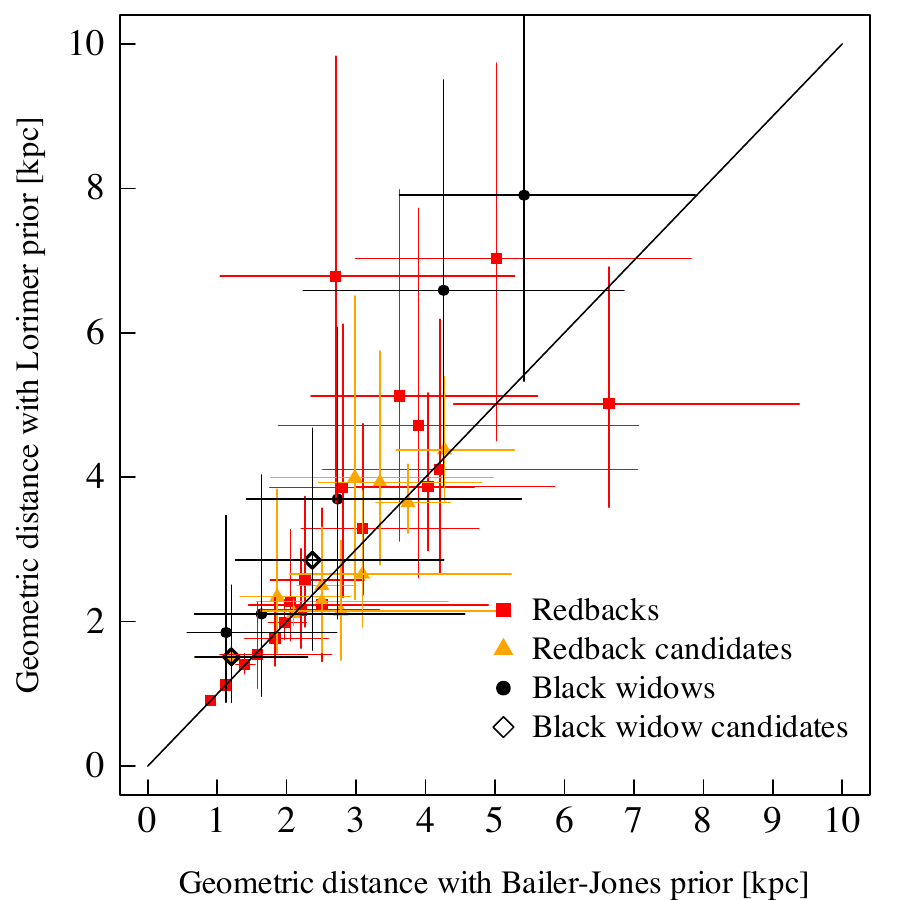}
    \caption{Comparison of \textit{Gaia} DR3 geometric distances of spiders using two different priors from \citet{bailerjones21} and \citet{lorimer06}.} 
    \label{fig:prior}
\end{figure}

We compared the distances using the different priors and found that they agree within errors (Fig. \ref{fig:prior}). For spiders located at greater distances, a subtle inclination towards larger distances is noticeable when utilizing the pulsar prior. Should this hold true, it could potentially amplify the disparity between the geometric distance and the dispersion measure. Nevertheless, the impact of the prior choice on the results presented in the paper is negligible.

\section{Spider distances}

Tables \ref{tab:distance} and \ref{tab:distance_cand} tabulate the DM, Gaia DR2, and Gaia DR3, as well as literature (radio, optical) distance estimates of the Galactic field spiders and spider candidates. 

\begin{table*} \centering
  \caption{DM, Gaia DR2, and Gaia DR3, as well as literature (radio, optical) distance estimates of the Galactic field spiders. The references correspond to the distances inferred using other distance methods ($d_{\rm other}$).}
  \label{tab:distance}
\begin{tabular}{@{\extracolsep{5pt}} lcccccc}
\\[-1.8ex]\hline
\hline \\[-1.8ex]
Source name & DM & $d_{\rm DM}$ & $d_{\rm DR2}$ & $d_{\rm DR3}$ & $d_{\rm other}$ & Ref. \\
& (cm$^{-3}$ pc) & (kpc) & (kpc) & (kpc) & (kpc) & \\
\hline \\[-1.8ex]
\textbf{Redbacks} \\
\hline \\[-1.8ex]
J0212+5320   & 25.7 & 1.26 & 1.11 & 1.09 & 0.92 & 1 \\
& & (1.24, 1.31) & (1.21, 0.84) & (1.15, 1.12) & (0.8, 1.08) \\
J1023$+$0038 & 14.3 & 1.11 & 1.27 & 1.39 & 1.30 & 2 \\
& & (1.04, 1.18) & (1.07, 1.54) & (1.27, 1.52) & (1.28, 1.32) \\
J1036$-$4353 & 61.1 & 0.41 & 1.82 & 2.87 & -- & -- \\
& & (0.21, 0.62) & (1.09, 3.23) & (1.79, 4.61) & & \\
J1048$+$2339 & 16.7 & 1.95 & 0.85 & 1.35 & -- & -- \\
& & (1.75, 2.24) & (0.54, 1.44) & (0.89, 1.96) & & \\
J1227$-$4853 & 43.4 & 1.25 & 1.58 & 2.44 & 1.8 & 2 \\
& & (1.17, 1.32) & (1.21, 2.24) & (1.85, 3.22) & (1.7, 1.9) \\
J1306$-$40   & 35.0 & 1.41 & 2.83 & 3.22 & 4.7 & 3 \\
& & (1.30, 1.51) & (1.84, 4.47) & (2.25, 5.19) & (4.2, 5.2) \\
J1431$-$4715 & 59.4 & 1.82 & 1.59 & 1.99 & -- & -- \\
& & (1.45, 2.22) & (1.20, 2.34) & (1.61, 2.58) & & \\
J1622$-$0315 & 21.4 & 1.13 & 2.74 & 5.84 & -- & -- \\
& & (1.05, 1.24) & (1.37, 5.32) & (1.66, 7.77) & & \\
J1628$-$3205 & 42.1 & 1.17 & 1.55 & 4.46 & -- & -- \\
& & (1.01, 1.41) & (0.69, 5.39) & (1.92, 8.10) & & \\
J1723$-$2837 & 19.7 & 0.72 & 0.91 & 0.90 & -- & -- \\
& & (0.69, 0.74) & (0.86, 0.96) & (0.87, 0.95) & & \\
J1803$-$6707 & 38.4 & 1.36 & 3.70 & 4.24 & -- & -- \\
& & (1.16, 1.55) & (2.18, 6.27) & (2.80, 6.29) & & \\
J1816$+$4510 & 38.9 & 4.27 & 2.67 & 4.09 & 4.5 & 4 \\
& & (3.88, 4.80) & (1.99, 3.72) & (3.12, 5.94) & (2.8, 7.2) \\
J1908$+$2105 & 61.9 & 2.55 & 3.75 & 4.12 & -- & --\\
& & (2.37, 2.77) & (1.87, 6.80) & (2.70, 6.91) & & \\
J1910$-$5320 & 24.4 & 0.98 & 3.67 & 7.01 & 4.06 & 5 \\
& & (0.91, 1.05) & (2.01, 6.59) & (4.41, 10.02) & (3.78, 4.41) \\
J1957$+$2516 & 44.1 & 2.66 & 2.40 & 1.86 & -- & -- \\
& & (2.55, 2.78) & (1.07, 4.79) & (0.67, 2.57) & & \\
J2039$-$5618 & 24.6 & 1.70 & 2.26 & 2.14 & 3.4 & 6 \\
& & (1.54, 1.87) & (1.48, 3.82) & (1.57, 3.02) & (3.0, 3.8) \\
J2129$-$0429 & 16.9 & 1.38 & 2.13 & 1.99 & 1.83 & 7 \\
& & (1.30, 1.45) & (1.78, 2.62) & (1.80, 2.30) & (1.72, 1.94) \\
J2215$+$5135 & 69.2 & 2.78 & 2.47 & 3.36 & 2.9 & 8 \\
& & (2.72, 2.84) & (1.46, 4.34) & (2.22, 4.75) & (2.8, 3.0) \\
J2339$-$0533 & 8.7 & 0.75 & 1.18 & 1.71 & 1.1 & 9 \\
& & (0.71, 0.80) & (0.90, 1.64) & (1.38, 2.40) & (0.8, 1.4) \\
\hline \\[-1.8ex]
\textbf{Black widows} \\
\hline \\[-1.8ex]
J1311-3430 & 37.8 & 2.40 & -- & 0.90 & -- & -- \\
& & (2.08 2.72) & -- & (0.50, 3.07) \\
J1653-0158 & -- & -- & 2.02 & 1.00 & 0.84 & 10 \\
& & & (0.61, 4.80) & (0.55, 2.32) & (0.80, 0.88) \\
J1810+1744 & 39.7 & 2.39 & 1.64 & 2.43 & -- & -- \\
& & (2.13, 2.64) & (0.76, 3.71) & (1.46, 3.69) \\
J1928+1245 & 179.2 & 6.08 & 3.65 & 5.39 & -- & -- \\
& & (5.81, 6.34) & (2.36, 5.95) & (3.43, 8.03) \\
B1957+20   & 29.1 & 1.73 & 3.08 & 3.74 & -- & -- \\
& & (1.66, 1.81) & (1.42, 5.80) & (1.54, 7.59) \\
\hline \\[-1.8ex]
\multicolumn{7}{|p{0.70\linewidth}|}{\textbf{References:} 1) \citet{linares17}, 2) \citet{stringer21}, 3) \citet{swihart19}, 4) \citet{kaplan13}, 5) \citet{au23}, 6) \citet{strader19}, 7) \citet{bellm16}, 8) \citet{linares18b}, 9) \citet{romani11}, 10) \citet{nieder20}}
\end{tabular}
\end{table*}

\begin{table*} \centering
  \caption{DM, Gaia DR2, and Gaia DR3, as well as literature (radio, optical) distance estimates of the Galactic field spider candidates. The references correspond to the distances inferred using other distance methods ($d_{\rm other}$).}
  \label{tab:distance_cand}
\begin{tabular}{@{\extracolsep{5pt}} lcccccc}
\\[-1.8ex]\hline
\hline \\[-1.8ex]
Source name & DM & $d_{\rm DM}$ & $d_{\rm DR2}$ & $d_{\rm DR3}$ & $d_{\rm other}$ & Ref. \\
& (cm$^{-3}$ pc) & (kpc) & (kpc) & (kpc) & (kpc) & \\
\hline \\[-1.8ex]
\textbf{Redback candidates} \\
\hline \\[-1.8ex]
J0407.7-5702 & -- & -- & 1.85 & 2.06 & -- & -- \\
& & & (1.27, 2.72) & (1.49, 3.02) \\
J0427.9-6704 & -- & -- & 2.27 & 2.50 & 2.40 & 1 \\
& & & (1.95, 2.69) & (2.17, 2.97) & (2.10, 2.70) \\
J0523-2529   & -- & -- & 2.12 & 2.16 & 1.10 & 2 \\
& & & (1.92, 2.38) & (1.98, 2.40) & (0.80, 1.40) \\
J0838.8-2829 & -- & -- & 1.67 & 2.25 & $<$1700 & 3 \\
& & & (0.92, 3.22) & (1.42, 3.69) \\
J0846.0+2820 & -- & -- & 3.52 & 3.76 & -- & -- \\
& & & (3.04, 4.14) & (3.28, 4.26) \\
J0935.3+0901 & -- & -- & -- & 1.17 & -- & -- \\
& & & & (0.72, 1.74) \\
J0940.3-7610 & -- & -- & 1.50 & 1.79 & 2.20 & 4 \\
& & & (1.03, 2.45) & (1.32, 2.42) & (1.90, 2.70) \\
J0954.8-3948 & -- & -- & 2.39 & 3.47 & -- & -- \\
& & & (1.712, 3.55) & (2.48, 4.96) \\
J1417.5-4402 & 55 & 2.25 & 3.72 & 4.24 & 3.10 & 5 \\
& & (1.74, 2.67) & (2.92, 4.97) & (3.54, 4.92) & (2.50, 3.70) \\
J1544-1128   & -- & -- & 1.87 & 2.81 & 3.80 & 6 \\
& & & (1.15, 3.77) & (2.01, 3.77) & (3.10, 4.50) \\
J2333.1-5527 & -- & -- & 1.96 & 2.93 & 3.10 & 7 \\
& & & (1.25, 3.06) & (1.71, 4.64) & (2.80, 3.40) \\
\hline \\[-1.8ex]
\textbf{Black widow candidates} \\
\hline \\[-1.8ex]
J0336.0+7505 & -- & -- & -- & 1.69 & -- & -- \\
& & & & (1.08, 2.73) \\
\hline \\[-1.8ex]
\multicolumn{7}{|p{0.70\linewidth}|}{\textbf{References:} 1) \citet{strader16}, 2) \citet{strader14}, 3) \citet{halpern17}, 4) \citet{swihart21}, 5) \citet{swihart18}, 6) \citet{britt17}, 7) \citet{swihart20}}
\end{tabular}
\end{table*}

\section{Optical and X-ray properties of spiders}

\begin{figure*}
    \includegraphics[width=\columnwidth]{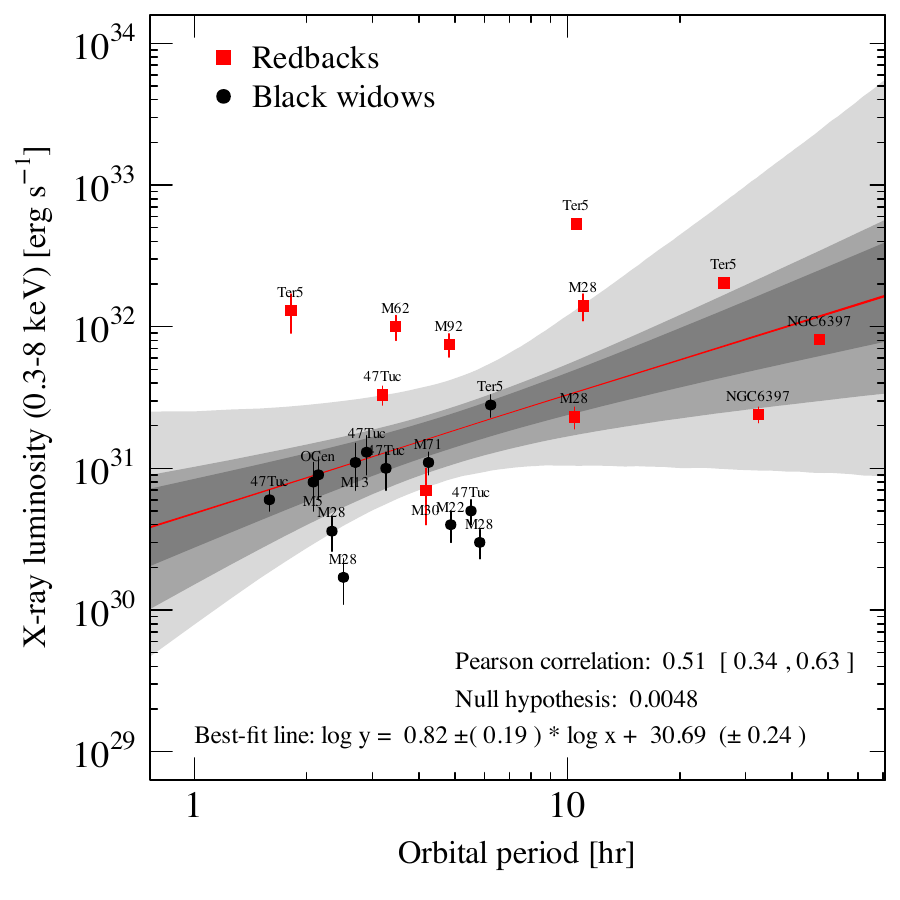}
    \includegraphics[width=\columnwidth]
    {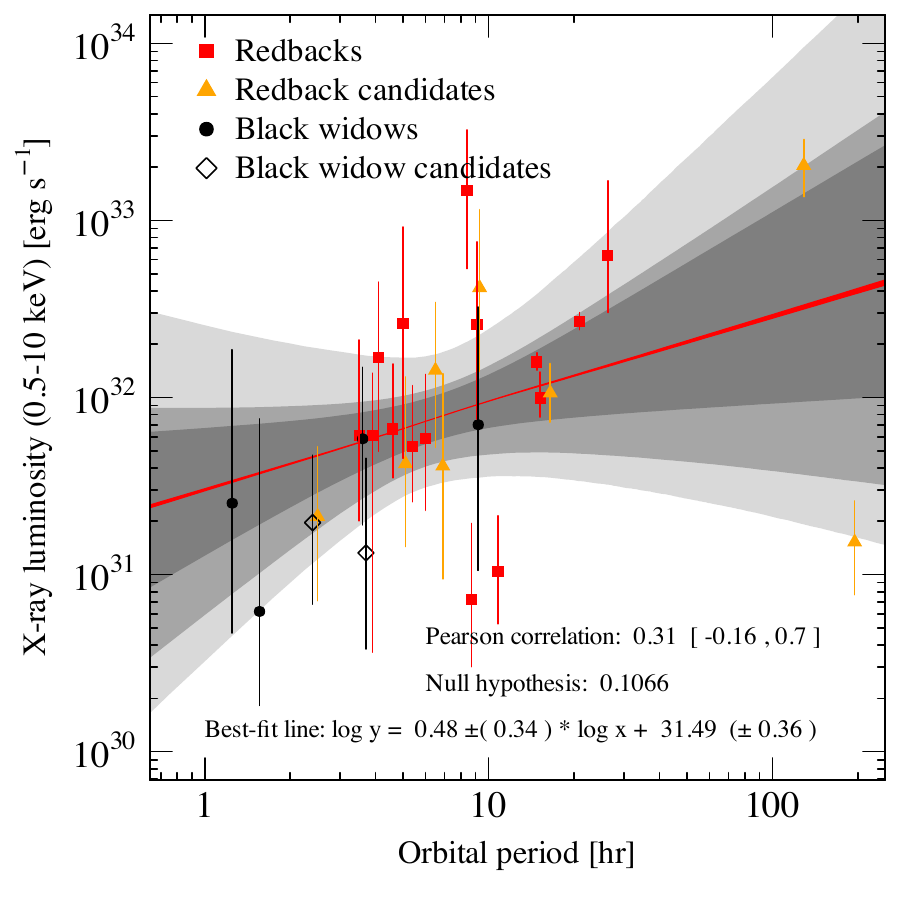}
    \caption{Unabsorbed X-ray luminosity as a function of the orbital period of the globular cluster (\textit{left}) and Galactic field (\textit{right}) spiders. We do not find a significant correlation between the X-ray luminosity and the orbital period.}
    \label{fig:gc_spider_orb_lum}
\end{figure*}

\begin{figure}
    \includegraphics[width=\columnwidth]{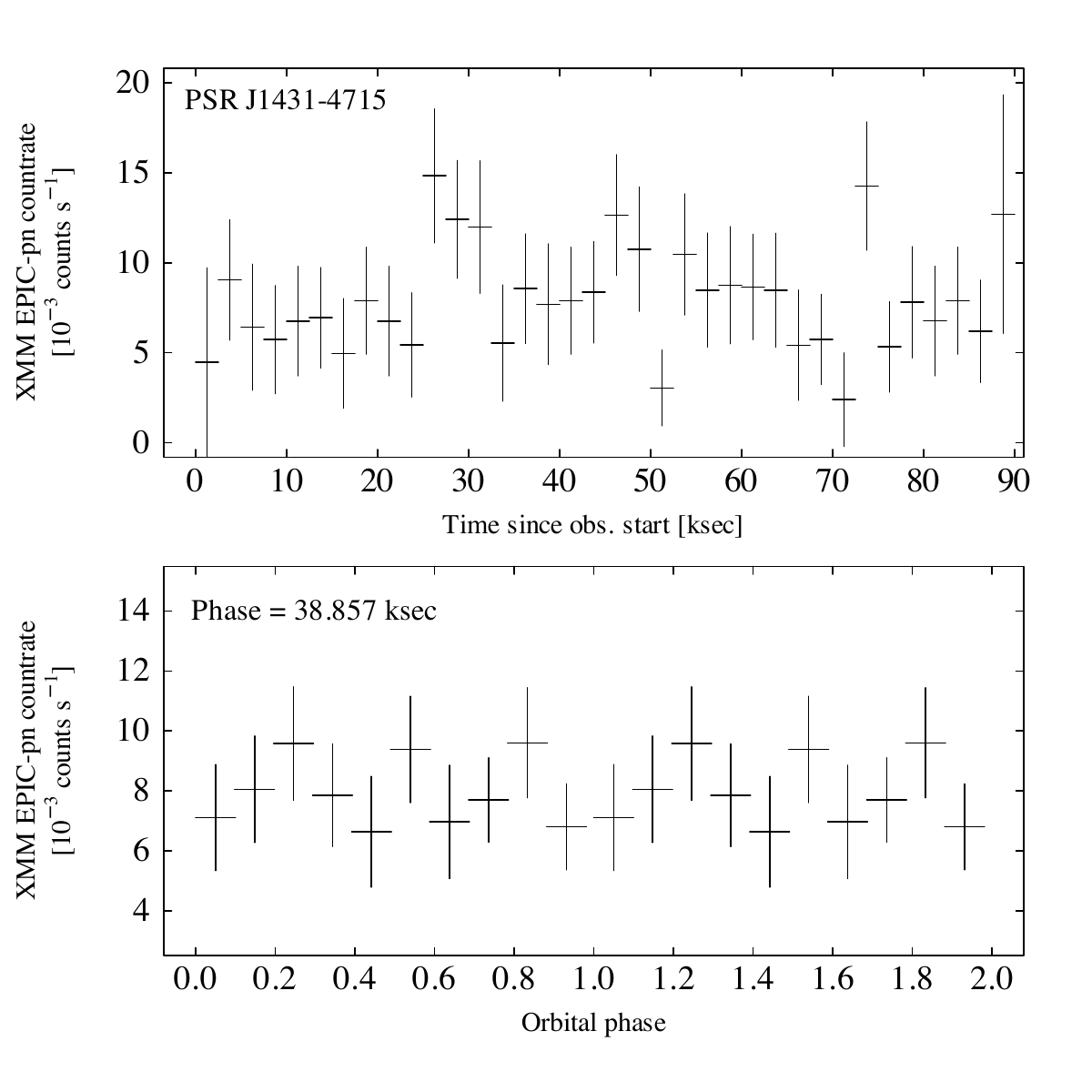}
    \caption{{\it Upper panel:} {\it XMM-Newton}/EPIC-pn light curve of PSR J1431$-$4715 with 2.5 ksec bins. {\it Lower panel:} The above light curve folded to the orbital period and binned to 10 bins over the orbit. We found that the count rate remained approximately constant over the full exposure spanning two orbits.}
    \label{fig:J1431_XMM}
\end{figure}

Tables \ref{tab:gal_spider_x} and \ref{tab:gal_spider_cand_x} tabulate the optical bolometric luminosities derived from \textit{Gaia} G-band magnitude and the corresponding extinction, spin-down and unabsorbed X-ray luminosities, and X-ray power law indices of the Galactic field spiders and spider candidates in the passive disk (pulsar) state. Fig. \ref{fig:gc_spider_orb_lum} shows the X-ray luminosity as a function of the orbital period of the Galactic field and globular cluster spiders. Fig. \ref{fig:J1431_XMM} shows the {\it XMM-Newton}/EPIC-pn light curve of PSR J1431$-$4715 extracted from pointing 0860430101 observed in May 2022.

\begin{table*} \centering
  \caption{Optical bolometric luminosities derived from \textit{Gaia} G-band magnitude and the corresponding extinction and bolometric correction used, spin-down and unabsorbed X-ray luminosities, and X-ray power law indices of the Galactic field spiders in the passive disk (pulsar) state assuming the parallax-derived distances. References correspond to the X-ray fluxes and power law indices found in the literature. The X-ray properties derived in this paper are marked with `T' in the reference column.}
  \label{tab:gal_spider_x}
\begin{tabular}{@{\extracolsep{5pt}} lccccccl}
\\[-1.8ex]\hline
\hline \\[-1.8ex]
Source name & $L_{\rm G}$ & A$_{\rm G}$ & BC$_{\rm G}$ & $\dot{E}$ & $L_{\rm X}$ & $\Gamma$ & Ref. \\
& (10$^{33}$ erg/s) & (mag) & (mag) & (10$^{34}$ erg/s) & (10$^{31}$ erg/s) \\
\hline \\[-1.8ex]
\textbf{Redbacks} \\
\hline \\[-1.8ex]
J0212+5320   & 11.6 & 0.68 & 0.09 & -- & 27.0 & 1.29 & 1 \\
& (11.0, 12.3) & & & & (24.2, 30.2) & (1.23, 1.35) \\
J1036$-$4353 & 0.4 & 0.35 & 0.08 & -- & -- & -- & -- \\
& (0.2, 1.0) & & & & & & \\
J1048$+$2339 & 0.11 & 0.07 & -0.37 & 0.94 & 5.9 & 1.2 & 2 \\
& (0.05, 0.24) & & & (0.87,1.02) & (2.3, 13.5) & (1.0, 1.4) \\
J1306$-$40   & 2.3 & 0.20 & -0.14 & -- & 63.2 & 1.31 & 3 \\
& (1.1, 6.0) & & & & (30.2, 167.4) & (1.27, 1.35) \\
J1431$-$4715 & 1.2 & 0.42 & 0.09 & 5.2 & 1.0 & 1.2 & T \\
& (0.8, 2.1) & & & (4.9, 5.5) & (0.5, 2.1) & (1.0, 1.4) \\
J1622$-$0315 & 2.1 & 0.15 & 0.09 & 0.09 & 6.1 & -- & 4 \\
& (0.2, 3.7) & & & (0.07, 0.11) & (0.4, 13.7) \\
J1628$-$3205 & 0.9 & 0.38 & 0.37 & -- & 26.2 & 1.6 & T \\
& (0.2, 3.0) & & & & (4.5, 91.8) & (1.2, 2.0) \\
J1723$-$2837 & 2.2 & 0.56 & 0.08 & 2.8 & 15.9 & 1.00 & 5 \\
& (2.0, 2.4) & & & (2.5, 3.2) & (14.3, 17.9) & (0.93, 1.07) \\
J1803$-$6707 & 0.9 & 0.15 & 0.08 & 6.5 & 25.9 & -- & 6 \\
& (0.4, 2.0) & & & (6.1, 6.8) & (7.5, 75.7) & \\
J1816$+$4510 & 4.6 & 0.22 & -0.44 & 5.17 & 0.7 & 2.6 & T \\
& (2.7, 9.6) & & & (5.15, 5.19) & (0.3, 1.9) & (2.1, 3.1) \\
J1908$+$2105 & 0.4 & 0.79 & 0.08 & 2.8 & 6.1 & 1.3 & 7 \\
& (0.2, 1.2) & & & (2.5, 3.1) & (2.0, 21.1) & (0.8, 1.8) \\
J1910$-$5320 & 3.3 & 0.13 & 0.08 & -- & 147.1 & 1.0 & 8 \\
& (1.3, 6.7) & & & & (53.4, 324.3) & (0.6, 1.4) \\
J1957$+$2516 & 0.18 & 1.02 & 0.08 & -- & -- & -- & -- \\
& (0.02, 0.34) \\
J2039$-$5618 & 0.6 & 0.09 & 0.00 & 2.5 & 5.3 & 1.36 & 9 \\
& (0.3, 1.1) & & & (2.3, 2.6) & (2.6, 11.7) & (1.27, 1.45) \\
J2129$-$0429 & 2.3 & 0.08 & -0.05 & -- & 9.9 & 1.03 & 10 \\
& (1.9, 3.1) & & & & (7.8, 13.9) & (0.93, 1.13) \\
J2215$+$5135 & 1.3 & 0.77 & 0.08 & 7.4 & 16.9 & 1.8 & 11 \\
& (0.6, 2.5) & & & (7.2, 7.6) & (5.0, 44.9) & (1.4, 2.2) \\
J2339$-$0533 & 0.3 & 0.05 & -0.04 & 2.1 & 6.6 & 1.07 & T \\
& (0.2, 0.5) & & & (2.0, 2.2) & (3.5, 15.5) & (0.95, 1.19) \\
\hline \\[-1.8ex]
\textbf{Black widows} \\
\hline \\[-1.8ex]
J1311$-$3430 & 0.02 & 0.17 & 0.08 & 4.85 & 0.62 & 1.67 & 12 \\
& (0.01, 0.19) & & & (4.80, 4.90) & (0.18, 7.6) & (1.58, 1.76) \\
J1653$-$0158 & 0.02 & 0.18 & 0.08 & 0.5 & 2.5 & 1.65 & 13 \\
& (0.01, 0.11) & & & (0.1, 0.8) & (0.48, 18.6) & (1.26, 1.99) \\
J1810$+$1744 & 0.2 & 0.30 & 0.08 & -- & 5.9 & 1.72 & 14 \\
& (0.1, 0.5) & & & & (1.9, 14.8) & (1.63, 1.81) \\
J1928$+$1245 & 22.0 & 1.35 & -0.47 & 2.28 & -- & -- & -- \\
& (9.0, 48.7) & & & (2.23, 2.33) & & & \\
B1957$+$20   & 0.5 & 0.44 & 0.08 & 0.67 & 7.0 & 1.56 & 15 \\
& (0.1, 1.9) & & & (0.66, 0.68) & (1.1, 32.4) & (1.35, 1.77) \\
\hline \\[-1.8ex]
\multicolumn{8}{|p{0.7\linewidth}|}{\textbf{References:} 1) \citet{linares17}, 2) \citet{cho18}, 3) \citet{linares18a}, 4) \citet{vahdat22}, 5) \citet{hui14}, 6) \citet{clark23}, 7) \citet{gentile18}, 8) \citet{au23}, 9) \citet{salvetti15}, 10) \citet{alnoori18}, 11) \citet{linares14}, 12) \citet{an17}, 13) \citet{romani14}, 14) \citet{boztepe20}, 15) \citet{kandel21}} \\
\end{tabular}
\end{table*}

\begin{table*} \centering
  \caption{Optical bolometric luminosities derived from \textit{Gaia} G-band magnitude and the corresponding extinction and bolometric correction used, unabsorbed X-ray luminosities, and X-ray power law indices of the Galactic field spider candidates in the passive disk (pulsar) state assuming the parallax-derived distances. References correspond to the X-ray fluxes and power law indices found in the literature. The X-ray properties derived in this paper are marked with `T' in the reference column.}
  \label{tab:gal_spider_cand_x}
\begin{tabular}{@{\extracolsep{5pt}} lcccccl}
\\[-1.8ex]\hline
\hline \\[-1.8ex]
Source name & L$_{\rm G}$ & A$_{\rm G}$ & BC$_{\rm G}$ & L$_{\rm X}$ & $\Gamma$ & Ref. \\
& (10$^{31}$ erg/s) & (mag) & (mag) & (10$^{31}$ erg/s) \\
\hline \\[-1.8ex]
\textbf{Redback candidates} \\
\hline \\[-1.8ex]
J0523-2529   & 3.9 & 0.10 & -0.10 & 10.6 & 1.5 & T \\
& (3.3, 4.8) & & & (7.3, 15.6) & (1.2, 1.8) \\
J0838.8-2829 & 0.2 & 0.51 & 0.08 & 4.2 & 1.6 & 1 \\
& (0.1, 0.6) & & & (1.4, 13.1) & (1.4, 1.8) \\
J0846.0+2820 & 25.8 & 0.24 & 0.05 & 1.5 & -- & -- \\
& (19.6, 33.0) & & & (0.8, 2.6) & \\
J0935.3+0901 & 0.02 & 0.12 & 0.08 & 2.1 & 1.9 & 2 \\
& (0.01, 0.05) & & & (0.7, 5.3) & (1.7, 2.1) \\
J0940.3-7610 & 0.4 & 0.58 & -0.27 & 14.2 & -- & 3 \\
& (0.2, 0.7) & & & (5.2, 34.4) & \\
J0954.8-3948 & 1.9 & 0.41 & 0.06 & 41.7 & 0.8 & 4 \\
& (1.0, 3.9) & & & (14.0, 114.9) & (0.3, 1.3) \\
J1417.5-4402 & 32.3 & 0.28 & -0.01 & 204.4 & 1.41 & 5 \\
& (22.5, 43.3) & & & (136.2, 286.5) & (1.32, 1.50) \\
J2333.1-5527 & 0.2 & 0.06 & -0.01 & 4.1 & 1.6 & 6 \\
& (0.1, 0.5) & & & (0.9, 13.6) & (1.3, 1.9) \\
\hline \\[-1.8ex]
\textbf{Black widow candidates} \\
\hline \\[-1.8ex]
J0336.0$+$7505 & 0.07 & 0.56 & 0.08 & 1.3 & 1.6 & 7 \\
& (0.03, 0.19) & & & (0.4, 4.5) & (0.9, 2.3) \\ 
J0935.3$+$0901 & 0.02 & 0.12 & 0.08 & 2.0 & 1.9 & 8 \\
& (0.01, 0.05) & & & (0.7, 4.7) & (1.7, 2.1) \\
\hline \\[-1.8ex]
\multicolumn{7}{|p{0.7\linewidth}|}{\textbf{References:} 1) \citet{rea17}, 2) \citet{zheng22}, 3) \citet{swihart21}, 4) \citet{li18}, 5) \citet{swihart18}, 6) \citet{swihart20}, 7) \citet{li21}, 8) \citet{zheng22}.}
\end{tabular}
\end{table*}

\section{Period searches for \textit{Gaia} epoch photometry}

For those \textit{Gaia} sources that have epoch photometry, we performed a period search (or folded the \textit{Gaia} data to a known period if no period was found) using the Lomb-Scargle periodogram from \textsc{gatspy} Python package with either one or two Fourier frequencies to be fitted in the model. In the figures below, the left panels show the Lomb-Scargle periodogram of the \textit{Gaia} lightcurve (black), a window function (red), the best/literature period (highlighted by arrows indicating the period peak), and the local false alarm probabilities (5\%, 1\%, 0.1\%) at the period peak (shown as dotted, dashed, and dot-dashed horizontal lines, respectively). These probability levels are derived by assuming a null hypothesis of non-varying data with Gaussian noise and the probability based on bootstrap resamplings of the input data. The right panels show the light curve folded to the best/literature period. In all sources excluding 4FGL J0427.9-6704, PSR J1023$+$0038 (both accreting systems), and 2FGL J0846.0$+$2820 (the longest period redback candidate), we detect a period consistent with the literature value.

\begin{figure*}
    \includegraphics[width=\textwidth]{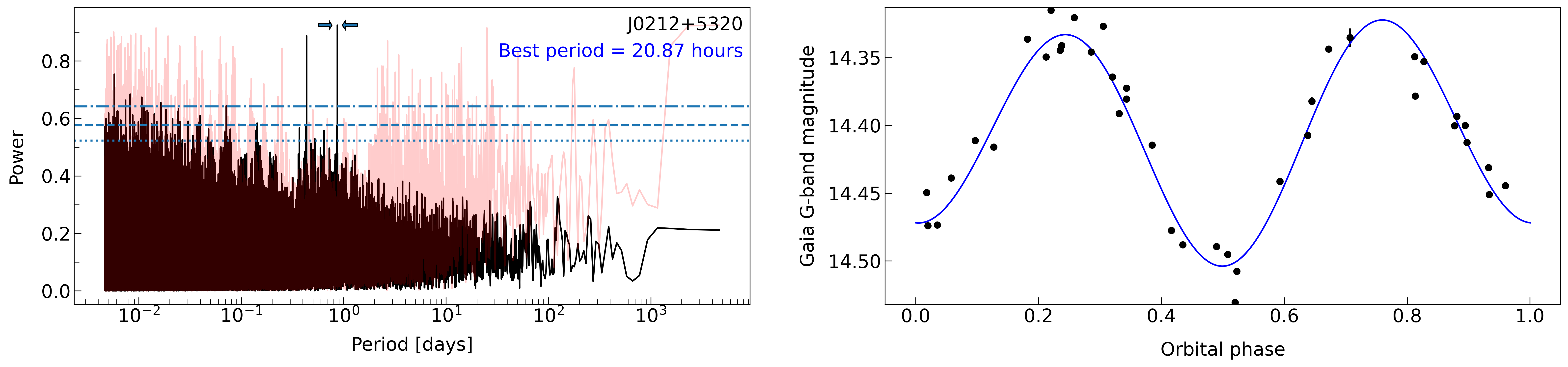}\hfill
    \includegraphics[width=\textwidth]{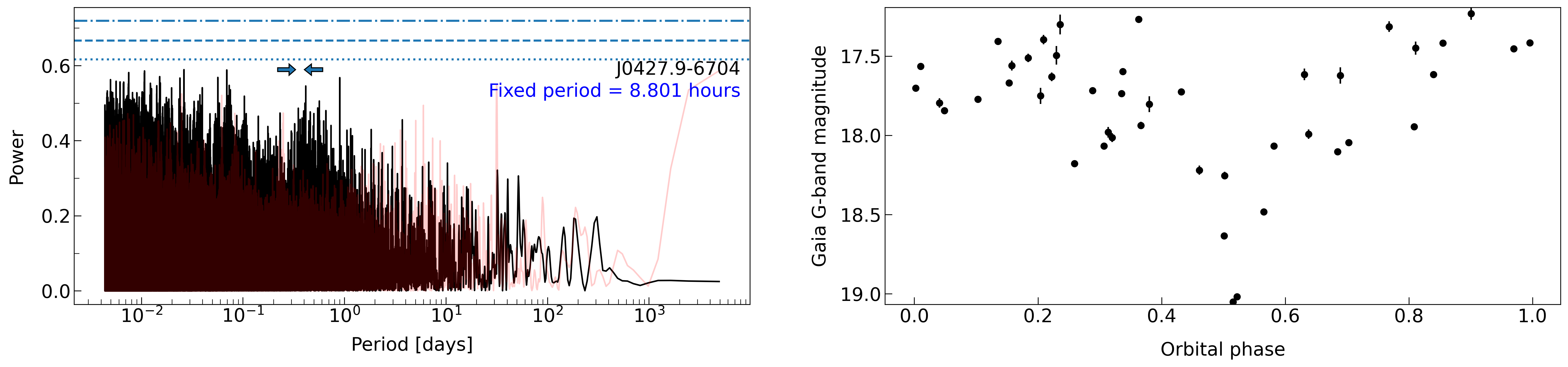}\hfill
    \includegraphics[width=\textwidth]{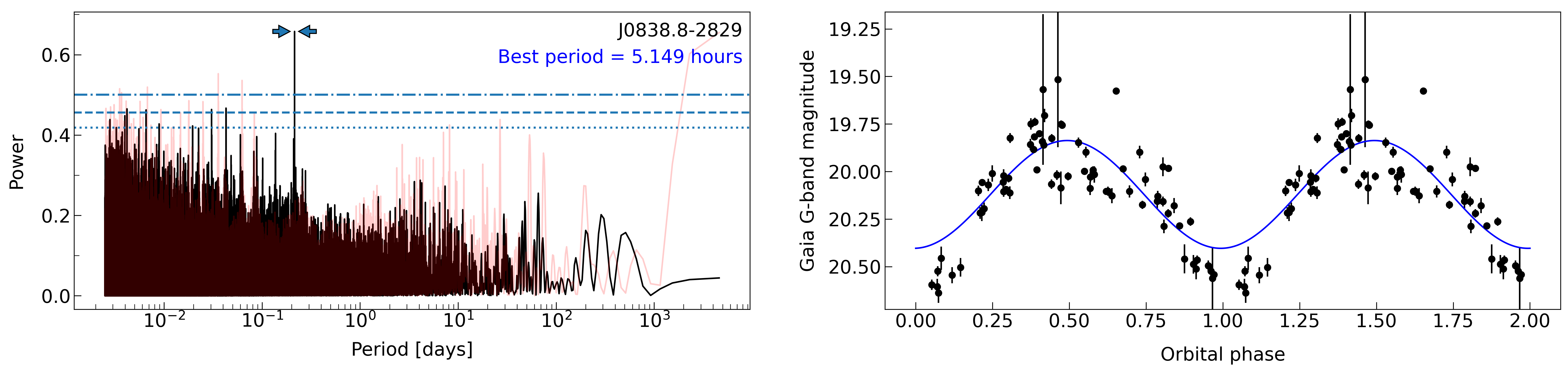}\hfill
    \includegraphics[width=\textwidth]{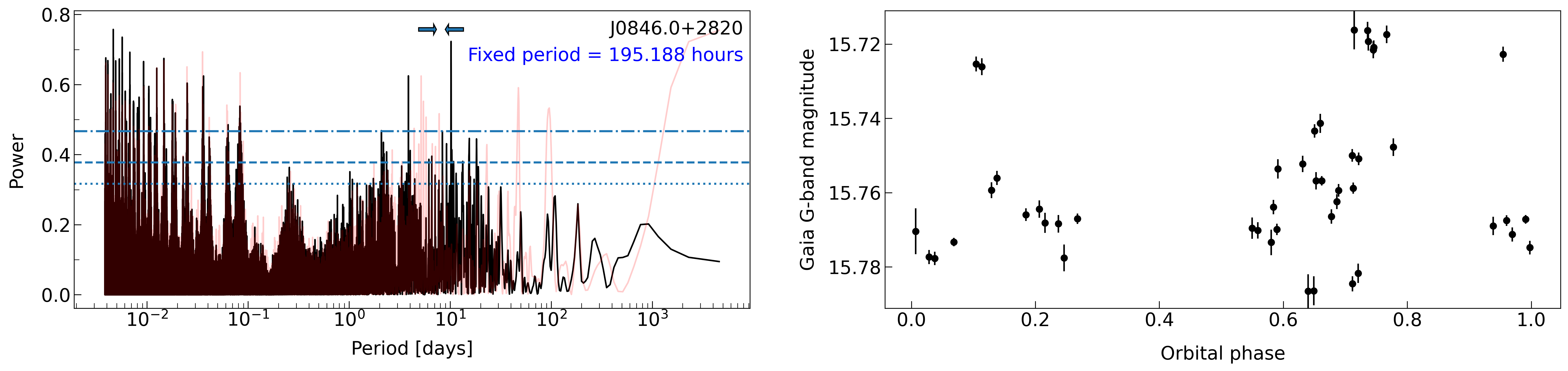}\hfill
    \includegraphics[width=\textwidth]{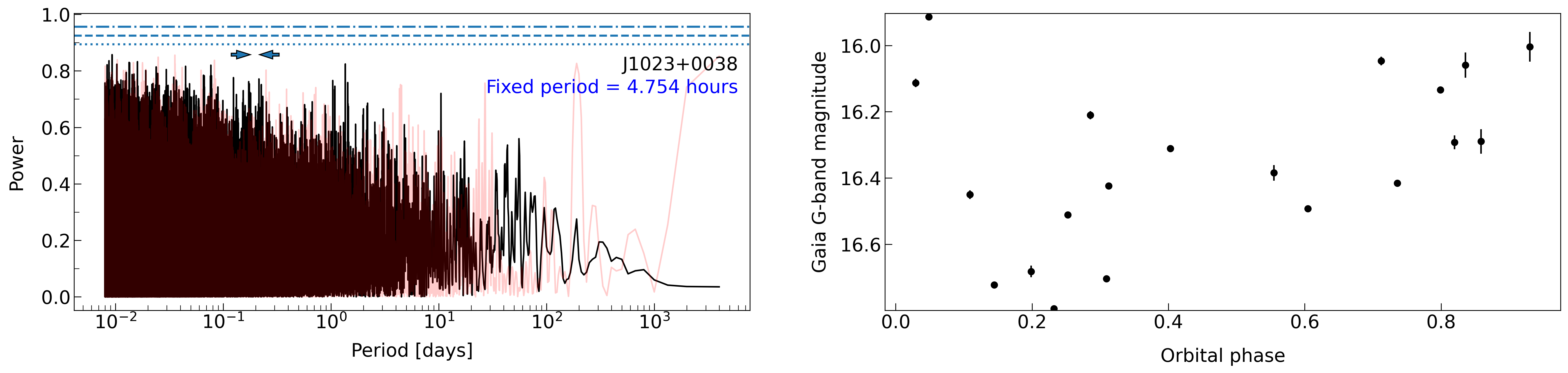}
    \caption{Lomb-Scargle periodograms of all spiders that have \textit{Gaia} epoch photometry available.}\label{fig:foobar1}
\end{figure*}

\begin{figure*}
    \includegraphics[width=\textwidth]{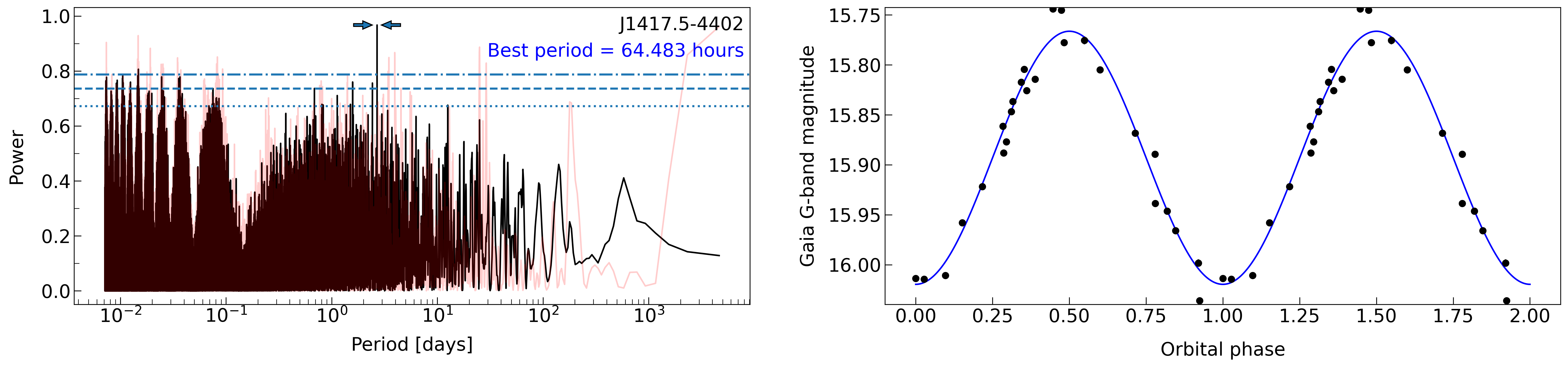}\hfill
    \includegraphics[width=\textwidth]{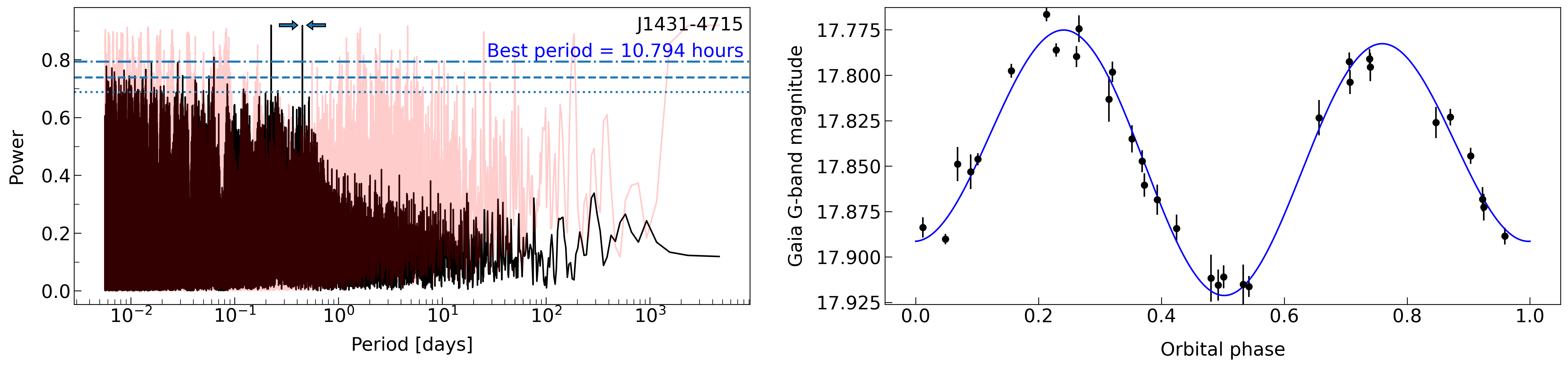}\hfill
    \includegraphics[width=\textwidth]{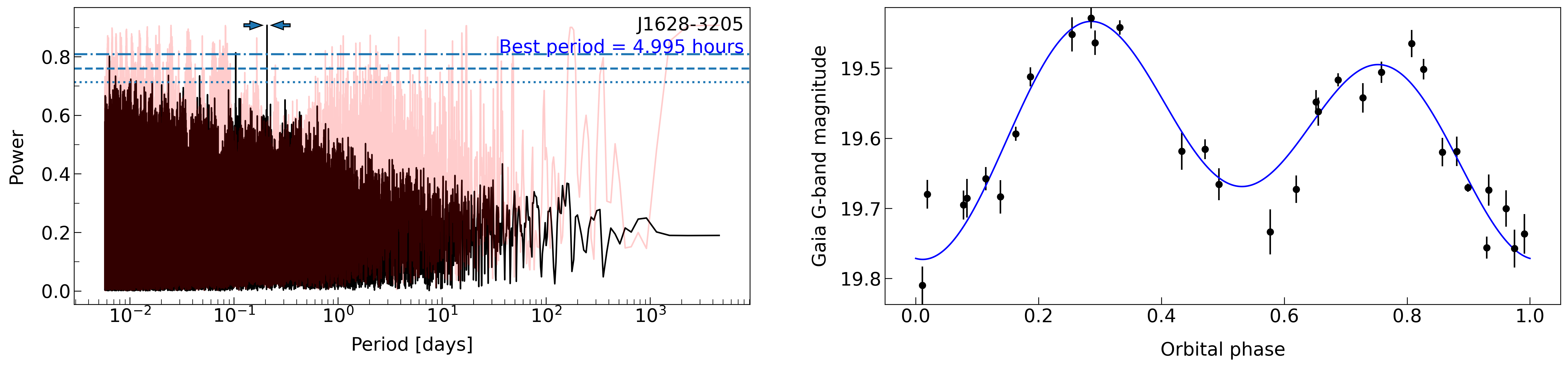}\hfill
    \includegraphics[width=\textwidth]{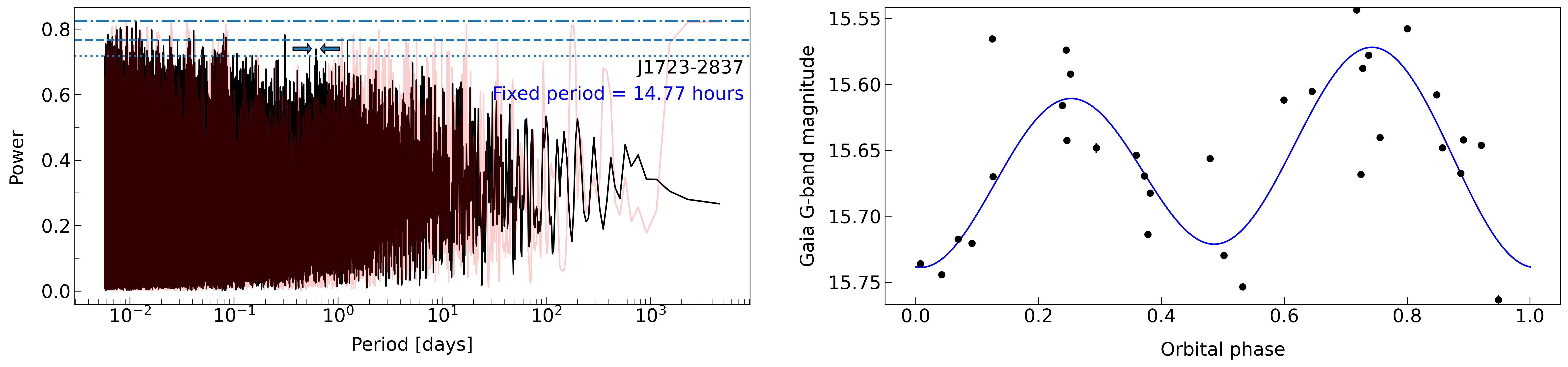}\hfill 
    \includegraphics[width=\textwidth]{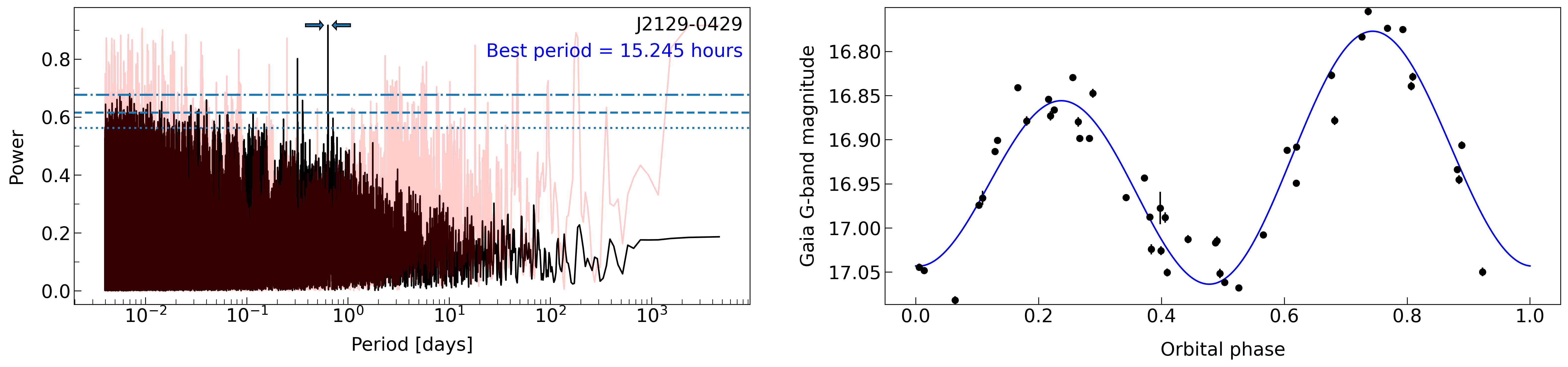}
    \caption{Continued.}\label{fig:foobar2}
\end{figure*}

\begin{figure*}
    \includegraphics[width=\textwidth]{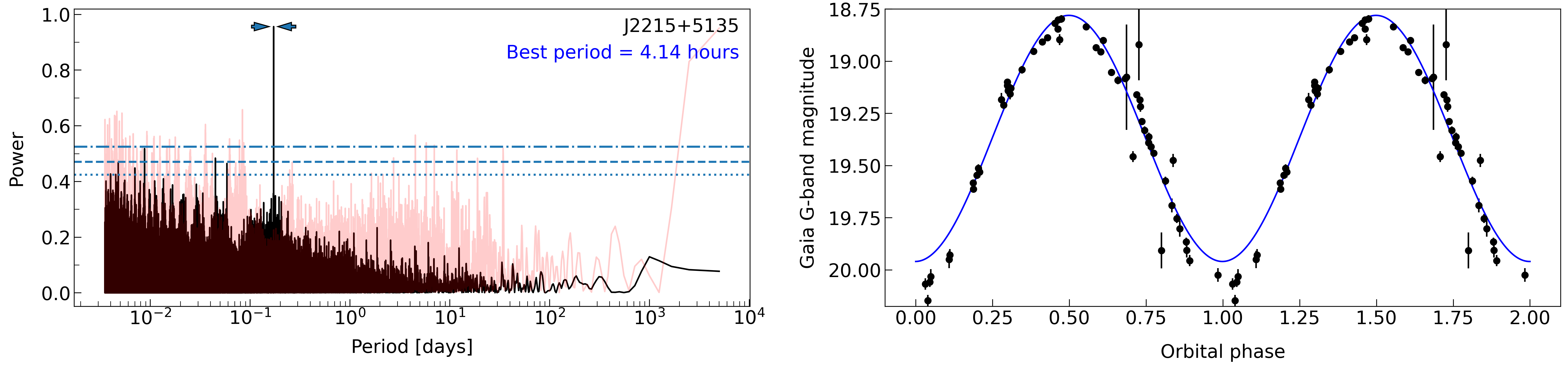}\hfill
    \includegraphics[width=\textwidth]{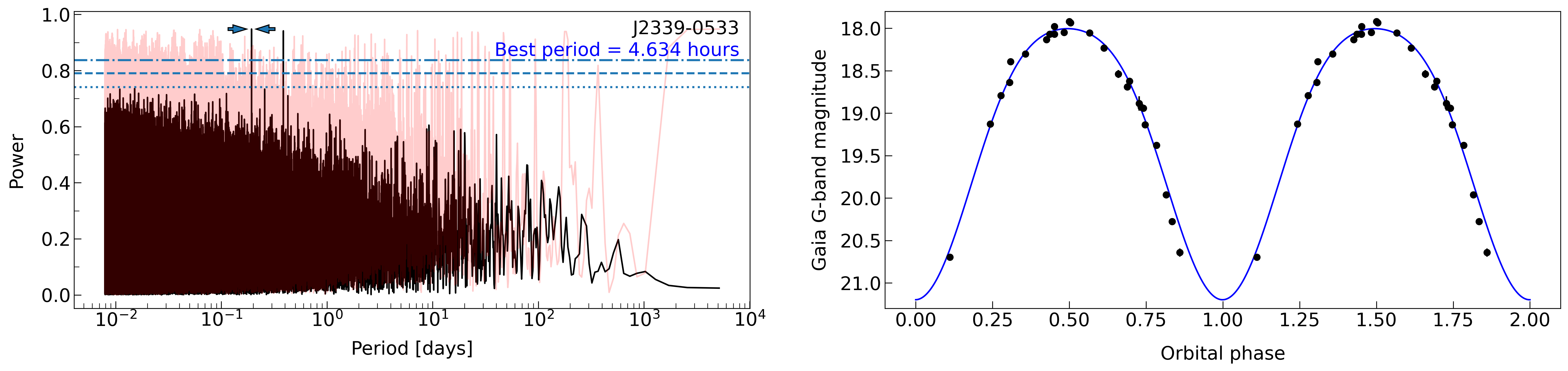}
    \caption{Continued.}\label{fig:foobar3}
\end{figure*}



\bsp	
\label{lastpage}
\end{document}